\PassOptionsToPackage{table}{xcolor}
\PassOptionsToPackage{noend}{algpseudocode}
\documentclass[10pt,journal,compsoc]{IEEEtran}

\usepackage[nocompress]{cite}
\usepackage{ntheorem}
\usepackage{amsmath,amssymb,amsfonts,mathtools}
\usepackage{graphicx}
\usepackage{xcolor}
\usepackage{adjustbox}
\usepackage{multirow,dcolumn,makecell,booktabs,array}
\usepackage{bm,mathrsfs}
\usepackage[caption=false,font=normalsize,labelfont=sf,textfont=sf]{subfig}
\usepackage{algorithm}
\usepackage{algpseudocode}
\usepackage{float,rotating,longtable,lscape}
\usepackage{tablefootnote}
\usepackage{verbatim,ltxtable}
\usepackage{chngcntr}
\usepackage{hyperref}
\hypersetup{
    colorlinks=true,
    linkcolor=cgblue,
    citecolor=cgblue,
    urlcolor=cgblue,
}

\definecolor{cgblue}{HTML}{2B6CB0}
\definecolor{cgorange}{HTML}{C66A1B}
\definecolor{cggreen}{HTML}{2F855A}
\definecolor{cgred}{HTML}{B33A3A}
\definecolor{cgpurple}{HTML}{6750A4}
\definecolor{cggray}{HTML}{5F6B7A}
\definecolor{cglightblue}{HTML}{EAF2FB}
\definecolor{cglightorange}{HTML}{FDF1E6}
\definecolor{cglightgreen}{HTML}{EAF6EF}
\definecolor{cglightpurple}{HTML}{F0ECFA}
\definecolor{cglightgray}{HTML}{F3F5F7}

\graphicspath{{./}{pics/}}

\newcommand{\argmin}{\operatornamewithlimits{arg\,min}}

\definecolor{LightPink}{RGB}{255,228,225}
\DeclareUnicodeCharacter{2011}{\nobreakdash-}
\DeclareUnicodeCharacter{2191}{\ensuremath{\uparrow}}
\DeclareUnicodeCharacter{2193}{\ensuremath{\downarrow}}

\theoremstyle{plain}
\theoremheaderfont{\hskip\parindent\normalfont\scshape}
\theorembodyfont{\itshape}
\theoremseparator{.}
\newtheorem{Lem}{Lemma}
\newtheorem{Mth}{Theorem}
\newtheorem{Cor}{Corollary}

\newtheorem{Def}{Definition}

\theoremheaderfont{\hskip\parindent\normalfont\itshape}
\theorembodyfont{\normalfont}
\newtheorem{Rem}{Remark}
\newtheorem{Con}{Condition}
\newtheorem*{Constar}{Condition}

\numberwithin{equation}{section}
\usepackage{etoolbox}
\newcommand{\TPAMImathsize}{9}
\newcommand{\TPAMIcompactmath}{%
  \fontsize{\TPAMImathsize}{10.2}\selectfont
  \setlength{\abovedisplayskip}{4pt plus 1pt minus 1pt}%
  \setlength{\belowdisplayskip}{4pt plus 1pt minus 1pt}%
  \setlength{\abovedisplayshortskip}{2pt plus 1pt}%
  \setlength{\belowdisplayshortskip}{3pt plus 1pt minus 1pt}%
  \setlength{\jot}{2pt}}
\BeforeBeginEnvironment{equation}{\begingroup\TPAMIcompactmath}
\AfterEndEnvironment{equation}{\endgroup}
\BeforeBeginEnvironment{equation*}{\begingroup\TPAMIcompactmath}
\AfterEndEnvironment{equation*}{\endgroup}
\BeforeBeginEnvironment{align}{\begingroup\TPAMIcompactmath}
\AfterEndEnvironment{align}{\endgroup}
\BeforeBeginEnvironment{align*}{\begingroup\TPAMIcompactmath}
\AfterEndEnvironment{align*}{\endgroup}
\BeforeBeginEnvironment{gather}{\begingroup\TPAMIcompactmath}
\AfterEndEnvironment{gather}{\endgroup}
\BeforeBeginEnvironment{multline}{\begingroup\TPAMIcompactmath}
\AfterEndEnvironment{multline}{\endgroup}

\begin{document}
\raggedbottom

\title{Model-agnostic information transfer and fusion for classification with label noise}

\author{Guojun~Zhu, Sanguo~Zhang, and Mingyang~Ren%
\IEEEcompsocitemizethanks{%
\IEEEcompsocthanksitem Guojun Zhu and Sanguo Zhang are with the School of Mathematical Sciences, University of Chinese Academy of Sciences, Beijing 100049, China.
\IEEEcompsocthanksitem Mingyang Ren is with the School of Mathematical Sciences, Shanghai Jiao Tong University, 800 Dongchuan RD, Minhang District, Shanghai, 200240, China. E-mail: mingyangren@sjtu.edu.cn.}}

\IEEEtitleabstractindextext{%
\begin{abstract}
Label noise presents a fundamental challenge in modern machine learning, especially when large-scale datasets are generated via automated processes. An increasingly common and important data paradigm, particularly in domains like medical imaging, involves learning from a large dataset with coarse, noisy labels supplemented by a small, expert-verified, clean dataset. This setting constitutes a typical information transfer and fusion problem. However, the significant distribution shift between the noisy and clean data violates the core overall parametric similarity assumptions of existing statistical transfer learning methods, while their reliance on parametric models is ill-suited for complex data like images. To address these limitations, this paper develops a generic model-agnostic nonparametric framework for binary classification with label noise, which applies to a broad class of classifiers. Our approach leverages the small clean dataset to ``purify'' the large noisy one and carefully manages the remaining ambiguous samples. This framework is underpinned by a rigorous statistical theory. Its empirical performance is demonstrated through simulations and a real-world application to medical image analysis for pneumonia diagnosis.
\end{abstract}

\begin{IEEEkeywords}
classification, distribution shift, mislabel, model-agnostic, transfer learning
\end{IEEEkeywords}}
\maketitle
\IEEEdisplaynontitleabstractindextext

\IEEEraisesectionheading{\section{Introduction}}
The explosion of data across scientific and industrial domains, from biomedical imaging and financial risk control to social media analytics, has fundamentally transformed modern statistical learning. However, this abundance of raw data stands in stark contrast to the scarcity of high-quality annotations. Obtaining accurate labels often demands significant expertise and resources; for instance, precise diagnosis in medical imaging requires time-consuming evaluation by specialized radiologists \cite{pesapane2018artificial}. Consequently, large-scale datasets collected in practice are frequently contaminated by label noise, where the observed labels deviate substantially from the underlying true labels. This noise is far from benign: it systematically misleads model training, leading to severe degradation in classifier performance and generalization. Even cutting-edge techniques, including large language models, are vulnerable, where noisy labels can propagate errors and contribute to issues like ``hallucination''  \cite{cheng2025exploring}. Addressing label noise is thus paramount for reliable inference. The statistics and machine learning communities have developed numerous methods to combat label noise, ranging from robust loss functions \cite{iscen2022learning} and sample reweighting  \cite{wang2017multiclass} to noise transition matrix estimation \cite{yang2022estimating, zhang2024cognition} and label correction \cite{hendrycks2018using,cheng2020learning}, with comprehensive reviews available \cite{song2022learning}.
PCSE estimates clean class means and covariances from noisy data \cite{pcse2025}.
Knockoffs-SPR uses knockoff filters to control the false selection rate in clean sample selection \cite{knockoffsspr}.
While effective under specific assumptions, these approaches typically operate within the confines of a single, noisy dataset.

\begin{figure}[htbp]
    \centering
    \includegraphics[width=\linewidth]{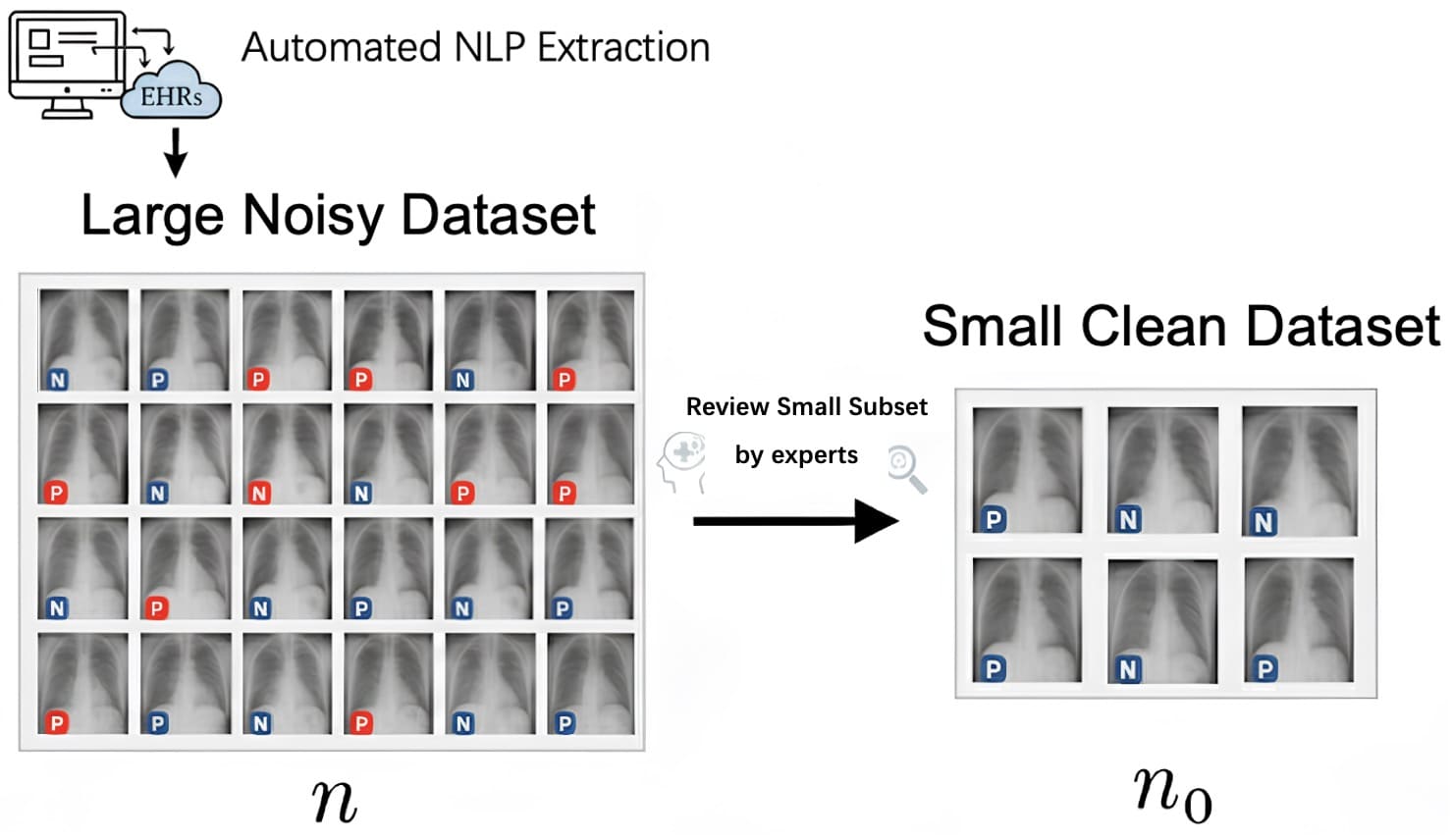}
    \caption{An illustration of the real data motivation. The red and blue icons represent incorrect and correct labels, respectively.}
    \label{fig:motivation}
\end{figure}

Motivated by practical data collection paradigms, we shift focus to a prevalent yet under-explored scenario: the co-existence of a large-scale but noisy dataset alongside a small-scale but meticulously curated clean dataset. Taking medical diagnosis as an example: 
Initial large-scale data collection often relies on automated pipelines, where labels are frequently extracted from Electronic Health Records (EHRs) using Natural Language Processing (NLP) techniques. While enabling the rapid assembly of massive datasets, these methods are inherently prone to inaccuracies, introducing significant label noise. Conversely, establishing a ground truth requires meticulous, resource-intensive review by expert radiologist panels. Such gold-standard annotation is prohibitively expensive and time-consuming to apply at scale. Consequently, to ensure quality, a small subset is typically verified by experts, yielding a small yet reliable clean dataset \cite{bernhardt2022active}, as shown in Figure \ref{fig:motivation}.
This data generation structure is ubiquitous: financial institutions generate massive volumes of automated fraud alerts, while expert investigators can only verify a small fraction; content moderation platforms rely on vast crowdsourced noisy labels, supplemented by limited expert audits. While methods leveraging small clean datasets to address label noise have also been studied \cite{hendrycks2018using,yu2023delving}, they often lack rigorous theoretical guarantees. This dual-data setting naturally constitutes an information transfer and fusion problem, where the core objective is to learn a better classifier by leveraging both the large noisy dataset and the small clean dataset, specifically to understand and mitigate the label noise process.
Statistical transfer learning has seen significant advances recently \cite{gao2022review}, particularly in contexts like classification \cite{cai2021transfer}, linear regression \cite{li2022transfer,tian2023transfer}, and graphical models \cite{li2023transfer}. 
However, most statistical transfer learning methods are developed in model-specific settings, and therefore typically adopt feature-level transfer strategies or require stringent parametric model specifications to derive theoretical guarantees and algorithms. Such methods face critical limitations when confronted with label noise, primarily due to their reliance on strong model assumptions and explicit overall similarity conditions.

While powerful within their domain, this model dependence in the existing statistical transfer methods severely limits their applicability to complex, high-order data like images or text, where flexible non-parametric models, such as deep neural networks, are essential. This necessitates the development of generic model-agnostic frameworks compatible with a broad class of complex classifiers.
More crucially for label noise, standard statistical transfer learning heavily relies on explicit similarity assumptions between domains, most commonly requiring the parameters, such as regression coefficients, distribution parameters, of the source and target models to be sufficiently ``close''  in some norm \cite{li2022transfer,tian2023learning}. This assumption is fundamentally mismatched to the label noise problem. The core challenge here is not transferring knowledge across inherently different tasks or feature distributions, but rather utilizing the clean data to characterize and correct the noise mechanism of the same but corrupted underlying task in the source noisy data. The distributions, especially the posterior distributions, between the clean and noisy datasets inherently exhibit significant drift due to the noise, contradicting the premise of overall parameter similarity. Pre-defining a meaningful and effective ``similarity''  metric, especially one capturing overall parameter similarity, for this specific correction task is both difficult and unnatural.  
This motivates a framework that substantially relaxes explicit overall similarity requirements and instead learns the relevant nonparametric transfer structure from the data itself, without imposing restrictive a priori distributional relationships. Our work relates to model-agnostic transfer learning like RECaST \cite{hickey2024transfer}, which calibrates a source model to the target population for uncertainty quantification. Unlike RECaST, we focus on classification with label noise by leveraging small clean data, not merely adapting a source predictor. Thus, the two approaches are complementary: RECaST offers calibration-based transfer, while ours provides a transfer-and-fusion mechanism tailored to label noise.

The core challenge of this work is therefore: How can we effectively leverage a small clean dataset to ``purify''  a large noisy dataset for classification, without relying on parametric model specifications or restrictive overall similarity assumptions between domains?
This challenge manifests in two key aspects. First, the process hinges on accurately estimating the noise characteristics. Given the inherent estimation error due to the small sample size of the clean dataset, a critical question is how to prevent this initial error from being catastrophically amplified when applied to the large noisy dataset, potentially leading to negative transfer where performance degrades compared to using only the clean data. Second, a portion of noisy data points will inevitably lie near the decision boundary, so-called boundary samples, making their true labels ambiguous. Simply discarding these points wastes valuable information, while naively trusting their noisy labels introduces bias. Effectively navigating this quality-quantity trade-off for boundary samples is paramount for achieving high final model performance.

To address these challenges, we propose a generic model-agnostic nonparametric framework for information transfer and fusion for classification with label noise.
Our method operates in two key stages.
First, we design a theoretically guaranteed Bayes Optimal (BO) data extraction module, which can be combined with a broad class of classifiers, with neural networks serving as a primary implementation in this paper.
Crucially, it incorporates a statistically justified safety threshold to identify a high-confidence subset of the noisy data whose corrected BO labels are highly reliable with guaranteed accuracy. Second, for the remaining ambiguous boundary samples, we propose two principled strategies: an active learning approach that selects a small number of the most informative boundary samples for expert annotation, maximizing performance gain per labeling cost; and a pseudo-tagging strategy employing an elegant assignment mechanism that utilizes these samples while preserving uncertainty about their labels.
This two-stage framework effectively transfers and fuses information about the noise structure from the ``small-but-clean''  dataset to the ``large-but-noisy''  dataset, ultimately constructing a large-scale, high-quality training set for the final classifier.
It shares a similar spirit with sample-level transfer, in that useful information is borrowed from abundant but imperfect data to improve learning with limited high-quality data. 
This perspective is also related to noisy-label learning methods that improve robustness by selecting, reweighting, or relabeling reliable examples, such as Co-teaching \cite{han2018co}, DivideMix \cite{li2020dividemix}, Confident Learning \cite{northcutt2021confident}, and meta-learning that use clean data to guide training \cite{li2019learning}. In contrast, our formulation explicitly targets two label-noise mechanisms, accommodates very general covariate shift, and uses the audited clean data to statistically certify BO data with a formal theoretical guarantee. Thus, purification is treated as an explicit information-fusion step, rather than only as an empirical robust training strategy.

Our primary contributions are multi-fold. First, we propose a unified, model-agnostic framework for information transfer and fusion under label noise. It can be combined with a broad class of complex learners, substantially relaxes explicit overall similarity requirements between the clean and noisy data distributions, and handles both class-dependent and instance-dependent noise mechanisms in a unified framework.
Second, we develop and theoretically analyze a novel Bayes Optimal data extraction method with a safety threshold, rigorously characterizing its extraction rate and problem complexity. For boundary samples, we introduce two efficient strategies, offering users a flexible choice between performance and annotation cost.
Third, we analyze BO-label extraction and the excess risk of the final classifier under abstract estimation error conditions. The resulting bounds show how the audited sample size and the choice of boundary-sample tagging affect classification error. We then derive explicit convergence rates for neural-network-based implementations.
Fourth, we demonstrate significant empirical improvements over state-of-the-art methods on the diagnosis of pneumonia based on medical imaging and provide powerful practical tools and ideas for this field.

\section{Data and Model Set-up}\label{section 2}

The real-world chest X-ray dataset analyzed in this paper exemplifies a common data paradigm prevalent in modern machine learning, particularly within medical imaging analysis, with two distinct but related datasets. First, a large dataset of size $n$ with noisy labels acquired through scalable automated methods. Second, a much smaller subset of size $n_0 \ll n$ with clean, ground-truth labels. These clean labels are the result of a meticulous expert verification process. The goal is to effectively combine these two datasets to train a highly accurate classifier that can generalize to the clean dataset. 

To formalize this, let $\bm{X} \in \mathcal{X} \subseteq \mathbb{R}^d$ be the $d$-dimensional feature vector, $Y \in \{-1, 1\}$ the underlying true label, and $\widetilde{Y} \in \{-1, 1\}$ the observed noisy label. From a statistical perspective, the clean data $(\bm{X}, Y)$ follows a distribution $D$, while the noisy data $(\bm{X}, \widetilde{Y})$ follows $D_\rho$. Our available data thus consists of a \textbf{large noisy dataset}, $D_{\rho,n} = \left\{\left(\bm{x}_i, \tilde{y}_i\right)\right\}_{i=1}^n$ drawn i.i.d. from $D_\rho$, and a \textbf{small corrected clean dataset}, 
$D_{n_0} = \left\{\left(\bm{x}_{0 i}, y_{0 i}, \tilde{y}_{0 i}\right)\right\}_{i=1}^{n_0}$, where $(\bm{x}_{0 i}, y_{0 i})$ are drawn i.i.d. from $D$. For each audited observation $\bm{x}_{0 i}$ we additionally observe its corresponding noisy label $\tilde{y}_{0 i}$. We denote the corresponding posterior probabilities as $\eta(\bm{x}) = P_D(Y=1|\bm{X}=\bm{x})$ and $\eta_{\rho}(\bm{x}) = P_{D_\rho}(\widetilde{Y}=1|\bm{X}=\bm{x})$, where $P_D(\cdot)$ represents the probability under distribution $D$. We consider the class probabilities are bounded away from the extremes of 0 or 1, that is, $\min\{P_D(Y=1), P_D(Y=-1), P_{D_\rho}(\tilde{Y}=1), P_{D_\rho}(\tilde{Y}=-1)\}\geq C_y$ for a positive constant $0<C_y<1$.
The $D_{n_0}$ and $D_{\rho,n}$ samples may have different covariate distributions, subject to the common-support, conditional-invariance, and density regularity conditions described in Section~\ref{section 4.1}.

\subsection{Label Noise Mechanism} 
A critical component in the classification with label noise is the formal characterization of the noise mechanism. We consider two widely-studied noise mechanisms. The first is class-dependent noise, which typically originates from systematic confusion between classes. This setting is common in tasks like crowdsourced image classification, where confusion is driven by inherent class similarity. For example, annotators may mislabel ``wolves''  as ``huskies''  at a relatively constant rate, no matter the background or angle of the particular image.
The second is instance-dependent noise, which assumes that the generation of noise is linked to the sample's individual features. This is common in scenarios where instance quality or difficulty dictates label accuracy, with medical image diagnosis being a canonical example. A blurry X-ray is more prone to mislabeling than a clear one \cite{bernhardt2022active}. We define the noise rates as the conditional probabilities $\rho_{+}(\bm{x})=P_D(\widetilde{Y}=-1 \mid Y=1, \bm{X}=\bm{x})$ and $\rho_{-}(\bm{x})=P_D(\widetilde{Y}=1 \mid Y=-1, \bm{X}=\bm{x})$. The same functions $\rho_{+}(\bm{x})$ and $\rho_{-}(\bm{x})$ apply under both $D$ and $D_\rho$, without loss of generality.
\begin{Con}\label{Con 1}
    {\bf (Noise Mechanism).} Assume one of the conditions holds:
    \begin{itemize}
        \item[(a)] \textbf{Class-dependent.} The noise rates $\rho_{-}(\bm{x}) \equiv \rho_{-}$ and $\rho_{+}(\bm{x}) \equiv \rho_{+}$, where $\rho_{-}$ and $\rho_{+}$ are constants independent of $\bm{x}$, and $\rho_{-} + \rho_{+} < 1$.
        \item[(b)] \textbf{Instance-dependent.} $\rho_{-}(\bm{x})$ and $\rho_{+}(\bm{x})$ are nontrivial functions of the instance $\bm{x}$, and $\rho_{-}(\bm{x}) + \rho_{+}(\bm{x}) < 1$ for all $\bm{x} \in \mathcal{X}$.
    \end{itemize}
\end{Con}
The condition $\rho_{-}(\bm{x})+\rho_{+}(\bm{x})<1$ requires the noisy label to remain informative about the true label conditional on $\bm{X}$, rather than merely agreeing with it on average \cite{cheng2020learning}. Our framework accommodates both class-dependent and instance-dependent noise, with different theoretical rates. Beyond the noise mechanism, classification difficulty also depends on the probability mass near the decision boundary. We therefore adopt the standard Tsybakov margin condition.
\begin{Con}\label{Con 2}
    {\bf (Tsybakov Margin).} There exist constants $C_q, q > 0$, such that for all $t\geq0$, we have $P_D( |\eta(\bm{X}) - \frac{1}{2}| \leqslant t)  \leqslant  C_q t^q$.
\end{Con}
This condition has a simple practical interpretation: when the two classes are meaningfully distinguishable, such as pneumonia versus non-pneumonia, only a limited fraction of samples are expected to lie very close to the decision boundary.
This assumption is widely used in the literature to derive optimal classification rates \cite{cai2021transfer,kim2021fast}. For completeness, we also consider another margin condition, the Massart Margin Condition, and provide a full theoretical analysis for it in Supplementary~\ref{Appendix C}.

\subsection{Bayes Optimal data}\label{BO-def}
The ultimate goal in binary classification is to find a classifier that minimizes the probability of making an incorrect prediction. This objective is formally captured by the 0-1 risk. 
For any given classifier $h: \mathcal{X} \rightarrow \{-1, 1 \}$, its 0-1 risk under the clean distribution $D$ is defined as $R^{0/1}_{D}(h) = \mathbb{E}_D [ \mathbb{I} (Y h(\bm{X}) < 0)]$, where $\mathbb{E}[\cdot]$ represents the expectation and $\mathbb{I}(\cdot)$ represents the indicator function. This risk is minimized by the well-known Bayes optimal classifier, $h^o_D(\bm{x}) = \operatorname{sign}(\eta(\bm{x}) - 1/2)$, where $\operatorname{sign}(t)=2\mathbb{I}(t\geq0)-1$. For a real-valued score $f$, we use $R_D^{0/1}(f):=R_D^{0/1}(\operatorname{sign}f)$, so a zero score predicts $+1$.

To facilitate our theoretical development, we introduce the concept of Bayes Optimal (BO) data \cite{cheng2020learning,yang2022estimating}, denoted by $(\bm{X}, Y^{\star})$, which is a theoretical construct where the label $Y^{\star}$ is deterministically assigned by the Bayes optimal classifier itself: $Y^{\star} = h^o_D(\bm{X})$. Let $D^{\star}$ be the distribution of the BO data $(\bm{X}, Y^{\star})$. The posterior probability for this distribution is then simply $\eta^{\star}(\bm{x}) = P_{D^\star}(Y^{\star}=1 \mid \bm{X}=\bm{x}) = \mathbb{I}[\eta(\bm{x}) \geqslant 1/2]$. 
The introduction of BO data is inspired by two observations. Firstly, BO data has the same good properties as clean data, that is, the Bayes optimal classifier for $D^{\star}$ is identical to that for $D$. By noting that $h^o_{D^\star}(\bm{x}) = 2\mathbb{I} [ \eta^\star(\bm{x}) \geqslant \frac{1}{2}]-1 = 2\mathbb{I} [ \mathbb{I}( h^o_{D}(\bm{x}) = 1 ) \geqslant \frac{1}{2}]-1 = h^o_{D}(\bm{x})$ for any $\bm{x} \in \mathcal{X}$, we can have the following claim: {\bf the Bayes optimal classifier under $D^\star$ coincides with that under $D$.} The significance of the claim is that if we can learn a classifier $\widehat{h}_m$ from $m$ BO samples $\left\{\left(\bm{x}_i^{\star}, y_i^\star\right)\right\}_{i=1}^m$ that converges to $h_{D^{\star}}^o$, then it also converges to the true Bayes optimal classifier $h_D^o$. Formally,
if $\widehat{h}_m \xrightarrow{m \rightarrow \infty} h_{D^{\star}}^o$, then $\widehat{h}_m \xrightarrow{m \rightarrow \infty} h_D^o$. Secondly, without additional investment, underlying clean labels of the larger noisy dataset are inaccessible; however, BO labels can be obtained with theoretical guarantees and high confidence solely using a carefully designed method according to the interaction of the two datasets. This constitutes the focal point of the methodological exploration in the subsequent sections of this paper.

\section{Methodology}\label{section 3}
We use the small audited dataset to estimate the label-noise mechanism and the large noisy dataset to estimate the noisy posterior probability. Together, these estimates determine a shifted threshold for BO-label extraction in Section \ref{section 3.1}. For the remaining boundary samples, Pro-AT queries expert labels and Pro-PT resamples pseudo-labels during final classifier training, as described in Section \ref{section 3.2}. Both methods combine the extracted BO samples with the tagged boundary samples to train a final classifier. Section \ref{section 3.3} summarizes the computational procedure. The procedure accommodates different classification learners, whose estimation rates enter the risk bounds in Section \ref{section 4}.

\subsection{Generic Extraction Method of Bayes Optimal data}\label{section 3.1}

Directly estimating the clean posterior probability $\eta(\bm{x})$ from the small clean dataset of size $n_0$ can be statistically challenging, often leading to unreliable results. In contrast, the large noisy dataset of size $n$ provides a much more stable basis for estimating the noisy posterior probability $\eta_{\rho}(\bm{x})$. Therefore, our core strategy is to avoid estimating $\eta(\bm{x})$ directly. Instead, we first establish a relationship between the $\eta(\bm{x})$ and $\eta_{\rho}(\bm{x}),\rho_+(\bm{x}), \rho_-(\bm{x})$. For any $\bm{x} \in \mathcal{X}$, we have
\begingroup\renewcommand{\TPAMImathsize}{8.5}
\begin{equation}\label{eta_null}
    \begin{aligned}
	\eta_{\rho}(\bm{x})
	& = P_D(\widetilde{Y}=1|Y = 1, \bm{X}=\bm{x})\,P_D(Y = 1 | \bm{X}=\bm{x})\\
        &\quad + P_D(\widetilde{Y}=1|Y = -1, \bm{X}=\bm{x})P_D(Y = -1 | \bm{X}=\bm{x}) \\
	& = [1 - \rho_{+}(\bm{x}) - \rho_{-}(\bm{x})] \eta(\bm{x}) + \rho_{-}(\bm{x}).
    \end{aligned} 
\end{equation}
\endgroup
Equation \eqref{eta_null} demonstrates how we can estimate the decision boundary of $\eta(\bm{x})$ by estimating $\eta_{\rho}(\bm{x})$, $\rho_{+}(\bm{x})$ and $\rho_{-}(\bm{x})$. By \eqref{eta_null}, if $\eta(\bm{x}) \geqslant \frac{1}{2}$, we have $\eta_\rho(\bm{x}) \geqslant \frac{1}{2}\left[1-\rho_+(\bm{x})-\rho_-(\bm{x})\right]+\rho_-(\bm{x})=\frac{1}{2}-\frac{1}{2}\left(\rho_+(\bm{x})-\rho_-(\bm{x})\right)$, and its contrapositive:
\begin{equation*}
\begin{aligned}
{}&\eta_\rho(\bm{x})<\frac{1}{2}-\frac{1}{2}\left(\rho_+(\bm{x})-\rho_-(\bm{x})\right)\\
&\quad\Longrightarrow\eta(\bm{x})<\tfrac{1}{2}\quad\Longrightarrow\text{BO label } y^\star=-1.
\end{aligned}
\end{equation*}
Similarly, $\eta_\rho(\bm{x})>\frac{1}{2}-\frac{1}{2}\left(\rho_+(\bm{x})-\rho_-(\bm{x})\right) \Longrightarrow \eta(\bm{x})>\frac{1}{2} \Longrightarrow \text{BO label } y^\star=1$. However, since the estimation of $\eta_{\rho}(\bm{x})$, $\rho_{+}(\bm{x})$ and $\rho_{-}(\bm{x})$ inherently introduces errors, a ``security margin''  $\tau > 0$ is introduced to guard against misclassifications when extracting BO data. We only assign a BO label if $\widehat{\eta}_{\rho}(\bm{x})$ is sufficiently far from the estimated decision threshold, ensuring that for any $\bm{x} \in \mathcal{X}$,
\begin{equation}\label{drift}
    \begin{aligned}
        {}& \widehat{\eta}_{\rho}(\bm{x}) > \tfrac{1}{2} - \tfrac{1}{2} (\widehat{\rho}_{+}(\bm{x}) - \widehat{\rho}_{-}(\bm{x})) + \tau \\ &\quad\Longrightarrow \eta_\rho(\bm{x})>\tfrac{1}{2}-\tfrac{1}{2}\left(\rho_+(\bm{x})-\rho_-(\bm{x})\right)\quad\Longrightarrow \eta(\bm{x}) > \tfrac{1}{2}, \\
        & \widehat{\eta}_{\rho}(\bm{x}) < \tfrac{1}{2} - \tfrac{1}{2} (\widehat{\rho}_{+}(\bm{x}) - \widehat{\rho}_{-}(\bm{x})) - \tau \\ &\quad\Longrightarrow \eta_\rho(\bm{x})<\tfrac{1}{2}-\tfrac{1}{2}\left(\rho_+(\bm{x})-\rho_-(\bm{x})\right) \quad\Longrightarrow\eta(\bm{x}) < \tfrac{1}{2}.
    \end{aligned} 
\end{equation}
The implications in \eqref{drift} are guaranteed to hold provided that the security margin $\tau$ is chosen to be sufficiently large to absorb the estimation errors. This brings us to the practical question of how to obtain the estimators $\widehat{\eta}_{\rho}(\bm{x})$, $\widehat{\rho}_{+}(\bm{x})$, and $\widehat{\rho}_{-}(\bm{x})$. 
More precisely, $\widehat{\eta}_{\rho}(\bm{x})$ can be obtained based on $\{(\bm{x}_{i},I(\widetilde{y}_{i}=1))\}_{i=1}^n$. For the instance-dependent noise probabilities, $\widehat{\rho}_{+}(\bm{x})$ and $\widehat{\rho}_{-}(\bm{x})$ can be obtained based on $\{(\bm{x}_{0i},I(\widetilde{y}_{0i}\neq y_{0i})):y_{0i}=1\}_{i=1}^{n_0}$ and $\{(\bm{x}_{0i},I(\widetilde{y}_{0i}\neq y_{0i})):y_{0i}=-1\}_{i=1}^{n_0}$, respectively. In the class-dependent case, they can simply be estimated by $\widehat{\rho}_{+}=\frac{\sum_{i=1}^{n_0}I(y_{0i}=1,\widetilde{y}_{0i}\neq y_{0i})}{\sum_{i=1}^{n_0}I(y_{0i}=1)}$ and $\widehat{\rho}_{-}=\frac{\sum_{i=1}^{n_0}I(y_{0i}=-1,\widetilde{y}_{0i}\neq y_{0i})}{\sum_{i=1}^{n_0}I(y_{0i}=-1)}$.
The proposed model-agnostic framework only requires access to these estimators together with suitable control of their estimation errors. Accordingly, let $\widehat{\eta}_{\rho}(\bm{x}), \widehat{\rho}_{+}(\bm{x}), \widehat{\rho}_{-}(\bm{x})$ denote generic estimators of $\eta_{\rho}(\bm{x})$, $\rho_{+}(\bm{x})$, and $\rho_{-}(\bm{x})$, respectively.
In the theoretical analysis, we mainly study neural networks and B-splines to provide learner-specific guarantees. In the numerical studies, we instantiate the framework with commonly used flexible classification learners.


Theoretically, to guarantee the derivation in \eqref{drift}, $\tau$ must be chosen to be larger than the supremum of the combined estimation errors $\epsilon_{n_0, n}$, defined as
\begin{equation}\label{epsilon n_0 n}
\begin{aligned}
\epsilon_{n_0,n}={}&\tfrac12\bigl(\sup\nolimits_{\bm{x}\in\mathcal{X}}|\widehat{\rho}_{+}(\bm{x})-\rho_{+}(\bm{x})|\\
&\qquad+\sup\nolimits_{\bm{x}\in\mathcal{X}}|\widehat{\rho}_{-}(\bm{x})-\rho_{-}(\bm{x})|\bigr)\\
&+\sup\nolimits_{\bm{x}\in\mathcal{X}}|\widehat{\eta}_{\rho}(\bm{x})-\eta_{\rho}(\bm{x})|.
\end{aligned}
\end{equation}
An appropriate learner can necessarily drive $\epsilon_{n_0, n}$ to zero, and its order depends on the specific classifier and is discussed in Section \ref{section 4.2}. In practice, $\tau$ can be treated as a hyperparameter and selected via cross-validation, with details in the Algorithm \ref{Algorithm 2} of Section \ref{section 3.3}.

A key question is the number of extracted BO samples, denoted by $m$, that can be extracted. With an appropriately chosen $\tau = C \epsilon_{n_0, n}$ for some sufficiently large constant $C>0$, the samples that are not assigned a BO label are confined to a boundary region where $|\eta(\bm{x}) - 1/2|$ is small, as detailed in \eqref{eq:BO-extraction-chain}.

\begin{figure*}[!t]
\begin{equation}\label{eq:BO-extraction-chain}
    \begin{aligned}
        &\widehat{\eta}_{\rho}(\bm{x})< \tfrac{1}{2}-\tfrac{1}{2}(\widehat{\rho}_+(\bm{x})-\widehat{\rho}_-(\bm{x}))-\tau\\ &\Rightarrow\eta_{\rho}(\bm{x}) < \tfrac{1}{2} - \tfrac{1}{2} (\widehat{\rho}_{+}(\bm{x}) - \widehat{\rho}_{-}(\bm{x})) + \sup_{\bm{x}} | \widehat{\eta}_{\rho}(\bm{x}) - \eta_{\rho}(\bm{x}) | - \tau\\
        &\Rightarrow \eta(\bm{x}) < \tfrac{1}{2} - \frac{2 \tau  + (\rho_{-}(\bm{x}) - \widehat{\rho}_{-}(\bm{x})) - (\rho_{+}(\bm{x}) - \widehat{\rho}_{+}(\bm{x}))  - 2 \sup_{\bm{x}} | \widehat{\eta}_{\rho}(\bm{x}) - \eta_{\rho}(\bm{x}) | }{2(1 - \rho_{+}(\bm{x}) - \rho_{-}(\bm{x}))} \leq \tfrac{1}{2} - C_\tau \tau, \\[3pt]
        &\widehat{\eta}_{\rho}(\bm{x})> \tfrac{1}{2}-\tfrac{1}{2}(\widehat{\rho}_+(\bm{x})-\widehat{\rho}_-(\bm{x}))+\tau\\ &\Rightarrow\eta_{\rho}(\bm{x}) > \tfrac{1}{2} - \tfrac{1}{2} (\widehat{\rho}_{+}(\bm{x}) - \widehat{\rho}_{-}(\bm{x})) - \sup_{\bm{x}} | \widehat{\eta}_{\rho}(\bm{x}) - \eta_{\rho}(\bm{x}) | + \tau\\
        &\Rightarrow \eta(\bm{x}) > \tfrac{1}{2} + \frac{2 \tau  + (\rho_{+}(\bm{x}) - \widehat{\rho}_{+}(\bm{x})) - (\rho_{-}(\bm{x}) - \widehat{\rho}_{-}(\bm{x})) - 2 \sup_{\bm{x}} | \eta_{\rho}(\bm{x}) - \widehat{\eta}_{\rho}(\bm{x}) | }{2(1 - \rho_{+}(\bm{x}) - \rho_{-}(\bm{x}))}  \geq \tfrac{1}{2} + C_\tau \tau.
    \end{aligned} 
\end{equation}
\end{figure*}

It can be shown that these boundary samples satisfy $|\eta(\bm{x}) - 1/2| \leq C_{\tau}\tau$ for some constant $C_{\tau}>0$. Under the Tsybakov Margin Condition \ref{Con 2}, the probability of this boundary region is bounded by $P_D(|\eta(\bm{X}) - 1/2| \leq C_{\tau}\tau) \leq C_q(C_{\tau}\tau)^q = O(\epsilon_{n_0, n}^q)$. Under the sample split specified in Section~\ref{section 4}, let $\mathcal{F}_{\mathrm{fit}}$ contain the nuisance-fitting data. Conditional on $\mathcal{F}_{\mathrm{fit}}$, the expected number of boundary samples satisfies $\mathbb{E}[n-m\mid\mathcal{F}_{\mathrm{fit}}]=nP_D(\bm{X}\in\mathrm{BS}\mid\mathcal{F}_{\mathrm{fit}})=O(n\epsilon_{n_0,n}^q)$ under the preceding boundary-probability bound. It is crucial to emphasize that $\epsilon_{n_0, n}^q$ is a vanishing term tending to zero as the sample sizes $n_0$ and $n$ increase. As a result, the number of extracted BO samples $m$, is of the same asymptotic order as the total sample size $n$, that is $\mathbb{E}[m]/n = 1-o(1)$. In other words, as more data becomes available, our method becomes increasingly efficient, successfully assigning the BO label to almost the entire dataset.

\subsection{Tagging Boundary Samples}\label{section 3.2}

Although the $n-m$ boundary samples $D^b_{n-m}=\{(\bm{x}_i^{b},\tilde{y}_i^{b})\}_{i=1}^{n-m}$ have not been identified with BO labels, they may still contain useful information for classification, and discarding them directly is generally suboptimal. Therefore, it is necessary to assign appropriate labels to the boundary samples in some principled way. In this section, we consider two tagging strategies for handling boundary samples: {\bf P}roposed-{\bf A}rtificial {\bf T}agging ({\bf Pro-AT}) and {\bf P}roposed-{\bf P}seudo {\bf T}agging ({\bf Pro-PT}).

\subsubsection{Artificial Tagging via Expert Querying}
For the boundary samples, our Pro-AT approach employs an expert-querying strategy, in the spirit of active learning. Specifically, this involves incurring annotation cost to query reliable experts for their labels \cite{ren2021survey}. We assume these actively acquired labels are perfect, drawn from the same true conditional distribution $P_D(Y|\bm{X}=\bm{x})$ as the original clean data. These newly labeled boundary samples are then combined with the extracted BO samples. 
This combined dataset is referred to as the mixed data $D_n^{\mathrm{AT}}:=\{(\bm{x}_i, y_i^{\mathrm{AT}})\}_{i=1}^n$ with underlying distribution $D^{\mathrm{AT}}$, where the marginal distribution $P_{D^{\mathrm{AT}}}(\bm{x})$ remains unchanged. We then focus on $P_{D^{\mathrm{AT}}}(y^{\mathrm{AT}}=1 \mid \bm{X}=\bm{x}) := \eta^{\mathrm{AT}}(\bm{x};\eta,\tau,\widehat{\eta}_\rho,\widehat{\rho}_{+},\widehat{\rho}_{-})$. To avoid ambiguity, we abbreviate $\eta^{\mathrm{AT}}(\bm{x};\eta,\tau,\widehat{\eta}_\rho,\widehat{\rho}_{+},\widehat{\rho}_{-})$ as $\eta^{\mathrm{AT}}(\bm{x})$, and define
\begingroup\renewcommand{\TPAMImathsize}{8.5}
\begin{equation*}
    \begin{aligned}{}&\eta^{\mathrm{AT}}(\bm{x})={}\mathbb{I}\left(\widehat{\eta}_\rho(\bm{x})>\tfrac{1}{2}-\tfrac{1}{2}\left(\widehat{\rho}_{+}(\bm{x})-\widehat{\rho}_{-}(\bm{x})\right)+\tau\right)\\ &\quad+\mathbb{I}\left(\left|\widehat{\eta}_\rho(\bm{x})-\left(\tfrac{1}{2}-\tfrac{1}{2}\left(\widehat{\rho}_{+}(\bm{x})-\widehat{\rho}_{-}(\bm{x})\right)\right)\right|\leq\tau\right)\eta(\bm{x}).\end{aligned}
\end{equation*}
\endgroup
$\eta^{\mathrm{AT}}(\bm{x})$ represents the posterior probability of the idealized dataset constructed for theoretical analysis, where boundary samples are assigned their true conditional probabilities. In practice, this corresponds to querying the true labels for these samples. It is important to note that $\eta^{\mathrm{AT}}(\bm{x})$ is a discontinuous function, which introduces technical challenges in the derivation of explicit convergence rates.

Given the mixed dataset $D_n^{\mathrm{AT}}$, we then train a final classifier using a suitable learner. Let $\widehat{f}_n^{\mathrm{AT}}$ denote the resulting scoring function learned from $D_n^{\mathrm{AT}}$, and let $\widehat{h}^{\mathrm{AT}}(\bm{x})=\operatorname{sign}(\widehat{f}_n^{\mathrm{AT}}(\bm{x}))$ be the corresponding classifier. A concrete neural-network-based result using specific surrogate losses will be introduced in Section \ref{section 4.2}.

While the Pro-AT approach provides a powerful mechanism to obtain high-quality labels, its primary drawback is the reliance on an additional annotation cost. This practical constraint motivates our exploration of a cost-free approach that can still effectively leverage the boundary samples. In the following, we introduce our Pro-PT method, designed precisely for this purpose.

\subsubsection{Pseudo Tagging with Uncertainty Preservation}
Without incurring manual annotation costs, we introduce a random pseudo-tagging step to leverage the boundary samples $D_{n-m}^b$. For each boundary sample $\left(\bm{x}_i^b, \cdot\right) \in D_{n-m}^b$, we first compute its estimated clean posterior probability $\eta^{\mathrm{pse}}\left(\bm{x}_i^b\right) := \Pi_{[0,1]}\left(\frac{\widehat{\eta}_\rho\left(\bm{x}_i^b\right)-\widehat{\rho}_{-}\left(\bm{x}_i^b\right)}{1-\widehat{\rho}_{+}\left(\bm{x}_i^b\right)-\widehat{\rho}_{-}\left(\bm{x}_i^b\right)}\right)$, where $\Pi_{[0,1]}(u)=\min \{1, \max (0, u)\}$.
A naive hard-labeling approach, for instance, would assign a fixed pseudo-label $y_i^{\mathrm{pse}} = \operatorname{sign}(\eta^{\mathrm{pse}}(\bm{x}_i^b) - 0.5)$. This approach is suboptimal because it discards the estimated label uncertainty. For example, a sample with an estimated posterior of 0.51, which is highly uncertain, is assigned a definitive hard label. This overconfident label can lead to an accumulation of errors \cite{arazo2020pseudo}.

To address this issue, we propose a dynamic pseudo-labeling scheme. Our approach shares a similar spirit with techniques such as label smoothing \cite{szegedy2016rethinking}, in that it preserves uncertainty and regularizes the learning procedure. However, in order to retain broad compatibility with downstream learners, including those that require hard labels, we continue to assign pseudo-labels from the discrete set $\{+1,-1\}$. We achieve this by independently resampling the pseudo-label for each boundary sample at every training epoch.

In practice, we generate a pseudo-label for the current training epoch via a Bernoulli trial. Specifically, we draw a random variable $u_i \sim U[0,1]$ and assign the temporary pseudo label $y_i^{\mathrm{pse}}$ as $y_i^{\mathrm{pse}} = 2\mathbb{I}(u_i < \eta^{\mathrm{pse}}(\bm{x}_i^b))-1$, which guarantees $\mathrm{Pr}(y_i^{\mathrm{pse}}=1\mid\bm{X}=\bm{x}_i^b)=\eta^{\mathrm{pse}}(\bm{x}_i^b)$.

Importantly, the approximation error between $\eta(\bm{x})$ and $\eta^{\mathrm{pse}}(\bm{x})$ can be controlled, in the sense that $\sup_{\bm{x}\in \mathcal{X}} \left|\eta(\boldsymbol{x})-\eta^{\mathrm{pse}}(\boldsymbol{x})\right|\lesssim \epsilon_{n_0,n}$ with more details in the Supplementary Material. This ensures that our sampling distribution is a high-fidelity approximation of the true posterior probability. Furthermore, the practice of resampling the pseudo-label at every training epoch provides a crucial regularization effect that prevents overfitting, which is a major pitfall of using fixed pseudo-labels with more details in Supplementary~\ref{Appendix C}.

This pseudo-tagged boundary samples $\{\bm{x}_i^b, y_i^{\mathrm{pse}}\}_{i=1}^{n-m}$, when combined with the extracted BO samples, forms the mixed data $D_n^{\mathrm{PT}}:=\left\{\bm{x}_i, y_i^{\mathrm{PT}}\right\}_{i=1}^n$, which is associated with the underlying distribution $D^{\mathrm{PT}}$. Similarly, we focus on $P_{D^{\mathrm{PT}}}\left(y^{\mathrm{PT}}=1 \mid \bm{x}\right):=\eta^{\mathrm{PT}}(\bm{x};\eta,\tau,\widehat{\eta}_\rho,\widehat{\rho}_{+},\widehat{\rho}_{-})$. Hereafter, to avoid ambiguity, we abbreviate $\eta^{\mathrm{PT}}(\bm{x};\eta,\tau,\widehat{\eta}_\rho,\widehat{\rho}_{+},\widehat{\rho}_{-})$ as $\eta^{\mathrm{PT}}(\bm{x})$, and
\begingroup\renewcommand{\TPAMImathsize}{8.5}
\begin{equation*}
    \begin{aligned}{}&\eta^{\mathrm{PT}}(\bm{x})={}\mathbb{I}\left(\widehat{\eta}_\rho(\bm{x})>\tfrac{1}{2}-\tfrac{1}{2}\left(\widehat{\rho}_{+}(\bm{x})-\widehat{\rho}_{-}(\bm{x})\right)+\tau\right)\\ &\quad+\mathbb{I}\left(\left|\widehat{\eta}_\rho(\bm{x})-\left(\tfrac{1}{2}-\tfrac{1}{2}\left(\widehat{\rho}_{+}(\bm{x})-\widehat{\rho}_{-}(\bm{x})\right)\right)\right|\leq\tau\right) \eta^{\mathrm{pse}}(\bm{x}).\end{aligned}
\end{equation*}
\endgroup
Given the mixed dataset $D_n^{\mathrm{PT}}$, we then use a suitable learning algorithm. Let $\widehat{f}_n^{\mathrm{PT}}$ denote the estimated scoring function, and let $\widehat{h}^{\mathrm{PT}}(\bm{x})=\operatorname{sign}(\widehat{f}_n^{\mathrm{PT}}(\bm{x}))$ be the associated classifier. A neural-network-based result is deferred to Section \ref{section 4.2}.

\begin{figure}[ht]
    \centering
    \includegraphics[width=\linewidth]{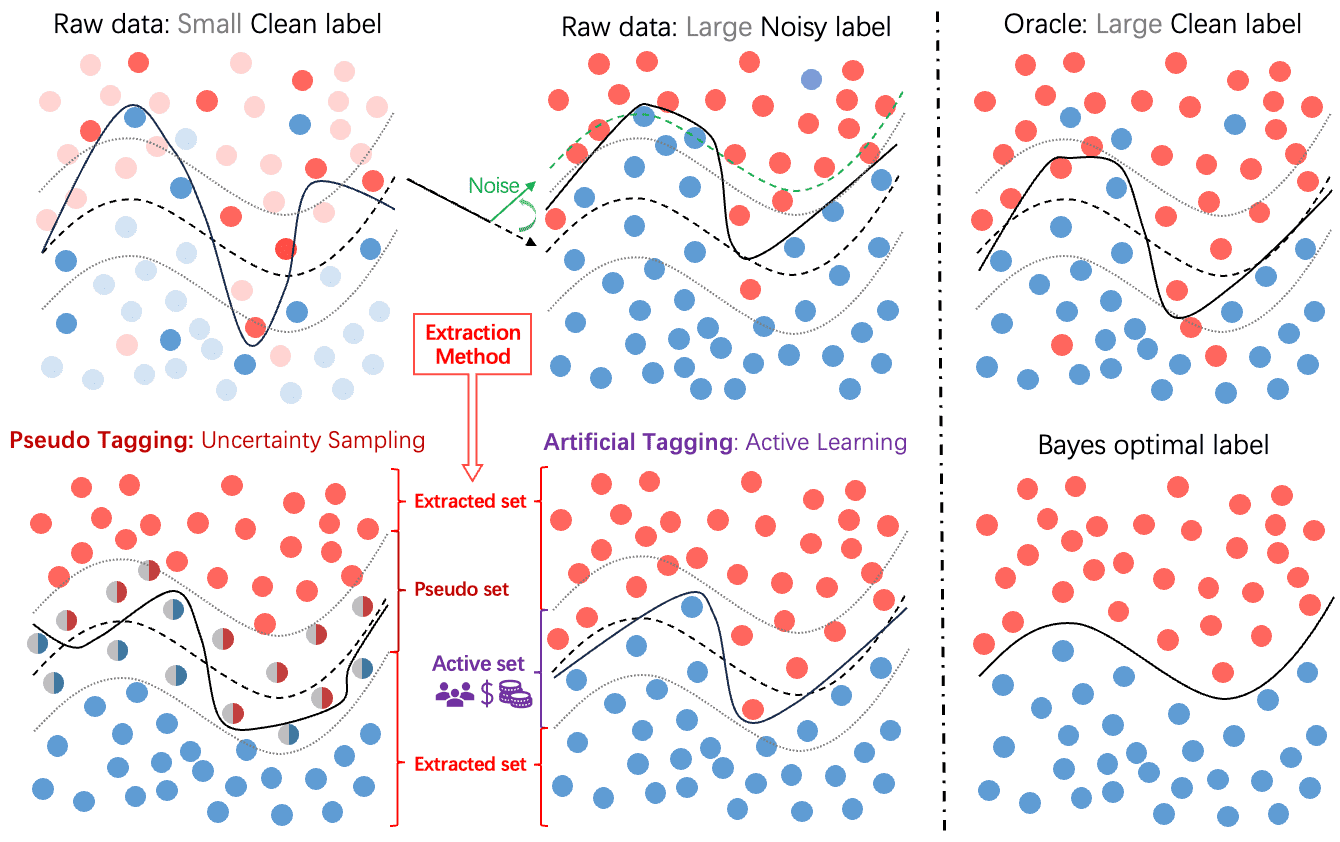}
    \caption{Illustration of BO-label extraction and boundary-sample tagging. Pro-PT assigns Bernoulli pseudo-labels to boundary samples, and Pro-AT queries their expert labels. Samples outside the boundary region receive extracted BO labels. The right-hand panels show the large clean-data oracle and ideal BO data.}
    \label{fig:method_summary}
\end{figure}

Figure~\ref{fig:method_summary} contrasts learning from the small clean or large noisy dataset alone with Pro-PT, Pro-AT and the large-clean oracle. Both proposed strategies move the estimated boundary closer to the Bayes boundary. They can also be combined by querying a subset of boundary samples and pseudo-tagging the remainder.

\subsection{Computation}\label{section 3.3}
The complete computational pipeline is summarized in Algorithm \ref{Algorithm 1} for the instance-dependent noise setting. The class-dependent case is handled similarly. Notably, the proposed model-agnostic framework does not require prior knowledge of the specific noise type in practice. One may apply both noise estimation strategies and select the one that yields better validation performance. Moreover, the framework itself does not depend on a specific classification learner; any suitable learner can be used in the final training stage once the mixed data are constructed.
\begin{algorithm}[ht]
    \caption{Unified Bayes Optimal Data Extraction and Pro-Tagging}
    \label{Algorithm 1}
    \begin{algorithmic}[1]
        \renewcommand{\algorithmicrequire}{\textbf{Require:}}
        \Require Datasets $D_{\rho,n}, D_{n_0}$, security margin $\tau$, tagging method $\in$ \{Pro-AT, Pro-PT\}, classification learner $\mathcal{A}$.
        \Statex {\bf Step 1. Extraction of Bayes Optimal Data.}
        \State Estimate $\widehat{\rho}_{+}(\bm{x}), \widehat{\rho}_{-}(\bm{x})$, and $\widehat{\eta}_\rho(\bm{x})$.
        \State Partition $D_{\rho,n}$ into extracted BO samples $D^\star_m$ and boundary samples $D^b_{n-m}$ according to \eqref{drift}.
        \Statex {\bf Step 2. Tagging Boundary Samples.}
        \If{tagging method is Pro-AT}
            \State Query labels for samples in $D^b_{n-m}$ and construct the mixed dataset $D_n^{\mathrm{AT}}$.
            \State Train the final classifier on $D_n^{\mathrm{AT}}$ using learner $\mathcal{A}$.
        \ElsIf{tagging method is Pro-PT}
            \State Estimate $\eta^{\mathrm{pse}}(\bm{x})$ and construct the mixed dataset $D_n^{\mathrm{PT}}$.
            \State Train the final classifier on $D_n^{\mathrm{PT}}$ using learner $\mathcal{A}$, with each boundary pseudo-label independently resampled from $\eta^{\mathrm{pse}}$ at the start of every training epoch.
        \EndIf
        \State \Return final classifier $\widehat{h}(\bm{x})$.
    \end{algorithmic}
\end{algorithm}

\noindent {\bf Data-Driven Selection of the Security Margin $\tau$.} $\tau$ is not merely a security constant but a hyperparameter that must be carefully tuned to balance this purity-quantity trade-off. An optimal $\tau$ must be large enough to ensure high-fidelity extraction while being small enough not to needlessly discard valuable data. Since the theoretical bounds required are often intractable in practice, a data-driven approach is essential. We propose a cross-validation procedure denoted as Algorithm \ref{Algorithm 2} to determine the optimal value $\tau^{\star}$. 
\begin{algorithm}[ht]
    \caption{5-Fold Cross-Validation for the Security Margin}
    \label{Algorithm 2}
    \small\renewcommand{\TPAMImathsize}{8.5}
\begin{algorithmic}[1]
        \renewcommand{\algorithmicrequire}{\textbf{Require:}}
        \renewcommand{\algorithmicensure}{\textbf{Ensure:}}
        \Require Candidate set $\mathcal{T}_{\mathrm{cand}}$, datasets $D_{\rho,n}, D_{n_0}$, tagging method $\in$ \{Pro-AT, Pro-PT\}, classification learner $\mathcal{A}$.
        \State Partition $D_{n_0}$ into 5 disjoint folds $F_1,\dots,F_5$.
        \State Estimate $\widehat{\eta}_\rho(\bm{x})$ on $D_{\rho,n}$.
        \For{$k = 1,\dots,5$}
            \State Set $D_{\mathrm{val}}^{(k)} = F_k$ and $D_{\mathrm{train}}^{(k)} = \bigcup_{i\neq k} F_i$.
            \State Estimate $\widehat{\rho}_{+}^{(k)}(\bm{x})$ and $\widehat{\rho}_{-}^{(k)}(\bm{x})$ on $D_{\mathrm{train}}^{(k)}$.
            \For{each candidate $\tau_j \in \mathcal{T}_{\mathrm{cand}}$}
                \State Obtain classifier $\widehat{h}^{(k,j)}$ by running Algorithm \ref{Algorithm 1} with inputs $D_{\rho,n}, D_{\mathrm{train}}^{(k)}, \tau_j$.
                \State \begin{minipage}[t]{\dimexpr\linewidth-\algorithmicindent-\algorithmicindent\relax}
\raggedright Compute
\begin{equation*}
\operatorname{Acc}^{(k)}(\tau_j)=\frac{1}{|D_{\mathrm{val}}^{(k)}|}\sum_{(\bm{x}_i,y_i)\in D_{\mathrm{val}}^{(k)}}\mathbb{I}\bigl(\widehat{h}^{(k,j)}(\bm{x}_i)=y_i\bigr).
\end{equation*}
\end{minipage}
            \EndFor
        \EndFor
        \State \begin{minipage}[t]{\linewidth}
\raggedright Compute average accuracy
\begin{equation*}
\overline{\operatorname{Acc}}(\tau_j)=\frac{1}{5}\sum_{k=1}^5 \operatorname{Acc}^{(k)}(\tau_j)
\end{equation*}
for each $\tau_j \in \mathcal{T}_{\mathrm{cand}}$.
\end{minipage}
        \State Select $\tau^\star = \arg\max_{\tau_j \in \mathcal{T}_{\mathrm{cand}}} \overline{\operatorname{Acc}}(\tau_j)$.
        \State \Return $\tau^\star$.
    \end{algorithmic}
\end{algorithm}

\begin{Rem}
    Section \ref{section 3} presents a model-agnostic framework. The proposed BO data extraction and tagging procedures only require estimators of $\eta_\rho,\rho_+,\rho_-$, together with a classification learner applied to the constructed mixed dataset. In the theoretical analysis, we mainly study neural networks and B-splines to provide learner-specific guarantees. In the numerical studies, we instantiate the framework with commonly used flexible classification learners, including neural networks, B-splines, random forests, and gradient boosting, to demonstrate its practical flexibility.
\end{Rem}

\section{Theoretical Results}\label{section 4}

In this section, we first establish a generic plug-in theory of the proposed procedure in Section \ref{section 4.1}, and then focus primarily on neural-network-based implementations in Section \ref{section 4.2}, which are suitable for complex data. 
A sieve-based instantiation using B-splines is also provided, with all technical details, in Supplementary~\ref{Appendix A}-\ref{Appendix C}. 
For the learner-specific analysis, nuisance fitting and mixed-sample construction use independent noisy subsamples of comparable sizes. Conditional on the nuisance-fitting data, the reconstructed observations are i.i.d. from the fitted mixed distribution. The theoretical results concern this sample-split variant, whereas Algorithm~1 describes full-sample reuse.
Let $\|\cdot\|_2$ denote the $L_2$ norm of a vector and $B(\bm{x}_0, r) = \left\{\bm{z} \mid\|\bm{x}_0 -\bm{z}\|_2 \leq r\right\}$ denote an $r$-radius ball centered at $\bm{x}_0$ with the Euclidean distance. $\mathbb{E}_{n,D}$ denotes the expectation with respect to $n$ training samples generated from the distribution $D$.
$\widetilde{O}(\cdot)$ is the Soft-O notation, where a function $f(n)$ is said to be $\widetilde{O}(g(n))$ if there exists a constant $k$ such that $f(n)=O\left(g(n) \log ^k n\right)$. Finally, the constants used here may be different from place to place.

\subsection{A Generic Plug-in Theory}\label{section 4.1}
This section presents a model-agnostic generic theory with two components: the error of the extraction and the error of the classification learner. Both can be characterized by learner-specific rates. For a given function $f$, its excess $0$-$1$ risk under the clean distribution $D$ is defined as $\mathcal{E}^{0/1}_D\left(f, h_D^o\right)=R^{0/1}_D(f)-R^{0/1}_D(h_D^o)$, where $R^{0/1}_D(\cdot)$ and $h_D^o$ are defined in Section \ref{BO-def}. Let $\varepsilon_{1,n_0}$, and $\varepsilon_{1,n}$ denote the expected error of different steps, such that  \begingroup\renewcommand{\TPAMImathsize}{8.5}
\begin{equation*}\begin{aligned}{}&\mathbb{E}_{n_0,D}\Bigl\{\sup_{\bm{x}\in\mathcal{X}}|\widehat{\rho}_{+}(\bm{x})-\rho_{+}(\bm{x})|+\sup_{\bm{x}\in\mathcal{X}}|\widehat{\rho}_{-}(\bm{x})-\rho_{-}(\bm{x})|\Bigr\}^2\\ &\quad\lesssim \varepsilon_{1,n_0},\end{aligned}\end{equation*}
\endgroup $\mathbb{E}_{n,D_{\rho}}\sup_{\bm{x}\in\mathcal{X}}(\widehat{\eta}_{\rho}(\bm{x})-\eta_{\rho}(\bm{x}))^2\lesssim \varepsilon_{1,n}$. Let $\varepsilon_{2,n}$ bound the expected excess risk of the final learner under the constructed mixed distribution, relative to that distribution's Bayes classifier.

\begin{Mth}\label{Mth generic}
    In the step of generic extraction of BO data,
    under Condition \ref{Con 1}(a), $\mathbb{E}_{n_0,D}\mathbb{E}_{n,D_{\rho}}\epsilon_{n_0,n}^2\lesssim n_0^{-1}+\varepsilon_{1,n}$, where $\epsilon_{n_0,n}$ is defined in \eqref{epsilon n_0 n}; under Condition \ref{Con 1}(b), $\mathbb{E}_{n_0,D}\mathbb{E}_{n,D_{\rho}}\epsilon_{n_0,n}^2\lesssim \varepsilon_{1,n_0}+\varepsilon_{1,n}$. For the final classifier, under condition~\ref{Con 2}, we have
    \begin{itemize}
        \item[(1)] For Pro-AT, {\small $\mathbb{E}_{n_0,D}\mathbb{E}_{n,D_{\rho}}\left[\mathcal{E}_D^{0/1}\bigl(\widehat{f}^{\mathrm{AT}},h_D^o\bigr)\right]\lesssim \varepsilon_{2,n}$}.
        \item[(2)] For Pro-PT, {\small $\mathbb{E}_{n_0,D}\mathbb{E}_{n,D_{\rho}}\left[\mathcal{E}_D^{0/1}\bigl(\widehat{f}^{\mathrm{PT}},h_D^o\bigr)\right]\lesssim \mathbb{E}_{n_0,D}\mathbb{E}_{n,D_{\rho}}\epsilon_{n_0,n}^{q+1}+\varepsilon_{2,n}$}.
    \end{itemize}
\end{Mth}
Intuitively, $\varepsilon_{1,n}$ quantifies the error of estimating the underlying function with a sample size of $n$. Under Condition \ref{Con 1}(b), $\rho_{+}$ and $\rho_{-}$ are nonparametric functions of $\bm{x}$ and contribute the rate $\varepsilon_{1,n_0}$ based on the audited sample. Similarly, $\varepsilon_{2,n}$ quantifies the downstream excess classification error of the learner trained on the mixed data. These quantities depend on the complexity of the underlying function class and the learner used in each stage. The Pro-AT bound shows that, after purification, the proposed procedure attains the same order as if a ``large-clean'' dataset with sample size $n$ is available. For Pro-PT, the extra term $\mathbb{E}_{n_0,D}\mathbb{E}_{n,D_{\rho}}\epsilon_{n_0,n}^{q+1}$ reflects the uncertainty. In many regimes, this term is larger than $\varepsilon_{2,n}$ and therefore becomes the leading component. For instance, when $q=1$, it is of order $n_0^{-1}+\varepsilon_{1,n}$ under Condition \ref{Con 1}(a) and of order $\varepsilon_{1,n_0}+\varepsilon_{1,n}$ under Condition \ref{Con 1}(b); when the ``small-clean'' sample size is the main bottleneck, this contribution is governed by $n_0^{-1}$ or $\varepsilon_{1,n_0}$.
For more complex learners such as neural networks, these rates are determined by the approximation and estimation errors of the network class, as analyzed in Section \ref{section 4.2}. This provides a modular framework that applies to a wide range of learners without requiring a specific parametric form.

We now relax this assumption to address the more general and practical setting of \textbf{covariate shift}. Let the small clean dataset $D^{\mathrm{cl}}_{n_0}:=\{(\bm{x}_{0i},\widetilde y_{0i},y_{0i})\}_{i=1}^{n_0}$ be drawn from a distribution $D^{\mathrm{cl}}$, whose marginal distribution $p_{D^{\mathrm{cl}}}(\bm{x})$ may differ from $p_{D_\rho}(\bm{x})$. 
The target covariate distribution is $p_D=p_{D_\rho}$, while the audited covariate distribution $p_{D^{\mathrm{cl}}}$ may differ. The extension retains common support, invariant conditional mechanisms, and the density lower bounds specified in Condition~\hyperref[Con:4_star]{4*}.
We use $\eta^{\mathrm{cl}}_\rho$, $\rho^{\mathrm{cl}}_+$, and $\rho^{\mathrm{cl}}_-$ to denote the noisy posterior and the two noise functions defined under $D^{\mathrm{cl}}$, in direct analogy to $\eta_\rho$, $\rho_+$, and $\rho_-$. The key requirement for our analysis is $\operatorname{Supp}(p_{D^{\mathrm{cl}}})=\operatorname{Supp}(p_{D_\rho})$, $\eta^{\mathrm{cl}}_{\rho}=\eta_{\rho}$, $\rho_{+}^{\mathrm{cl}}=\rho_{+}$, and $\rho_{-}^{\mathrm{cl}}=\rho_{-}$, where $\operatorname{Supp}(\cdot)$ denotes the support. Under this condition, the core proof remains unchanged; the details are provided in Supplementary~~\ref{Appendix C}. 
For notational simplicity, Section~\ref{section 4.2} presents the theory under $D^{\mathrm{cl}}=D$. This differs from standard covariate-shift learning, where the source-to-target discrepancy is typically handled through the density ratio $p_{D_\rho}(\bm{x})/p_{D^{\mathrm{cl}}}(\bm{x})$. Recent work has relaxed the classical bounded-ratio requirement, but such relaxations still impose direct technical control. For example, Ma et al.~\cite{ma2023optimally} allows unbounded likelihood ratios, but requires finite second moments; and Xu et al.~\cite{xu2025estimating} studies unbounded density-ratio estimation under explicit tail and regularity conditions. In contrast, our audited sample is not used to construct an importance-weighted target risk. 

\subsection{Neural-Network-Specific Properties} \label{section 4.2}
In this section, we focus on properties of neural-network-based implementations leveraging the generic theory established in Section \ref{section 4.1}.
As discussed in Section \ref{section 3.1}, the extraction rate is determined by $P_D(|\eta(\bm{x})-1/2| \le C_\tau \tau )$. Under Condition \ref{Con 2}, this probability is $O(\tau^q)$, and hence $O(\epsilon_{n_0,n}^q)$ when $\tau=C\epsilon_{n_0,n}$ for a fixed $C>1$, as in Section \ref{section 3.1}. The generic framework in Section \ref{section 3.1} only requires suitable estimators of $\eta_\rho(\bm{x})$, $\rho_{+}(\bm{x})$, and $\rho_{-}(\bm{x})$ together with appropriate control of their estimation errors. We now derive the rate of $\epsilon_{n_0,n}^q$ for neural-network-based estimators. Before doing so, we introduce two conditions. These conditions are standard in nonparametric statistics \cite{cai2021transfer}. 
\begin{Con}\label{Con 3}
    {\bf (Smooth Conditional Distribution).} Let the conditional distribution $\eta(\bm{x}), \eta_\rho(\bm{x}), \rho_{-}(\bm{x}), \rho_{+}(\bm{x})$ be contained in the function class $\mathcal{H}_{l, \mathcal{P}}$ for some $l \in \mathbb{N}$ and $\mathcal{P} \subseteq[1, \infty) \times \mathbb{N}$, where $\mathcal{H}_{l, \mathcal{P}}$ and $\mathcal{P}$ are defined in Supplementary~\ref{Appendix B}. 
\end{Con}
$\mathcal{H}_{l, \mathcal{P}}$ generalizes many other function class assumptions in the literature \cite{Johannes2020Nonparametric}. Condition \ref{Con 3} imposes a smoothness assumption on the unknown functions we aim to estimate.
\begin{Con}\label{Con 4}
    {\bf (Marginal Distribution Boundedness).} Let $\mathcal{X} \subset[0,1]^d$ be the closed support of $\bm{X}$ and $P_D(\bm{x})$ is absolutely continuous with respect to the Lebesgue measure $\lambda$ on $[0,1]^d$. There exist constants $r_{\bm{x}}, C_{\bm{x}}, C^{\prime}_{\bm{x}}, C_y, \mu_{-}>0$ such that
    \begin{align}
        \lambda[\mathcal{X} \cap B(\bm{x}, r)] & \geq C_{\bm{x}} \lambda[B(\bm{x}, r)], \notag\\ &\quad \forall \bm{x}\in \mathcal{X},\quad\forall 0<r \leq r_{\bm{x}}, \label{eq:cone_condition} \\
        \int_{\mathcal{X}\cap B(\bm{x}_0, \frac{N}{L})}R(\bm{x};&\bm{x}_0)\mathrm{d}\bm{x} \geq C^{\prime}_{\bm{x}}\int_{B(\bm{x}_0, \frac{N}{L})}R(\bm{x};\bm{x}_0)\mathrm{d}\bm{x},\label{eq:geom_condition}\\
        p_D(\bm{x})&=p_{D_{\rho}}(\bm{x})\geq \mu_{-}, \label{eq:density_condition}
    \end{align}
    where $R(\bm{x};\bm{x}_0)=N-L\left\|\bm{x}-\bm{x}_0\right\|_2$, and \eqref{eq:geom_condition} holds for all $\bm{x}_0\in\mathcal{X}$ and $N,L>0$ with $N/L\leq r_{\bm{x}}$.
\end{Con}

Condition \ref{Con 4} ensures the data distribution is well-behaved. Specifically, \eqref{eq:cone_condition} and \eqref{eq:geom_condition} are geometric constraints on the support $\mathcal{X}$, which prevent singular regions like isolated points. \eqref{eq:density_condition} assumes that the marginal density is bounded away from zero, which guarantees that no region within the support has a vanishingly small probability. Finally, the condition for controlling the size of the neural network is detailed in Condition \ref{Con S1} of the Supplementary Material, a common condition in deep learning.
Under these conditions, we first analyze the class-dependent noise case. 
\begin{Lem}\label{Lem 1}
    Under the Condition \hyperref[Con 1]{1(a)}, we have
    $\mathbb{E}_{n_0,D}\left|\rho_{+}-\widehat{\rho}_{+}\right|^2 \lesssim n_0^{-1}.$ Symmetrically, the same bound holds for $\mathbb{E}_{n_0,D}\left|\rho_{-}-\widehat{\rho}_{-}\right|^2$.
\end{Lem}
In this case, the audited clean data are used only to estimate two scalar quantities, which leads to the fast parametric rate $n_0^{-1}$. Hence, this $n_0$ term is usually not the dominant error. We next turn to the more challenging instance-dependent setting, where $\widehat{\rho}_{-}(\bm{x})$ and $\widehat{\rho}_{+}(\bm{x})$ must be estimated nonparametrically. In this case, we instantiate the generic estimators in Section \ref{section 3.1} using neural networks. Let $\mathcal{L}_{\text{MSE}}(y, f(\bm{x})) = (\frac{y+1}{2}-f(\bm{x}))^2, R^{\text{MSE}}_{D_\rho}(f)=\mathbb{E}_{D_\rho} \mathcal{L}_{\text{MSE}}(\widetilde{Y}, f(\bm{X})), \eta_\rho=\arg\min _{f \in \mathcal{H}_{l,\mathcal{P}}} R^{\text{MSE}}_{D_\rho}(f)$, and define $\widehat{\eta}_\rho(\bm{x})$, $\widehat{\rho}_{-}(\bm{x})$, and $\widehat{\rho}_{+}(\bm{x})$ by the neural-network-based procedure. We then have the following explicit sup-norm error bound.
\begin{Lem}\label{Lem 2}
    Suppose that Conditions \hyperref[Con 1]{1(b)}, \ref{Con 3}, \ref{Con 4}, and Condition \ref{Con S1} hold. Then
    \begin{equation*}
        \begin{aligned}
            {}&\mathbb{E}_{n,D_{\rho}}\sup_{\bm{x}\in\mathcal{X}} \left(\widehat{\eta}_\rho\left(\bm{x}\right)-\eta_\rho\left(\bm{x}\right)\right)^2\\ &\quad\lesssim \varepsilon_{1,n}= \widetilde{O}(\max _{(p, k) \in \mathcal{P}}n^{-\frac{2p}{(2p+k)(d+1)}}),\\
            &\mathbb{E}_{n_0,D}\sup_{\bm{x}\in\mathcal{X}} \Bigl\{\left|\widehat{\rho}_{-}\left(\bm{x}\right)-\rho_{-}(\bm{x})\right|+\left|\widehat{\rho}_{+}(\bm{x})-\rho_{+}(\bm{x})\right|\Bigr\}^2\\ &\quad\lesssim \varepsilon_{1,n_0}= \widetilde{O}(\max _{(p, k) \in \mathcal{P}}n_{0}^{-\frac{2p}{(2p+k)(d+1)}}),\\
        \end{aligned}
    \end{equation*}
    where $p$ is the smoothness constant of the functions $\eta_\rho(\bm{x}), \rho_{+}(\bm{x}), \rho_{-}(\bm{x})$, and $k$ is a constant such that $k \ll d$, as specified in Condition \ref{Con S1} of the Supplementary Material.
\end{Lem}
Lemma \ref{Lem 2} shows that the uniform estimation error bound achieves the rate $n^{-\frac{2p}{(2p+k)(d+1)}}$ up to logarithmic factors. This rate depends on the intrinsic complexity $k\leq d$. The additional dependence on $d$ is consistent with the technical difficulty of establishing sup-norm guarantees for neural-network estimators. Furthermore, for the noise under Condition \hyperref[Con 1]{1(a)}, the estimation is more efficient, since $n_0^{-1} \ll n_0^{-\frac{2 p}{(2p+k)(d+1)}}$.

As discussed before, the proportion of boundary samples is of the order $\epsilon_{n_0, n}^q$. The following theorem provides an analysis of the statistical convergence rate. 
\begin{Mth}\label{Mth 1}
    Suppose that Conditions \hyperref[Con 1]{1(a)}, \ref{Con 3}, \ref{Con 4} and Condition \ref{Con S1}  of the Supplementary Material hold, then $\mathbb{E}_{n_0,D}\mathbb{E}_{n,D_\rho}\epsilon_{n_0, n}^q= \widetilde{O}(n_0^{-\frac{q\wedge2}{2}}+\max _{(p, k) \in \mathcal{P}} n^{-\frac{p(q\wedge2)}{(2 p+k)(d+1)}})$. However, if the conditions of Lemma \ref{Lem 2} hold, we have
    \begin{equation*}
        \mathbb{E}_{n_0,D}\mathbb{E}_{n,D_\rho}\epsilon_{n_0, n}^q=\widetilde{O}(\max _{(p, k) \in \mathcal{P}}n_0^{-\frac{p(q\wedge2)}{(2 p+k)(d+1)}}),
    \end{equation*}
    where $p$ and $k$ are defined in Lemma \ref{Lem 2}, and $q$ is the Tsybakov Margin exponent. 
\end{Mth}
Under the simpler class-dependent noise setting of Condition \hyperref[Con 1]{1(a)}, the error contribution from the clean data diminishes at the fast parametric rate $(q\wedge2)/2$.
In contrast, under the instance-dependent noise setting of Condition \hyperref[Con 1]{1(b)}, the overall error is dictated by $\widetilde{O}\!\left(\max _{(p, k) \in \mathcal{P}}n_0^{-\frac{p(q\wedge2)}{(2 p+k)(d+1)}}\right)$ because $n_0\ll n$. This highlights a key bottleneck: estimating the instance-dependent noise rates is a challenging nonparametric task that relies entirely on the small audited clean dataset, thereby emphasizing the need for a sufficiently large $n_0$ in this more complex setting.

In this subsection, we specialize the generic mixed-data constructions in Section \ref{section 3.2} to specific learners. We first focus on neural networks; rather than estimating the conditional probability directly, we estimate a scoring function and classify by its sign. This is standard in statistical learning and can lead to faster convergence rates \cite{audibert2007fast}. We consider the hinge loss and the sigmoid loss, defined respectively by $\mathcal{L}_{\mathrm{Hinge}}(y,f(\bm{x}))=\max(1-yf(\bm{x}),0), \mathcal{L}_{\mathrm{Sigmoid}}(y,f(\bm{x}))=1-\tanh(yf(\bm{x}))$, where $\tanh(z)=(e^z-e^{-z})/(e^z+e^{-z})$ and $f(\bm{x})$ denotes the scoring function. The predicted label is given by $\operatorname{sign}(f(\bm{x}))$. Let $\mathcal{F}_{\mathrm{NN}}$ denote the neural-network function class defined in Definition \ref{Def S3} of the Supplementary Material. For the Pro-AT mixed data $D_n^{\mathrm{AT}}$, define the empirical risks and estimated scoring function,
\begin{equation*}
    \begin{aligned}
        \widehat{R}^{\mathrm{Hinge}}_{D^{\mathrm{AT}}}(f)
        &=\frac{1}{n}\sum_{i=1}^n \mathcal{L}_{\mathrm{Hinge}}(y_i^{\mathrm{AT}}, f(\bm{x}_i)),
        \\
        &\widehat{f}^{\mathrm{AT,H}}_{n} \in \arg\min_{f\in\mathcal{F}_{\mathrm{NN}}}\widehat{R}^{\mathrm{Hinge}}_{D^{\mathrm{AT}}}(f),\\
        \widehat{R}^{\mathrm{Sigmoid}}_{D^{\mathrm{AT}}}(f)
        &=\frac{1}{n}\sum_{i=1}^n \mathcal{L}_{\mathrm{Sigmoid}}(y_i^{\mathrm{AT}}, f(\bm{x}_i)),
        \\
        &\widehat{f}^{\mathrm{AT,S}}_{n} \in \arg\min_{f\in\mathcal{F}_{\mathrm{NN}}}\widehat{R}^{\mathrm{Sigmoid}}_{D^{\mathrm{AT}}}(f).
    \end{aligned}
\end{equation*}
Similarly, we define $\widehat{f}^{\mathrm{PT,H}}_{n}, \widehat{f}^{\mathrm{PT,S}}_{n}$ for $D_n^{\mathrm{PT}}$. The corresponding classifier is defined by $\widehat{h}^{\mathrm{AT,H}}(\bm{x})=\operatorname{sign}(\widehat{f}^{\mathrm{AT,H}}_{n}(\bm{x}))$, and the remaining classifiers are defined analogously.

We present the explicit analysis for the hinge loss. The corresponding results for the sigmoid loss are deferred to Supplementary~\ref{Appendix C}. To state the resulting rates, we introduce the Complexity Assumption on the Decision set (CAD, \cite{audibert2007fast, kim2021fast}), which characterizes the smoothness of the Bayes decision boundary. Let $\mathcal{C}^{p_h,C,K,T}$ denote the class of classifiers whose positive decision region is a union of $T$ disjoint pieces. Each piece is defined by $K$ inequalities of the form $x_j\geq g(\bm{x}_{-j})$, where $\bm{x}_{-j}$ omits the $j$th component and $g$ has H\"older smoothness $p_h$ and H\"older norm bounded by $C$. The full definitions are given in Definitions~\ref{Def 1}--\ref{Def 2} of the Supplementary Material.
\begin{Con}\label{Con 5}
    {\bf (CAD).} The classifiers $h_{D^{\mathrm{AT}}}^o$ and $h_{D^{\mathrm{PT}}}^o$ belong to $\mathcal{C}^{p_h,C,K,T}$.
\end{Con}
Recall that $h_{D^\star}^o=h_D^o$. Furthermore, from the definition of $\eta^{\mathrm{AT}}(\bm{x})$ and the property of $\tau$ in \eqref{drift}, we have $\operatorname{sign}\left(\eta(\bm{x})-\frac{1}{2}\right)=\operatorname{sign}\left(\eta^{\mathrm{AT}}(\bm{x})-\frac{1}{2}\right)$, for $\forall \bm{x} \in \mathcal{X}$, which implies that $h_{D^{\mathrm{AT}}}^o=h_D^o$. For Pro-PT, the assumption that $h_{D^{\mathrm{PT}}}^o$ has a smooth decision boundary is a standard technical condition and is crucial for the following analysis. The next two theorems instantiate the generic Pro-AT and Pro-PT constructions in Section \ref{section 3.2} under the neural-network-based learner above. 
For Pro-PT, we additionally require the fitted mixed distribution to satisfy the posterior-margin condition in Lemma~\ref{Lem S2} and, when $q>1$, its geometric-margin condition, with constants uniform over nuisance estimates.
\begin{Mth}\label{Mth 2}
    Suppose that Conditions \ref{Con 1}-\ref{Con 5} and Condition \ref{Con S1} of the Supplementary Material hold, $\gamma=q\vee 1$, and $\tau=C(n_0^{-1/2} + n^{-c})$ for class-dependent noise or $\tau=Cn_0^{-c}$ for instance-dependent noise, where $C>0$ is a sufficiently large constant and $0 < c < 1$ is a sufficiently small constant. Then, the excess 0-1 error of the estimator is bounded by
    \begin{equation*}
\begin{aligned}{}&\mathbb{E}_{n_0, D}\mathbb{E}_{n, D_{\rho}}\left[\mathcal{E}^{0/1}_D\left(\widehat{f}_{n}^{\mathrm{AT,H}}, h_D^o\right)\right]\\ &\quad\lesssim \varepsilon_{2,n}= \widetilde{O}\left(n^{-\frac{p_h(q+1)}{p_h(q+2)+(d-1)(q+1) / \gamma}}\right),\end{aligned}
\end{equation*}
    where $q$ is the Tsybakov Margin exponent, and $p_h$ is the constant of the boundary.
\end{Mth}
The result in Theorem \ref{Mth 2} applies to both class-dependent and instance-dependent noise. The key reason is that, through expert querying, the boundary samples are relabeled using clean labels, while the extracted BO samples are already correctly labeled by construction. In the proof, the contribution from the BO samples yields a beneficial non-positive term whose magnitude depends on the extracted sample size $m$ and hence on the specific noise model. To derive a general rate that holds uniformly across noise settings, we conservatively bound this beneficial term to zero. Although this simplification may be loose, it leads to a robust worst-case guarantee that is agnostic to the specific noise type. A further feature of Theorem \ref{Mth 2} is its explicit dependence on the ambient dimension $d$. Since $(q+1)/\gamma \in [1,2]$, the dimension appears in the denominator of the exponent, indicating that the curse of dimensionality is not avoided. This limitation appears intrinsic to the hinge-loss analysis under such general assumptions, and removing the dimension dependence would likely require stronger structural assumptions on the data-generating process, such as Gaussian mixture models \cite{zhou2024classification}.
\begin{Rem}\label{rem:benchmarks}
    To contextualize the performance of our methods, we establish two crucial benchmarks under the conditions of Theorem \ref{Mth 2}. The first is a baseline classifier trained exclusively on the small clean dataset of size $n_0$, which achieves an error rate of $\widetilde{O}(n_0^{-\frac{p_h(q+1)}{p_h(q+2)+(d-1)(q+1) / \gamma}})$. The second is a hypothetical Oracle classifier trained on a fully clean dataset of size $n$, which has an excess-risk upper bound of $\widetilde{O}(n^{-\frac{p_h(q+1)}{p_h(q+2)+(d-1)(q+1) / \gamma}})$.
\end{Rem}
Remark \ref{rem:benchmarks} provides a useful interpretation of Theorem \ref{Mth 2}. Compared with the baseline trained only on the small clean dataset, Pro-AT yields a substantial gain by effectively exploiting the large noisy dataset of size $n$. Compared with the fully clean oracle benchmark, the upper bound for Pro-AT has the same convergence order. This comparison uses the boundary-sample bound above, under which the proportion of queried labels tends to zero. Furthermore, we introduce a third benchmark that trains exclusively on the extracted BO samples while discarding all boundary samples; its theoretical performance is analyzed in Corollary \ref{Cor S1} of the Supplementary Material. For Pro-PT, we also provide the analysis as follows.
\begin{Mth}\label{Mth 3}
    Suppose that Conditions \hyperref[Con 1]{1(a)}, \ref{Con 2}-\ref{Con 5}, and Condition \ref{Con S1}  of the Supplementary Material hold, $\gamma = q\vee 1$, and $\tau=C(n_0^{-\frac{1}{2}}+n^{-c})$ for a sufficiently large constant $C>0$ and sufficiently small constant $0<c<1$, we have $\varepsilon_{2,n}=\widetilde{O}(n^{-\frac{p_h(q+1)}{p_h(q+2)+(d-1)(q+1)/\gamma}})$, and
    \begin{equation*}
\begin{aligned}{}&\mathbb{E}_{n_0, D}\mathbb{E}_{n, D_{\rho}}\left[\mathcal{E}^{0/1}_D\left(\widehat{f}_{n}^{\mathrm{PT,H}}, h_D^o\right)\right]\\ &\quad\lesssim \widetilde{O}(n_0^{-\frac{(q+1)\wedge2}{2}}+\max _{(p, k) \in \mathcal{P}}n^{-\frac{p((q+1)\wedge2)}{(2p+k)(d+1)}})+\varepsilon_{2,n},\end{aligned}
\end{equation*}
    where $p_h$ is the smoothness constant of boundary, $q$ is the Tsybakov Margin exponent, $p$ and $k$ are defined in Lemma \ref{Lem 2}. Moreover, if Condition \hyperref[Con 1]{1(b)} holds instead of \hyperref[Con 1]{1(a)}, $\tau=Cn_0^{-c}$ for some sufficiently large $C>0$ and sufficiently small $0<c<1$, we have
    \begin{equation*}
\begin{aligned}{}&\mathbb{E}_{n_0, D}\mathbb{E}_{n, D_{\rho}}\left[\mathcal{E}^{0/1}_D\left(\widehat{f}_{n}^{\mathrm{PT,H}}, h_D^o\right)\right]\\ &\quad\lesssim \widetilde{O}(\max _{(p, k) \in \mathcal{P}}n_0^{-\frac{p((q+1)\wedge2)}{(2p+k)(d+1)}})+\varepsilon_{2,n}.\end{aligned}
\end{equation*}
\end{Mth}
A detailed comparison with Theorem \ref{Mth 2} reveals the trade-off underlying Pro-PT. The error bound in Theorem \ref{Mth 3} can be decomposed into two components. The first is the classification error $\widetilde{O}(n^{-\frac{p_h(q+1)}{p_h(q+2)+(d-1)(q+1)/\gamma}})$, which is identical to the rate achieved by Pro-AT and reflects the intrinsic difficulty of learning the classification boundary from $n$ samples. The second is the pseudo-labeling error, which under class-dependent noise is $\widetilde{O}(n_0^{-\frac{(q+1)\wedge2}{2}}+\max _{(p, k) \in \mathcal{P}}n^{-\frac{p((q+1)\wedge2)}{(2p+k)(d+1)}})$ and under instance-dependent noise is dominated by $\widetilde{O}(\max _{(p, k) \in \mathcal{P}}n_0^{-\frac{p((q+1)\wedge2)}{(2p+k)(d+1)}})$. This additional term, corresponding to $\mathbb{E}[\epsilon_{n_0, n}^{q+1}]$, is the price paid for avoiding manual annotation. 
This term can be small when the nuisance functions are estimated accurately, but it may become non-negligible when the audited clean sample size $n_0$ is limited or when uniform estimation is difficult. Therefore, Pro-PT should be viewed as a cost-efficient alternative to Pro-AT: it reduces the need for additional manual annotation while still allowing the large noisy sample to contribute to downstream learning.
\begin{Rem}\label{rem:other_learners}
The proofs of Theorems \ref{Mth 2} and \ref{Mth 3} isolate two learner-specific ingredients: the extraction rate from Section \ref{section 4.2} and the learning rate of the final classifier on the mixed data $D_n^{\mathrm{AT}}$ or $D_n^{\mathrm{PT}}$. Therefore, the same proof strategy can be adapted to other learners. For example, in simpler structured settings, one may replace the neural-network learner with other learners, such as B-spline. Details are deferred to Supplementary~\ref{Appendix C}.
\end{Rem}

\section{Simulation}\label{section 5}
We consider three simulation examples, with $\sigma(t)=(1+e^{-t})^{-1}$. Unless otherwise stated, we use $(n_0,n)=(500,5000)$. \textbf{Example 1.} (Low-dimensional model) Let $\bm{X}=(X_1,\ldots,X_{10})\in\mathbb{R}^{10}$, with $\eta(\bm{X})=\sigma\{2\sin(2X_1)+X_2\}$. 
\textbf{Example 2.} (High-dimensional sparse model) We retain the posterior in Example 1 and increase the dimension to $100$, with $\bm{X}=(X_1,\ldots,X_{100})\in\mathbb{R}^{100}$. 
\textbf{Example 3.} (High-dimensional model) Let $\bm{X}=(X_1,\ldots,X_{100})\in\mathbb{R}^{100}$, with $\eta(\bm{X})=\sigma\{6(\|\bm{X}\|_2-1)\}$.

The full covariate distributions and auxiliary sampling variables are specified in Supplementary~\ref{Appendix D.1}. For class-dependent noise, we incorporate two distinct categories of label noise: low noise, characterized by $\rho_{-}(\bm{x})\equiv0.1$ and $\rho_{+}(\bm{x})\equiv0.3$, and high noise, defined by $\rho_{-}(\bm{x})\equiv0.15$ and $\rho_{+}(\bm{x})\equiv0.35$. Besides, we investigate the more challenging scenario of instance-dependent noise. For this setting, we use $n_0\in\{1000,2000\}$. The corresponding noise generation process is detailed in Supplementary~\ref{Appendix D.1}.

We consider the following methods for a comprehensive comparison: 
(1) Vanilla-$D$, a baseline trained exclusively on the small clean dataset; 
(2) Vanilla-$D_\rho$, a baseline trained on the entire large noisy dataset; 
(3) Co-teaching \cite{han2018co}, a robust loss method that employs dual-network mutual learning; 
(4) NCR \cite{iscen2022learning}, a robust loss method that leverages neighbor consistency regularization; 
(5) ALC(10\%) \cite{bernhardt2022active}, an active learning-based method that adaptively corrects labels by ranking instances. To ensure a fair comparison under budget constraints, we set its relabeling budget to be $10\%n$; 
(6) Our proposed Pro-PT; 
(7) Our proposed Pro-AT(10\%), a budget-constrained variant of our active tagging method, where the number of queries to the expert is also fixed at $10\% n$; 
(8) Our proposed Pro-AT; 
(9) The Oracle model, a benchmark trained on the large dataset with fully clean labels.

For comparison, we set $N = 20000$ to build $D_{\text{eval}}$ and  employ a Monte Carlo method to compute the empirical excess 0-1 risk $\widehat{\mathcal{E}}^{0/1}_D(f, h_D^o)$. In addition, we use two complementary metrics: Area Under the Receiver Operating Characteristic Curve (AUC) and $F_1$-score ($F_1$). 
For Pro-AT(10\%), where the relabeling budget must be $10\%$ of the dataset, we use binary search to adjust $\tau$. 
Other details are deferred to Supplementary~\ref{Appendix D.1}.

We conduct 500 independent trials for each experiment and present a representative subset of our findings in Tables \ref{tab:result_Simulation_class}-\ref{tab:result_Simulation_instance}. The complete results, including those based on three additional learners other than neural networks, are detailed in Supplementary~\ref{Appendix D.1}. Results are presented as Mean(Standard deviation), with the best result (excluding the oracle) highlighted in bold. For class-dependent noise as shown in Table \ref{tab:result_Simulation_class}, Pro-AT shows stability against increased noise levels. For instance, while the performance of competing methods like Vanilla-$D_\rho$ degrades in the high noise condition, Pro-AT maintains an almost perfectly stable $F_1$ score, declining negligibly from 0.8488 to 0.8487. 
\begin{table*}[ht]
    \small
    \centering
    \caption{Performance Comparison (Class-dependent Noise)}
    \resizebox{0.9\textwidth}{!}{
    \begin{tabular}{l|lcccccc}
        \toprule  
        \multirow{2}{*}{Example} & Noise level & \multicolumn{3}{c}{Low} & \multicolumn{3}{c}{High} \\ \cline{3-8} 
        & Metric & $\widehat{\mathcal{E}}^{0/1}_D(f, h_D^o)$ (↓) & AUC (↑) & $F_1$ (↑) & $\widehat{\mathcal{E}}^{0/1}_D(f, h_D^o)$  (↓) & AUC (↑) & $F_1$ (↑)\\
        \hline
        \multirow{9}{*}{Example 1} 
        & Vanilla-$D$  & 0.1383(0.0713) & 0.8481(0.0505) & 0.7403(0.1116) & 0.1383(0.0713) & 0.8481(0.0505) & 0.7403(0.1116)\\ 
        & Vanilla-$D_\rho$  & 0.0749(0.0068) & 0.9047(0.0023) & 0.7905(0.0104)  & 0.0920(0.0099) & 0.9033(0.0024) & 0.7641(0.0157)\\ 
        & Co-teaching & 0.0598(0.0068) & 0.9042(0.0024) & 0.8142(0.0104) & 0.0674(0.0091) & 0.9023(0.0027) & 0.8034(0.0141)\\
        & NCR & 0.0714(0.0063) & 0.9061(0.0022) & 0.7955(0.0095)  & 0.0881(0.0093) & 0.9055(0.0023) & 0.7696(0.0145)\\
        & ALC(10\%) & 0.0713(0.0063) & 0.9054(0.0022) & 0.7959(0.0097)  & 0.0849(0.0088) & 0.9047(0.0023) & 0.7750(0.0138)\\
        \cline{2-8}
        & \cellcolor{LightPink}Pro-PT & \cellcolor{LightPink}0.0425(0.0040) & \cellcolor{LightPink}0.9061(0.0022) & \cellcolor{LightPink}0.8465(0.0047)  & \cellcolor{LightPink}0.0456(0.0063) & \cellcolor{LightPink}0.9056(0.0023)  & \cellcolor{LightPink}0.8442(0.0066)\\
        & \cellcolor{LightPink}Pro-AT(10\%)  & \cellcolor{LightPink}0.0398(0.0030)  & \cellcolor{LightPink}{\bf 0.9066(0.0021)} & \cellcolor{LightPink}0.8487(0.0038)  & \cellcolor{LightPink}0.0416(0.0044) & \cellcolor{LightPink}0.9063(0.0022)  & \cellcolor{LightPink}0.8487(0.0038)\\
        & \cellcolor{LightPink}Pro-AT  & \cellcolor{LightPink}{\bf 0.0391(0.0023)}  & \cellcolor{LightPink}{\bf 0.9066(0.0021)} & \cellcolor{LightPink}{\bf 0.8488(0.0032)}  & \cellcolor{LightPink}{\bf 0.0392(0.0024)} & \cellcolor{LightPink}{\bf 0.9066(0.0021)}  & \cellcolor{LightPink}{\bf 0.8487(0.0029)}\\
        \cline{2-8}
        & Oracle & 0.0385(0.0021)  & 0.9067(0.0021) & 0.8492(0.0029) & 0.0385(0.0021) & 0.9067(0.0021) & 0.8492(0.0029)\\
        \bottomrule
    \end{tabular}
    }
    \label{tab:result_Simulation_class}
\end{table*}

Under instance-dependent noise, Pro-AT achieves the lowest mean excess 0-1 risk and the highest mean AUC and $F_1$ among the non-Oracle methods in all settings in Table \ref{tab:result_Simulation_instance}. For Example 3 with $n_0=2000$, its mean $F_1$ score is 0.8739. Its performance approaches the fully clean-data Oracle benchmark. 
Pro-PT and Pro-AT(10\%) also achieve lower mean excess 0-1 risk than both Vanilla baselines in all these settings.
\begin{table*}[ht]
    \centering
    \caption{Performance Comparison (Instance-dependent Noise)}
    \resizebox{0.85\textwidth}{!}{
    \begin{tabular}{l|lcccccc}
        \toprule  
        \multirow{2}{*}{Example} & $n_0$ & \multicolumn{3}{c}{1000} & \multicolumn{3}{c}{2000} \\ \cline{3-8} 
        &Metric & $\widehat{\mathcal{E}}^{0/1}_D(f, h_D^o)$ (↓) & AUC (↑) & $F_1$ (↑) & $\widehat{\mathcal{E}}^{0/1}_D(f, h_D^o)$ (↓) & AUC (↑) & $F_1$ (↑)\\
        \hline
        \multirow{9}{*}{Example 2} 
        & Vanilla-$D$ & 0.1563(0.0313) & 0.8059(0.0327) & 0.7294(0.0398) & 0.0859(0.0083) & 0.8744(0.0073) & 0.8016(0.0089)\\ 
        & Vanilla-$D_\rho$ & 0.0873(0.0061) & 0.8871(0.0035) & 0.7799(0.0094) & 0.0873(0.0061) & 0.8871(0.0035) & 0.7799(0.0094)\\ 
        & Co-teaching & 0.0813(0.0057) & 0.8880(0.0034) & 0.7903(0.0087) & 0.0813(0.0057) & 0.8880(0.0034) & 0.7903(0.0087)\\
        & NCR & 0.0847(0.0060) & 0.8876(0.0036) & 0.7838(0.0089) & 0.0847(0.0060) & 0.8876(0.0036) & 0.7838(0.0089)\\
        & ALC(10\%) & 0.0750(0.0047) & 0.8935(0.0028) & 0.7966(0.0072) & 0.0750(0.0047) & 0.8935(0.0028) & 0.7966(0.0072)\\
        \cline{2-8}
        & \cellcolor{LightPink}Pro-PT& \cellcolor{LightPink}0.0826(0.0226) & \cellcolor{LightPink}0.8873(0.0058) & \cellcolor{LightPink}0.7925(0.0398) & \cellcolor{LightPink}0.0720(0.0114) & \cellcolor{LightPink}0.8893(0.0045) & \cellcolor{LightPink}0.8097(0.0199)\\
        & \cellcolor{LightPink}Pro-AT(10\%) & \cellcolor{LightPink}0.0742(0.0193) & \cellcolor{LightPink}0.8918(0.0044) & \cellcolor{LightPink}0.8035(0.0341) & \cellcolor{LightPink}0.0644(0.0077) & \cellcolor{LightPink}0.8939(0.0037) & \cellcolor{LightPink}0.8189(0.0129)\\
        & \cellcolor{LightPink}Pro-AT& \cellcolor{LightPink}{\bf 0.0665(0.0146)} & \cellcolor{LightPink}{\bf 0.8947(0.0042)} & \cellcolor{LightPink}{\bf 0.8138(0.0247)} & \cellcolor{LightPink}{\bf 0.0599(0.0064)} & \cellcolor{LightPink}{\bf 0.8984(0.0029)} & \cellcolor{LightPink}{\bf 0.8244(0.0104)}\\ 
        \cline{2-8}
        & Oracle & 0.0541(0.0027) & 0.8985(0.0025) & 0.8332(0.0037) & 0.0541(0.0027) & 0.8985(0.0025) & 0.8332(0.0037)\\
        \hline
        \multirow{9}{*}{Example 3} 
        & Vanilla-$D$ & 0.1280(0.0840) & 0.9226(0.0027) & 0.8032(0.0495) & 0.0299(0.0336) & 0.9258(0.0022) & 0.8658(0.0224)\\ 
        & Vanilla-$D_\rho$ & 0.0791(0.0134) & 0.9206(0.0028) & 0.7808(0.0200) & 0.0791(0.0134) & 0.9206(0.0028) & 0.7808(0.0200)\\ 
        & Co-teaching & 0.0666(0.0115) & 0.9209(0.0042) & 0.7995(0.0163) & 0.0666(0.0115) & 0.9209(0.0042) & 0.7995(0.0163)\\
        & NCR & 0.0592(0.0079) & 0.9213(0.0023) & 0.8101(0.0113) & 0.0592(0.0079) & 0.9213(0.0023) & 0.8101(0.0113)\\
        & ALC(10\%) & 0.0762(0.0104) & 0.9216(0.0025) & 0.7851(0.0154) & 0.0762(0.0104) & 0.9216(0.0025) & 0.7851(0.0154)\\
        \cline{2-8}
        & \cellcolor{LightPink}Pro-PT& \cellcolor{LightPink}0.0747(0.0520) & \cellcolor{LightPink}0.9156(0.0094) & \cellcolor{LightPink}0.7835(0.0799) & \cellcolor{LightPink}0.0281(0.0157) & \cellcolor{LightPink}0.9230(0.0037) & \cellcolor{LightPink}0.8527(0.0210)\\
        & \cellcolor{LightPink}Pro-AT(10\%) & \cellcolor{LightPink}0.0463(0.0387) & \cellcolor{LightPink}0.9215(0.0071) & \cellcolor{LightPink}0.8258(0.0553) & \cellcolor{LightPink}0.0156(0.0096) & \cellcolor{LightPink}0.9264(0.0026) & \cellcolor{LightPink}0.8684(0.0127)\\
        & \cellcolor{LightPink}Pro-AT& \cellcolor{LightPink}{\bf 0.0232(0.0155)} & \cellcolor{LightPink}{\bf 0.9261(0.0025)} & \cellcolor{LightPink}{\bf 0.8583(0.0204)} & \cellcolor{LightPink}{\bf 0.0114(0.0050)} & \cellcolor{LightPink}{\bf 0.9272(0.0021)} & \cellcolor{LightPink}{\bf 0.8739(0.0068)}\\
        \cline{2-8}
        & Oracle & 0.0054(0.0019) & 0.9282(0.0021) & 0.8815(0.0030) & 0.0054(0.0019) & 0.9282(0.0021) & 0.8815(0.0030)\\
        \bottomrule
    \end{tabular}
    }
    \label{tab:result_Simulation_instance}
\end{table*}

\begin{figure*}[ht]
    \centering
    \subfloat[Cost-efficiency Tradeoff]{
    \includegraphics[width=0.44\linewidth]{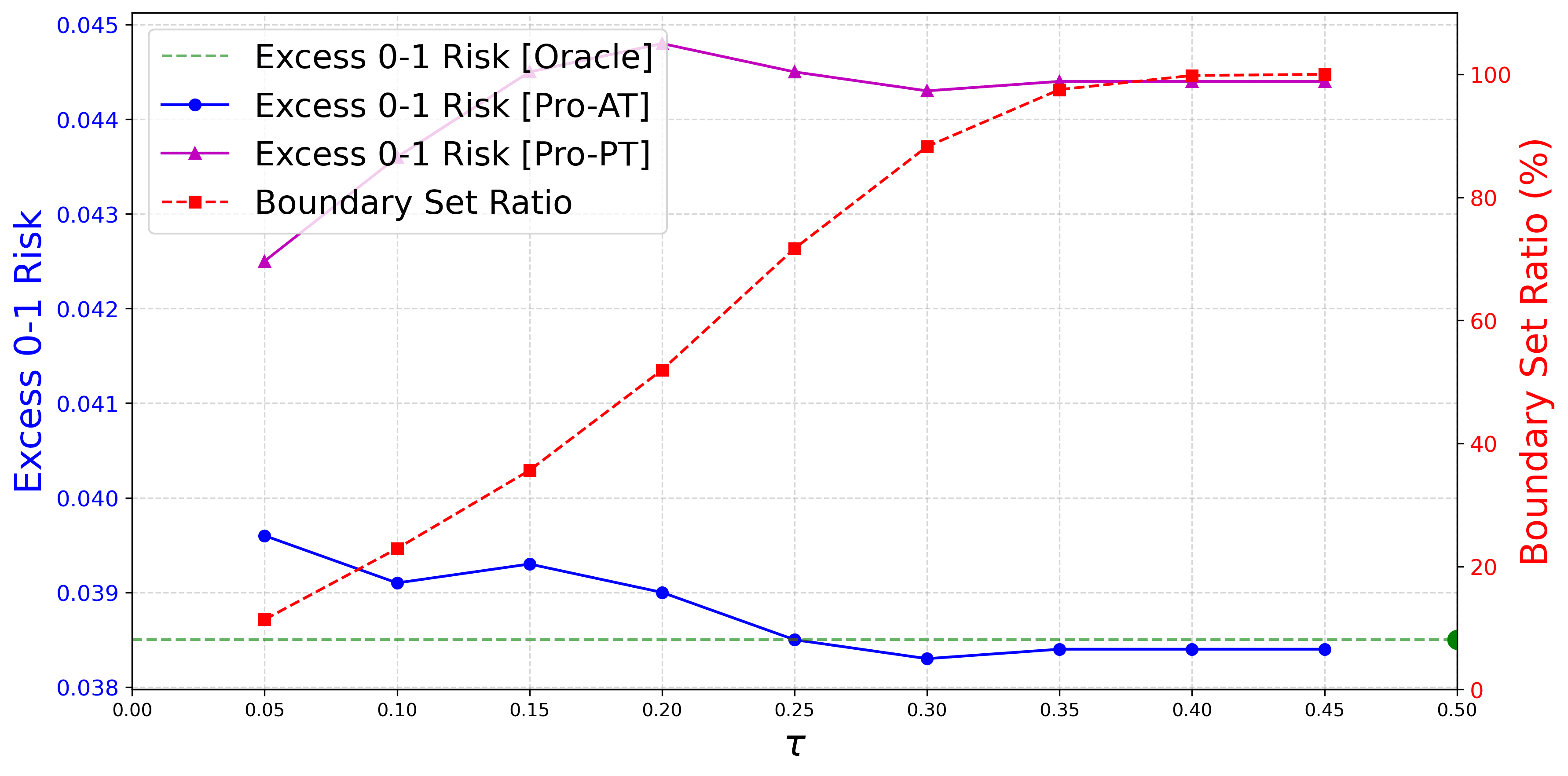}
    \label{Fig: AL_cost}
    }
    \hfill
    \subfloat[Extraction between ALC and Pro-AT]{
    \includegraphics[width=0.41\linewidth]{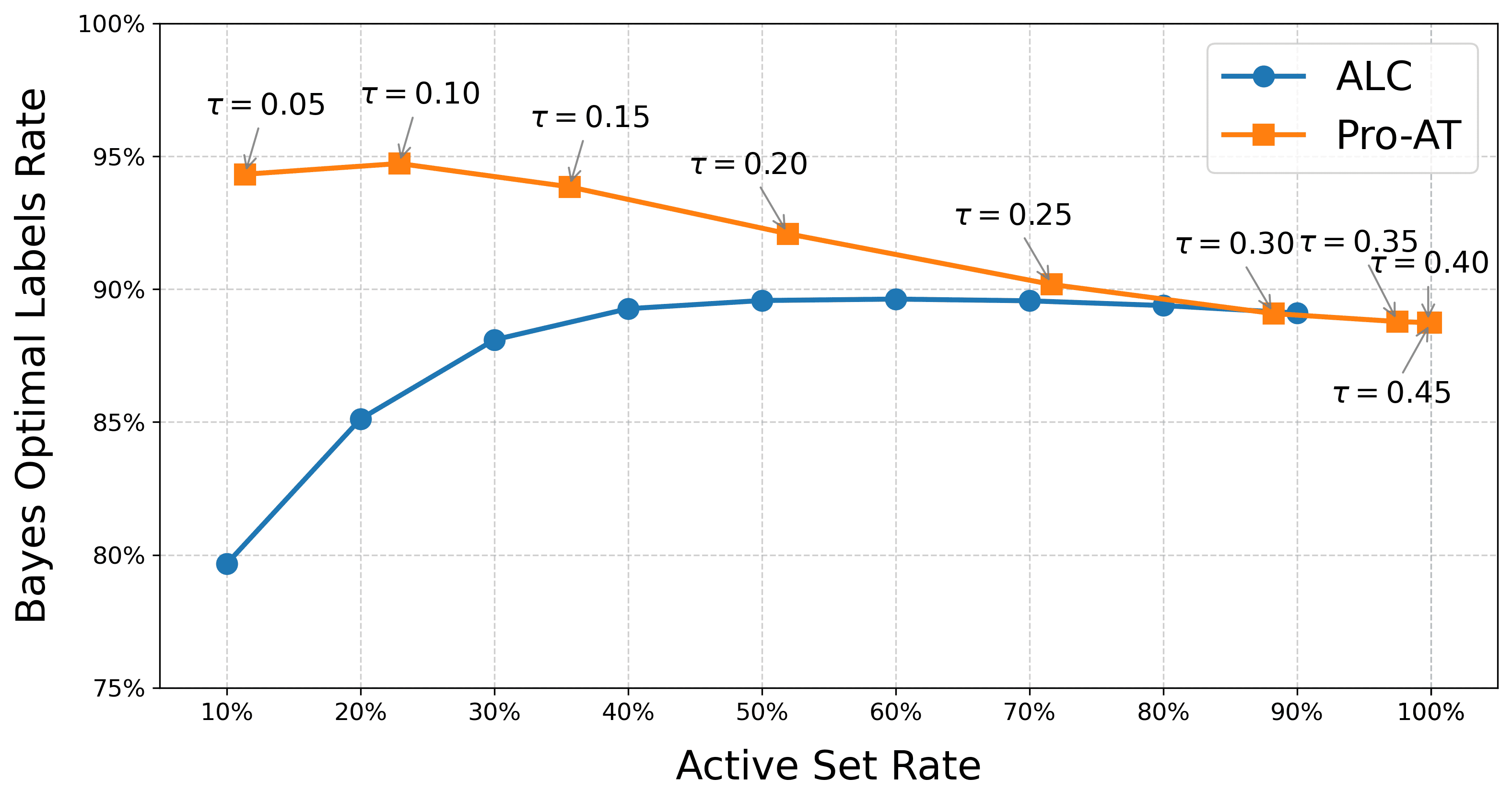}
    \label{Fig: Pro-ATvsALC}
    }
    \caption{Analysis of Extraction Method on Example 1 (Class-dependent Low Noise).}
\end{figure*}

Figure \ref{Fig: AL_cost} illustrates the impact of the hyperparameter $\tau$. Excess 0-1 risk of Pro-AT rapidly decreases as $\tau$ increases, achieving near-oracle performance for $\tau\geq 0.25$. 
This cost-efficiency is particularly evident when compared to ALC (Figure \ref{Fig: Pro-ATvsALC}), where Pro-AT consistently outperforms, especially with smaller Active Set sizes, corresponding to lower re-labeling costs. Pro-AT excels at minimizing low-probability mismatches (e.g., assigning $y=-1$ to samples where $\eta(x)>0.5$),  allowing for efficient extraction at low annotation costs.

\begin{figure*}[ht]
    \centering
    \subfloat[Excess 0-1 Risk]{
    \includegraphics[width=0.45\linewidth]{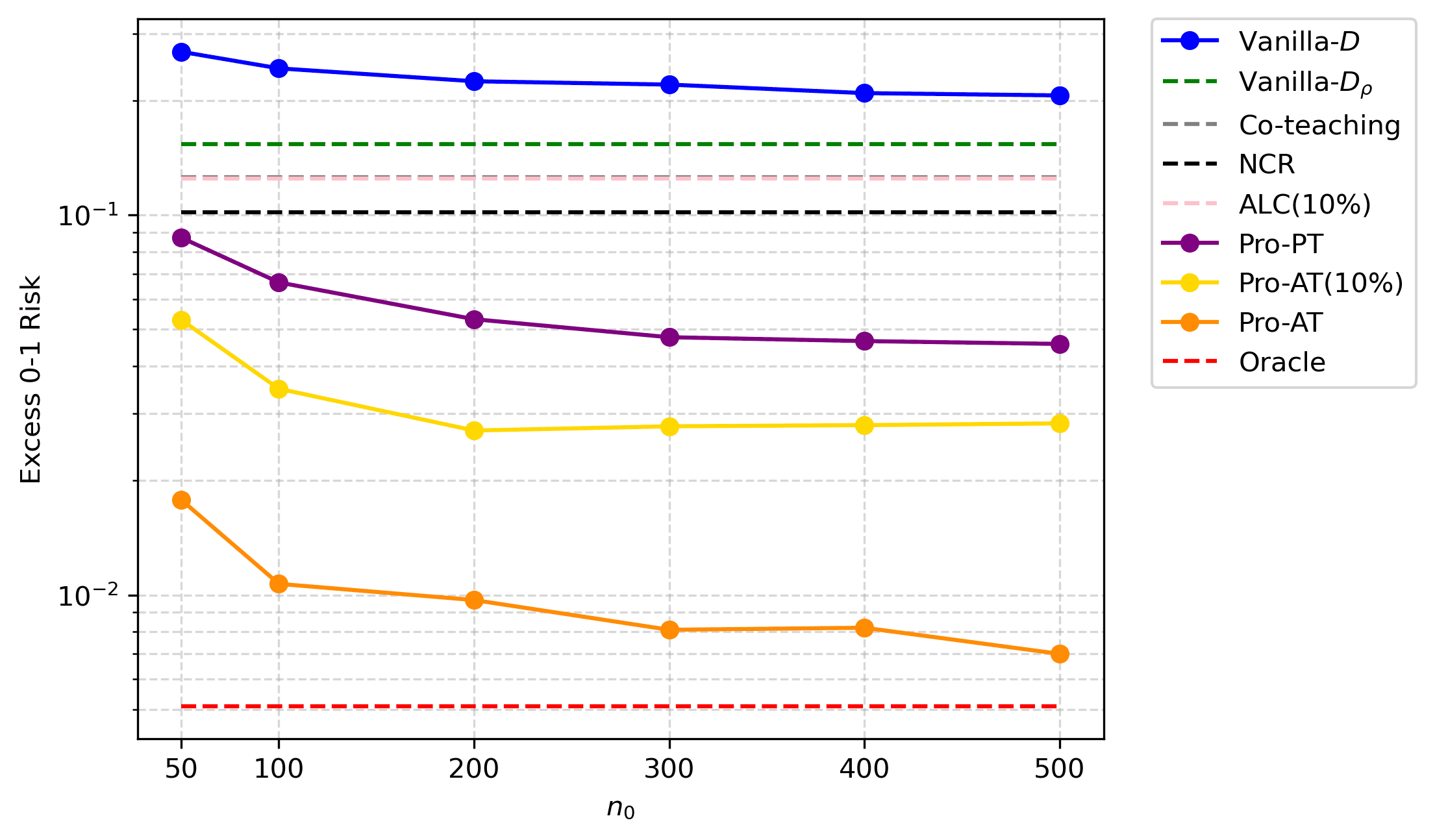}
    \label{Fig: n_0 Excess Risk}
    }
    \hfill
    \subfloat[$F_1$ Score]{
    \includegraphics[width=0.45\linewidth]{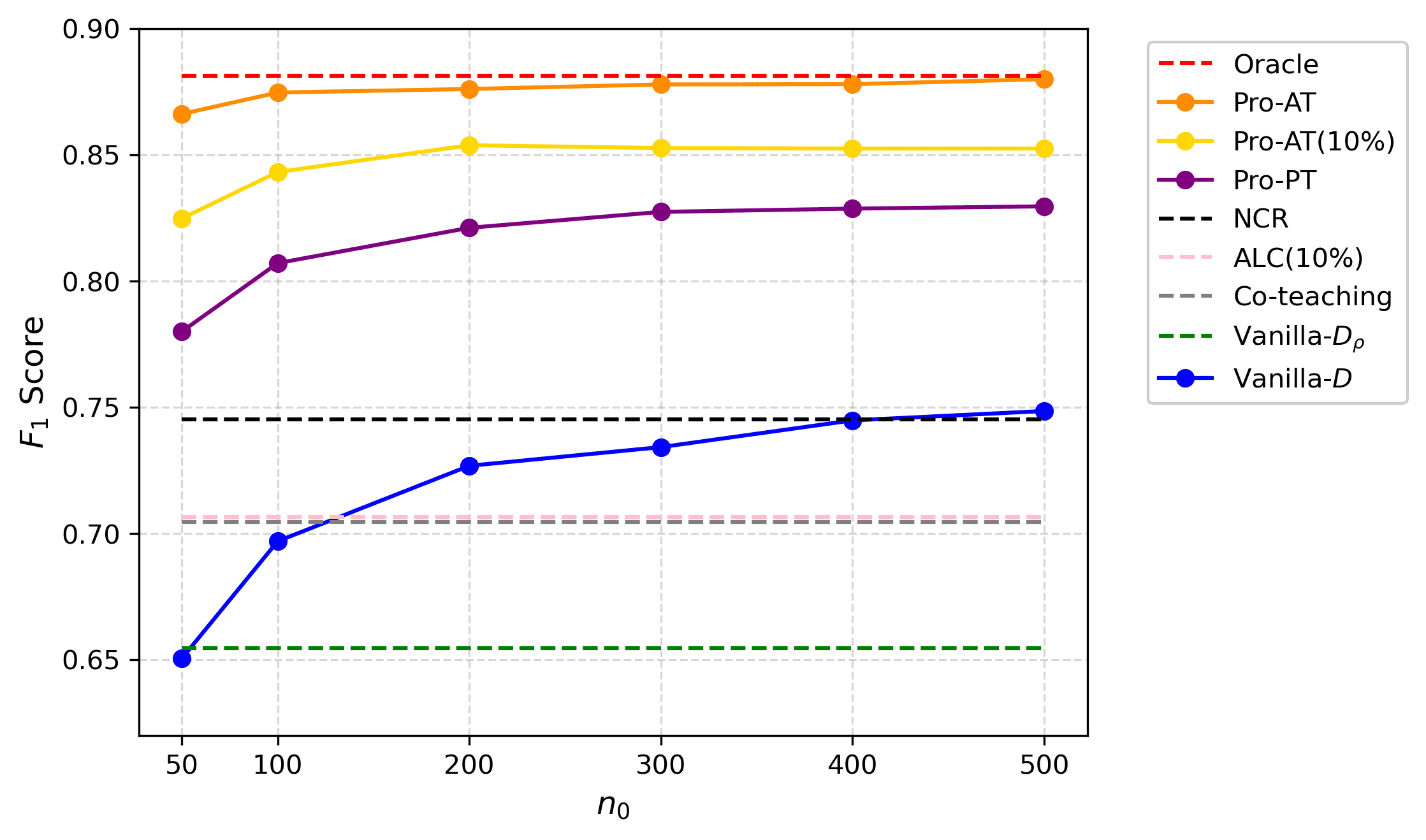}
    \label{Fig: n_0 f1}
    }
    \caption{Analysis of the Impact of $n_0$ in Example 3 (Class-dependent Low noise).}
\end{figure*}

We evaluated the impact of $n_0$ across different values shown in Figure \ref{Fig: n_0 Excess Risk}, \ref{Fig: n_0 f1}. As illustrated, the performance of all our proposed methods improves with a larger $n_0$, evidenced by a corresponding reduction in excess 0-1 risk and an increase in the $F_1$ Score. 
Notably, even with a minimal initial sample of $n_0=50$, which can represent as little as 1\% of the total dataset, our methods  substantially outperform all competing baselines. 

\section{Real Application}\label{section 6}
To demonstrate the practical utility of our proposed framework, we apply it to a critical real-world problem: the diagnosis of pneumonia from chest radiographs. A common workaround is to use NLP to automatically extract labels from vast Electronic Health Records. This process generates exactly the kind of large, noisy datasets. This application serves as a perfect testbed to evaluate our method's ability.

Our study initially utilizes a well-known public dataset, the ChestX-ray dataset\footnote{\url{https://nihcc.app.box.com/v/ChestXray-NIHCC/folder/36938765345}}, from the National Institutes of Health (NIH), which contains 112,120 chest X-ray images from over 30,000 patients \cite{wang2017hospital}. In this original dataset, labels are automatically generated via NLP, which is noisy. The clean labels for our task come from a subsequent initiative by the Radiological Society of North America (RSNA) for its Pneumonia Detection Challenge. 
For this challenge, a subset of the ChestX-ray dataset is meticulously re-reviewed and annotated by expert radiologists, specifically for the presence of pneumonia-like lung opacities\footnote{\url{https://www.kaggle.com/competitions/rsna-pneumonia-detection-challenge/data}}. By combining these two sources, we construct a benchmark dataset which we refer to as Noisy Chest X-Ray datasets (NoisyCXR). The procedure for creating this pairing of noisy labels and expert-verified clean labels follows the methodologies established in prior works \cite{bernhardt2022active}. Samples are hence classified into two categories: ``pneumonia-like''  and ``no pneumonia-like''. The final dataset contains 26,684 images with a noise rate of 12.6\%, partly visualized in Table \ref{tab:NoisyCXRFig}, with more details in Supplementary~\ref{Appendix D.2}.

To prepare the data for our framework, we follow the established practice of preprocessing the raw images into features using a powerful pre-trained deep learning model \cite{yao2024eva}. The resulting dataset is then partitioned using random sampling into three subsets in each trial: 70\% for the large noisy training set, 10\% for the small clean training set, and 20\% for evaluation. All specific training details are deferred to Supplementary~\ref{Appendix D.2}. We evaluate our methods against the same baseline approaches. Given that the true data distribution is unknown, Accuracy (ACC) serves as a practical substitute for $\widehat{\mathcal{E}}_D\left(f, h_D^o\right)$. We also report the Area Under the Precision-Recall Curve (PR-AUC).

For our proposed methods, we primarily report the results using the class-dependent noise estimation procedure in Table \ref{tab:result_NoisyCXR}. The results for the instance-dependent case are comparable, which are deferred to the Supplementary Material for completeness.
\begin{table*}[ht] 
     \centering 
     \caption{Performance Comparison in NoisyCXR.} 
     \resizebox{0.7\textwidth}{!}{
     \begin{tabular}{l|cccc} 
         \toprule  Method & ACC (↑) & AUC (↑) & $F_1$ (↑) & PR-AUC (↑)\\ 
         \hline 
         Vanilla-$D$ & 0.8248(0.0057) & 0.8151(0.0070) & 0.5043(0.0306) & 0.6275(0.0159) \\ 
         Vanilla-$D_\rho$ & 0.8151(0.0067) & 0.8630(0.0049) & 0.3726(0.0436) & 0.6530(0.0138) \\ 
         Co-teaching  & 0.8169(0.0091) & 0.8645(0.0049) & 0.3870(0.0680) & 0.6562(0.0136) \\ 
         NCR  & 0.8160(0.0082) & 0.8642(0.0049) & 0.3782(0.0636) & 0.6557(0.0138) \\ 
         ALC(10\%)  & 0.8219(0.0088) & 0.8665(0.0048) & 0.4259(0.0639) & 0.6609(0.0136) \\ 
         \cline{2-5}  
         Pro-PT & \cellcolor{LightPink}0.8288(0.0061) & \cellcolor{LightPink}0.8600(0.0050) & \cellcolor{LightPink}0.5646(0.0420) & \cellcolor{LightPink}0.6475(0.0139)\\ 
         Pro-AT(10\%)  & \cellcolor{LightPink}0.8312(0.0050) & \cellcolor{LightPink}0.8619(0.0050) & \cellcolor{LightPink}0.5680(0.0335) & \cellcolor{LightPink}0.6525(0.0138)\\ 
         Pro-AT & \cellcolor{LightPink}{\bf 0.8338(0.0047)}  &\cellcolor{LightPink}{\bf 0.8667(0.0054)} & \cellcolor{LightPink}{\bf 0.5720(0.0323)} & \cellcolor{LightPink}{\bf 0.6616(0.0142)}\\ 
         \cline{2-5} 
         Oracle & 0.8350(0.0050) & 0.8703(0.0047) & 0.5739(0.0319) & 0.6682(0.0134)\\ 
         \bottomrule 
    \end{tabular} 
    }
    \label{tab:result_NoisyCXR} 
\end{table*}
Due to the significant class imbalance, metrics like the $F_1$ are more informative indicators of model performance than ACC, as they provide a better assessment of a model's ability to correctly identify the minority class. In this critical regard, our proposed Pro-AT model demonstrates outstanding results. It achieves an $F_1$ of 0.5720, substantially outperforming all competing methods. 
Furthermore, Pro-AT achieves an $F_1$ score close to the fully clean-data Oracle benchmark of 0.5739. Among all non-oracle methods, Pro-AT also achieves the highest scores across the remaining metrics, validating its effectiveness and robustness for application.

\section{Discussion}\label{section 7}
This work presents a unified nonparametric information transfer-and-fusion framework for learning from a large noisy dataset and a small clean one. Its core components, Bayes Optimal data extraction and boundary-sample handling via artificial or pseudo-tagging, are supported by theory and experiments. While focused on binary classification, the strategy of partitioning data into high-confidence and ambiguous subsets extends to multi-class settings. More broadly, the work reframes label noise not as a single-dataset issue but as a transfer-and-fusion problem tailored to the ``large noisy, small clean'' scenario. Strong results on NoisyCXR demonstrate practical relevance, offering a foundation for future methodological and theoretical advances.


\medskip
\noindent\textbf{Disclosure Statement.} We declare no conflicts of interest.

\medskip
\noindent\textbf{Supplementary Material.} Additional theoretical and numerical results are provided after the references.

\medskip
\noindent\textbf{Acknowledgements.} 
Mingyang Ren is the corresponding author.
This work was partially supported by the National Natural Science Foundation of China (12501373), Shenzhen Medical Research Fund (E250200620, E250200621), the Shanghai Sailing Program (24YF2721900).

\bibliographystyle{IEEEtran}
\bibliography{reference}

\clearpage
\onecolumn
\phantomsection
\pdfbookmark[0]{Supplementary Material}{supplement}
\begin{center}
{\Large\bfseries Supplementary Material}
\end{center}
\medskip
\appendices
\providecommand{\theHDef}{}
\renewcommand{\theHDef}{supp.\arabic{Def}}
\providecommand{\theHCon}{}
\renewcommand{\theHCon}{supp.\arabic{Con}}
\providecommand{\theHCor}{}
\renewcommand{\theHCor}{supp.\arabic{Cor}}
\providecommand{\theHRem}{}
\renewcommand{\theHRem}{supp.\arabic{Rem}}
\providecommand{\theHMth}{}
\renewcommand{\theHMth}{supp.\arabic{Mth}}
\providecommand{\theHLem}{}
\renewcommand{\theHLem}{supp.\arabic{Lem}}
\providecommand{\theHPro}{}
\renewcommand{\theHPro}{supp.\arabic{Pro}}
\providecommand{\theHRes}{}
\renewcommand{\theHRes}{supp.\arabic{Res}}
\providecommand{\theHProof}{}
\renewcommand{\theHProof}{supp.\arabic{Proof}}
\renewcommand{\theHfigure}{\thesection.\arabic{figure}}
\renewcommand{\theHtable}{\thesection.\arabic{table}}

\renewcommand{\theDef}{S\arabic{Def}}
\renewcommand{\theCon}{S\arabic{Con}}
\renewcommand{\theCor}{S\arabic{Cor}}
\renewcommand{\theRem}{S\arabic{Rem}}
\renewcommand{\theMth}{S\arabic{Mth}}
\renewcommand{\theLem}{S\arabic{Lem}}
\setcounter{Def}{0}
\setcounter{Con}{0}
\setcounter{Cor}{0}
\setcounter{Rem}{0}
\setcounter{Mth}{0}
\setcounter{Lem}{0}

This supplementary material is organized as follows. Section~\ref{Appendix A} proves the generic plug-in result in Theorem~\ref{Mth generic} of the main text. Section~\ref{Appendix B} provides the technical details and proofs for the neural-network results in Section~\ref{section 4.2} of the main text. Section~\ref{Appendix C} presents the covariate-shift extension, a discussion of pseudo-label resampling, additional corollaries, and a B-spline implementation. Section~\ref{Appendix D} gives further details of the simulations and the NoisyCXR application.

\renewcommand{\theequation}{A.\arabic{equation}}
\renewcommand{\thetable}{A.\arabic{table}}
\renewcommand{\thefigure}{A.\arabic{figure}}
\setcounter{equation}{0} 
\setcounter{table}{0} 
\setcounter{figure}{0}

\section{Proof of Theorem~\ref{Mth generic}}\label{Appendix A}

Under Condition~\ref{Con 1}(a) with class-dependent noise, recall that 
\begin{equation*}
    \epsilon_{n_0,n} = \frac{1}{2}\left(\left|\rho_{+}-\widehat{\rho}_{+}\right|+\left|\rho_{-}-\widehat{\rho}_{-}\right|\right)+\sup _{\bm{x} \in \mathcal{X}}\left|\eta_\rho(\bm{x})-\widehat{\eta}_\rho(\bm{x})\right|.
\end{equation*}
The first two terms are controlled by Lemma~\ref{Lem 1} of the main text, restated below and proved in Section~\ref{sec:A.1}.

\noindent {\bf Lemma~\ref{Lem 1}.} {\it Under Condition~\ref{Con 1}(a), we have $\mathbb{E}_{n_0,D}\left|\rho_{+}-\widehat{\rho}_{+}\right|^2 \lesssim n_0^{-1}.$ Symmetrically, the same bound holds for $\mathbb{E}_{n_0,D}\left|\rho_{-}-\widehat{\rho}_{-}\right|^2$.}

\noindent By Lemma~\ref{Lem 1} and the definition of $\varepsilon_{1,n}$, we have
\begin{equation*}
    \mathbb{E}_{n_0,D}\mathbb{E}_{n,D_{\rho}}\epsilon_{n_0,n}^2 \lesssim n_0^{-1}+\varepsilon_{1,n}.
\end{equation*}
Next, under Condition~\ref{Con 1}(b) with instance-dependent noise, recall that 
\begin{equation*}
    \epsilon_{n_0,n} = \frac{1}{2}\left(\sup _{\bm{x} \in \mathcal{X}}\left|\rho_{+}(\bm{x})-\widehat{\rho}_{+}(\bm{x})\right|+\sup _{\bm{x} \in \mathcal{X}}\left|\rho_{-}(\bm{x})-\widehat{\rho}_{-}(\bm{x})\right|\right)+\sup _{\bm{x} \in \mathcal{X}}\left|\eta_\rho(\bm{x})-\widehat{\eta}_\rho(\bm{x})\right|.
\end{equation*}
By the definitions of $\varepsilon_{1,n}$ and $\varepsilon_{1,n}$, we have
\begin{equation*}
    \mathbb{E}_{n_0,D}\mathbb{E}_{n,D_{\rho}}\epsilon_{n_0,n}^2 \lesssim \varepsilon_{1,n_0}+\varepsilon_{1,n}.
\end{equation*}

We next consider the tagging step. For Pro-AT, since $\widehat{f}_{n}^{\mathrm{AT}}$ is trained on $D^{\mathrm{AT}}$, we first need to decompose the error,
\begin{equation*}
    \mathcal{E}^{0/1}_D\left(\widehat{f}_{n}^{\mathrm{AT}}, h_D^o\right)=\mathcal{E}^{0/1}_D\left(\widehat{f}_{n}^{\mathrm{AT}}, h_D^o\right) - \mathcal{E}^{0/1}_{D^{\mathrm{AT}}}\left(\widehat{f}_{n}^{\mathrm{AT}}, h_D^o\right) + \mathcal{E}^{0/1}_{D^{\mathrm{AT}}}\left(\widehat{f}_{n}^{\mathrm{AT}}, h_D^o\right).
\end{equation*}
It is noted that
\begin{equation*}
    \begin{aligned}
        \mathbb{E}_D&[\mathbb{I}(Y\operatorname{sign}(\widehat{f}_{n}^{\mathrm{AT}}(\bm{X}))<0)] = \mathbb{E}_{\bm{X}}\mathbb{E}_{Y\mid\bm{X}}\mathbb{I}(Y\operatorname{sign}(\widehat{f}_{n}^{\mathrm{AT}}(\bm{X}))<0) \\
        &=\mathbb{E}_{\bm{X}}\left[\mathbb{I}(\widehat{f}_{n}^{\mathrm{AT}}(\bm{X})<0)P_D(Y=1|\bm{X})+\mathbb{I}(\widehat{f}_{n}^{\mathrm{AT}}(\bm{X})\geq0)P_D(Y=-1|\bm{X})\right]\\
        &=\mathbb{E}_{\bm{X}}\left[\mathbb{I}(\widehat{f}_{n}^{\mathrm{AT}}(\bm{X})<0)\eta(\bm{X})+\mathbb{I}(\widehat{f}_{n}^{\mathrm{AT}}(\bm{X})\geq0)(1-\eta(\bm{X}))\right].\\
        \mathbb{E}_D&[\mathbb{I}(Y h_D^o(\bm{X})<0)] = \mathbb{E}_{\bm{X}}\mathbb{E}_{Y\mid\bm{X}}\mathbb{I}(Yh_D^o(\bm{X})<0) \\
        &=\mathbb{E}_{\bm{X}}\left[\mathbb{I}(h_D^o(\bm{X})<0)P_D(Y=1|\bm{X})+\mathbb{I}(h_D^o(\bm{X})>0)P_D(Y=-1|\bm{X})\right]\\
        &=\mathbb{E}_{\bm{X}}\left[\mathbb{I}(\eta(\bm{X})<\frac{1}{2})\eta(\bm{X})+\mathbb{I}(\eta(\bm{X})\geq\frac{1}{2})(1-\eta(\bm{X}))\right].
    \end{aligned}
\end{equation*}
We expand the first two term, $\mathcal{E}^{0/1}_D\left(\widehat{f}_{n}^{\mathrm{AT}}, h_D^o\right)$ and $\mathcal{E}^{0/1}_{D^{\mathrm{AT}}}\left(\widehat{f}_{n}^{\mathrm{AT}}, h_D^o\right)$,
\begin{equation*}
    \begin{aligned}
        \mathcal{E}^{0/1}_D\left(\widehat{f}_{n}^{\mathrm{AT}}, h_D^o\right)= & \mathbb{E}_{\bm{X}}\left[\eta(\bm{X})\left(\mathbb{I}(\widehat{f}_{n}^{\mathrm{AT}}(\bm{X})<0)-\mathbb{I}(\eta(\bm{X})<\frac{1}{2})\right)\right.\\
            &\left.+(1-\eta(\bm{X}))\left(\mathbb{I}(\widehat{f}_{n}^{\mathrm{AT}}(\bm{X})\geq0)-\mathbb{I}(\eta(\bm{X})\geq\frac{1}{2})\right)\right].
    \end{aligned}
\end{equation*}
Similarly, we have 
\begin{equation*}
    \begin{aligned}
        \mathcal{E}^{0/1}_{D^{\mathrm{AT}}}\left(\widehat{f}_{n}^{\mathrm{AT}}, h_D^o\right)=& \mathbb{E}_{\bm{X}}\left[\eta^{\mathrm{AT}}(\bm{X})\left(\mathbb{I}(\widehat{f}_{n}^{\mathrm{AT}}(\bm{X})<0)-\mathbb{I}(\eta(\bm{X})<\frac{1}{2})\right)\right.\\
        & \left.+(1-\eta^{\mathrm{AT}}(\bm{X}))\left(\mathbb{I}(\widehat{f}_{n}^{\mathrm{AT}}(\bm{X})\geq0)-\mathbb{I}(\eta(\bm{X})\geq\frac{1}{2})\right)\right],
    \end{aligned}
\end{equation*}
Thus, we obtain that
\begin{equation}\label{Generic decompose AT}
    \begin{aligned}
        \mathcal{E}^{0/1}_D&\left(\widehat{f}_{n}^{\mathrm{AT}}, h_D^o\right)=\mathcal{E}^{0/1}_D\left(\widehat{f}_{n}^{\mathrm{AT}}, h_D^o\right) - \mathcal{E}^{0/1}_{D^{\mathrm{AT}}}\left(\widehat{f}_{n}^{\mathrm{AT}}, h_D^o\right) + \mathcal{E}^{0/1}_{D^{\mathrm{AT}}}\left(\widehat{f}_{n}^{\mathrm{AT}}, h_D^o\right) \\
        &=2\mathbb{E}_{\bm{X}}\left\{(\eta(\bm{X})-\eta^{\mathrm{AT}}(\bm{X}))\left[\mathbb{I}(\eta(\bm{X})\geq\frac{1}{2})-\mathbb{I}(\widehat{f}_{n}^{\mathrm{AT}}(\bm{X})\geq0)\right]\right\} + \mathcal{E}_{D^{\mathrm{AT}}}^{0/1}\left(\widehat{f}_{n}^{\mathrm{AT}}, h_D^o\right) \\
        &\leq \mathcal{E}_{D^{\mathrm{AT}}}^{0/1}\left(\widehat{f}_{n}^{\mathrm{AT}}, h_D^o\right).
    \end{aligned}
\end{equation}
As for the inequality, recalling the definition of $\eta^{\mathrm{AT}}(\bm{x})$,
\begin{equation*}
    \begin{aligned}
        \eta^{\mathrm{AT}}(\bm{x})=&\mathbb{I}\left(\widehat{\eta}_\rho(\bm{x})>\frac{1}{2}-\frac{1}{2}\left(\widehat{\rho}_{+}(\bm{x})-\widehat{\rho}_{-}(\bm{x})\right)+\tau\right)\\
        &+\mathbb{I}\left(\left|\widehat{\eta}_\rho(\bm{x})-\left(\frac{1}{2}-\frac{1}{2}\left(\widehat{\rho}_{+}(\bm{x})-\widehat{\rho}_{-}(\bm{x})\right)\right)\right|\leq\tau\right) \eta(\bm{x}).
    \end{aligned}
\end{equation*}
Given that $\tau$ is chosen to satisfy the implications in \eqref{drift}, we can observe that when $\eta(\bm{x})\geq\frac{1}{2}$, we have $ \eta(\bm{x})-\eta^{\mathrm{AT}}(\bm{x}) \leq 0$. On the other hand, when $\eta(\bm{x}) < \frac{1}{2}$, we have $\eta(\bm{x})-\eta^{\operatorname{\mathrm{AT}}}(\bm{x}) \geq 0$. Therefore, the expectation term is a non-positive value. Based on the existing theory of various learner to control $\mathcal{E}_{D^{\mathrm{AT}}}^{0/1}\left(\widehat{f}_{n}^{\mathrm{AT}}, h_D^o\right)$, we finally obtain
\begin{equation*}
    \mathbb{E}_{n_0,D}\mathbb{E}_{n,D_\rho}\left\{\mathcal{E}^{0/1}_D(\widehat f^{\mathrm{AT}},h_D^o)\right\} \lesssim \varepsilon_{2,n}.
\end{equation*}
\begin{Rem}
    While dropping the non-positive term in \eqref{Generic decompose AT} provides a valid upper bound, retaining this term allows for a more insightful analysis. We can achieve this by taking the Oracle Bayes Optimal dataset $D^{\star}$ where labels are assigned by the Oracle Bayes classifier, i.e., its corresponding conditional probability is $\eta^{\star}(\bm{x})=\mathbb{I}(\eta(\bm{x})>1 / 2)$. By replacing the active learning dataset $D^{\mathrm{AT}}$ with $D^{\star}$, the expectation term becomes:
    \begin{equation*}
        2 \mathbb{E}_{\bm{X}}\left\{\left(\eta(\bm{X})-\eta^{\star}(\bm{X})\right)\left[\mathbb{I}\left(\eta(\bm{X})\geq\frac{1}{2}\right)-\mathbb{I}\left(\widehat{f}_n^{\star}(\bm{X})\geq0\right)\right]\right\}.
    \end{equation*}
    This expression remains non-positive and $\left|\eta(\bm{x})-\eta^{\star}(\bm{x})\right|\geq \left|\eta(\bm{x})-\eta^{\mathrm{AT}}(\bm{x})\right|$. By analyzing its value instead of eliminating it, we can develop a more optimized and descriptive upper bound for the excess error $\mathcal{E}_D^{0 / 1}\left(\hat{f}_n^{\star}, h_D^o\right)$. 
\end{Rem}

For Pro-PT, we face a new challenge $h_{D^{\mathrm{PT}}}^o\neq h_D^o$. So, we decompose the error as
\begin{equation}\label{Generic decompose}
    \begin{aligned}
        \mathcal{E}_D^{0 / 1}\left(\widehat{f}_n^{\mathrm{PT}}, h_D^o\right)=&\mathcal{E}_D^{0 / 1}\left(\widehat{f}_n^{\mathrm{PT}}, h_D^o\right)-\mathcal{E}_{D^{\mathrm{PT}}}^{0 / 1}\left(\widehat{f}_n^{\mathrm{PT}}, h_D^o\right) \\
        &+\mathcal{E}_{D^{\mathrm{PT}}}^{0 / 1}\left(\widehat{f}_n^{\mathrm{PT}}, h_D^o\right) - \mathcal{E}_{D^{\mathrm{PT}}}^{0 / 1}\left(\widehat{f}_n^{\mathrm{PT}}, h_{D^{\mathrm{PT}}}^o\right) + \mathcal{E}_{D^{\mathrm{PT}}}^{0 / 1}\left(\widehat{f}_n^{\mathrm{PT}}, h_{D^{\mathrm{PT}}}^o\right).
    \end{aligned}
\end{equation}
We define the pseudo-labeling domain, denoted by $\mathrm{BS}$, as
\begin{equation*}
    \mathrm{BS}:=\{\bm{x}:\left|\widehat{\eta}_\rho(\bm{x})-\left(\frac{1}{2}-\frac{1}{2}\left(\widehat{\rho}_{+}(\bm{x})-\widehat{\rho}_{-}(\bm{x})\right)\right)\right| \leq \tau\}.
\end{equation*}
With a sufficiently large $\tau$, we have seen the advantage of BO data, which makes an associated error term non-positive. We now turn to our pseudo-labeling domain $\mathrm{BS}$. Under Condition~\ref{Con 2}, we obtain $P_D(\bm{X}\in\mathrm{BS})\lesssim \epsilon_{n_0,n}^q$. Besides, unlike the BO case, this approach preserves uncertainty by sampling from the estimated probabilities at each epoch, which introduces some errors. 
\begin{equation*}
    \begin{aligned}
        \eta^{\mathrm{PT}}(\bm{x})= & \mathbb{I}\left(\widehat{\eta}_\rho(\bm{x})>\frac{1}{2}-\frac{1}{2}\left(\widehat{\rho}_{+}(\bm{x})-\widehat{\rho}_{-}(\bm{x})\right)+\tau\right) \\
        & +\mathbb{I}\left(\left|\widehat{\eta}_\rho(\bm{x})-\left(\frac{1}{2}-\frac{1}{2}\left(\widehat{\rho}_{+}(\bm{x})-\widehat{\rho}_{-}(\bm{x})\right)\right)\right| \leq \tau\right) \eta^{\mathrm{pse}}(\bm{x}) .
    \end{aligned}
\end{equation*}
In terms of the first part of \eqref{Generic decompose}, we denote it as posterior drift error. We obtain
\begin{equation}\label{Generic PT 1}
    \begin{aligned}
        &\quad\mathcal{E}_D^{0 / 1}\left(\widehat{f}_n^{\mathrm{PT}}, h_D^o\right)-\mathcal{E}_{D^{\mathrm{PT}}}^{0 / 1}\left(\widehat{f}_n^{\mathrm{PT}}, h_D^o\right) \\
        & =2\mathbb{E}_{\bm{X}}\left\{(\eta(\bm{X})-\eta^{\mathrm{PT}}(\bm{X}))\left[\mathbb{I}(\eta(\bm{X})\geq\frac{1}{2})-\mathbb{I}(\widehat{f}_{n}^{\mathrm{PT}}(\bm{X})\geq0)\right]\right\}\\
        & \leq 2\mathbb{E}_{\bm{X}}\left\{(\eta(\bm{X})-\eta^{\mathrm{PT}}(\bm{X}))\left[\mathbb{I}(\eta(\bm{X})\geq\frac{1}{2})-\mathbb{I}(\widehat{f}_{n}^{\mathrm{PT}}(\bm{X})\geq0)\right]\mid \bm{X}\in \mathrm{BS}\right\}P_D(\bm{X}\in \mathrm{BS})\\
        & \leq 2\mathbb{E}_{\bm{X}}\left\{\left|(\eta(\bm{X})-\eta^{\mathrm{PT}}(\bm{X}))\left[\mathbb{I}(\eta(\bm{X})\geq\frac{1}{2})-\mathbb{I}(\widehat{f}_{n}^{\mathrm{PT}}(\bm{X})\geq0)\right]\right|\mid \bm{X}\in \mathrm{BS}\right\}P_D(\bm{X}\in \mathrm{BS})\\
        & \leq 2\mathbb{E}_{\bm{X}}\left\{\left|\eta(\bm{X})-\eta^{\mathrm{PT}}(\bm{X})\right|\mid \bm{X}\in \mathrm{BS}\right\}P_D(\bm{X}\in \mathrm{BS})\\
        &\leq 2P_D(\bm{X}\in \mathrm{BS})\sup_{\bm{x}\in \mathrm{BS}}\left|\eta(\bm{x})-\eta^{\mathrm{pse}}(\bm{x})\right|.
    \end{aligned}
\end{equation}
where the first inequality holds because the terms $\eta(\bm{x})-\eta^{\mathrm{PT}}(\bm{x})$ and $\mathbb{I}(\eta(\bm{X})\geq\frac{1}{2})-\mathbb{I}(\widehat{f}_{n}^{\mathrm{PT}}(\bm{X})\geq0)$ have opposite signs for any $\bm{x}\notin \mathrm{BS}$ and the third inequality is based on $|\mathbb{I}(\cdot)-\mathbb{I}(\cdot)|\leq 1$.  Then, aiming to scale $\sup_{\bm{x}\in \mathrm{BS}}\left|\eta(\bm{x})-\eta^{\mathrm{pse}}(\bm{x})\right|$, we expand it as
\begin{equation*}
    \begin{aligned}
        &\quad \sup_{\bm{x}\in \mathrm{BS}}\left|\eta(\bm{x})-\eta^{\mathrm{pse}}(\bm{x})\right|\\
        &\leq \sup_{\bm{x}\in \mathrm{BS}}\left|\frac{\eta_\rho(\bm{x})-\rho_{-}(\bm{x})}{1-\rho_{+}(\bm{x})-\rho_{-}(\bm{x})}-\frac{\widehat{\eta}_\rho(\bm{x})-\widehat{\rho}_{-}(\bm{x})}{1-\widehat{\rho}_{+}(\bm{x})-\widehat{\rho}_{-}(\bm{x})}\right|\\
        &= \sup_{\bm{x}\in \mathrm{BS}}\left|\frac{(\eta_\rho(\bm{x})-\rho_{-}(\bm{x}))(1-\widehat{\rho}_{+}(\bm{x})-\widehat{\rho}_{-}(\bm{x}))-(\widehat{\eta}_\rho(\bm{x})-\widehat{\rho}_{-}(\bm{x}))(1-\rho_{+}(\bm{x})-\rho_{-}(\bm{x}))}{(1-\rho_{+}(\bm{x})-\rho_{-}(\bm{x}))(1-\widehat{\rho}_{+}(\bm{x})-\widehat{\rho}_{-}(\bm{x}))}\right|\\
        &\leq \sup_{\bm{x}\in \mathrm{BS}}\left|\frac{(\eta_\rho(\bm{x})-\rho_{-}(\bm{x}))(\rho_{+}(\bm{x})+\rho_{-}(\bm{x})-\widehat{\rho}_{+}(\bm{x})-\widehat{\rho}_{-}(\bm{x}))}{(1-\rho_{+}(\bm{x})-\rho_{-}(\bm{x}))(1-\widehat{\rho}_{+}(\bm{x})-\widehat{\rho}_{-}(\bm{x}))}\right| \\
        & \qquad +\sup_{\bm{x}\in \mathrm{BS}}\left|\frac{(\eta_\rho(\bm{x})-\rho_{-}(\bm{x})-\widehat{\eta}_\rho(\bm{x})+\widehat{\rho}_{-}(\bm{x}))(1-\rho_{+}(\bm{x})-\rho_{-}(\bm{x}))}{(1-\rho_{+}(\bm{x})-\rho_{-}(\bm{x}))(1-\widehat{\rho}_{+}(\bm{x})-\widehat{\rho}_{-}(\bm{x}))}\right|\\
        & \leq \frac{1}{C_\rho^{\prime}}\left(\sup_{\bm{x}\in \mathrm{BS}}\left|\rho_{+}(\bm{x})+\rho_{-}(\bm{x})-\widehat{\rho}_{+}(\bm{x})-\widehat{\rho}_{-}(\bm{x})\right| + \sup_{\bm{x}\in \mathrm{BS}}\left|\eta_\rho(\bm{x})-\rho_{-}(\bm{x})-\widehat{\eta}_\rho(\bm{x})+\widehat{\rho}_{-}(\bm{x})\right|\right)\\
        & \leq \frac{1}{C_\rho^{\prime}}\left(\sup_{\bm{x}\in \mathcal{X}}\left|\rho_{+}(\bm{x})-\widehat{\rho}_{+}(\bm{x})\right| + 2\sup_{\bm{x}\in \mathcal{X}}\left|\rho_{-}(\bm{x})-\widehat{\rho}_{-}(\bm{x})\right| + \sup_{\bm{x}\in \mathcal{X}}\left|\eta_\rho(\bm{x})-\widehat{\eta}_\rho(\bm{x})\right|\right) \\
        & \lesssim \epsilon_{n_0, n},
    \end{aligned}
\end{equation*}
where the first inequality follows from projection onto $[0,1]$ and the second inequality holds because $\eta(\bm{x})\leq 1$ and $1-\widehat{\rho}_{+}(\bm{x})-\widehat{\rho}_{-}(\bm{x})\geq C_{\rho}^\prime$. Then
\begin{equation*}
    \mathcal{E}_D^{0 / 1}\left(\widehat{f}_n^{\mathrm{PT}, \mathrm{H}}, h_D^o\right)-\mathcal{E}_{D^{\mathrm{PT}}}^{0 / 1}\left(\widehat{f}_n^{\mathrm{PT}, \mathrm{H}}, h_D^o\right)\lesssim \epsilon_{n_0,n}^{q+1}.
\end{equation*}
Next, we denote the second part of \eqref{Generic decompose} as boundary shift error. We have
\begin{equation*}
    \begin{aligned}
        \mathcal{E}_{D^{\mathrm{PT}}}^{0 / 1}\left(\widehat{f}_n^{\mathrm{PT}, \mathrm{H}}, h_D^o\right)&= \mathbb{E}_{\bm{X}}\left[\eta^{\mathrm{PT}}(\bm{X})\left(\mathbb{I}(\widehat{f}_{n}^{\mathrm{PT,H}}(\bm{X})<0)-\mathbb{I}(\eta(\bm{X})<\frac{1}{2})\right)\right.\\
        & \left.+(1-\eta^{\mathrm{PT}}(\bm{X}))\left(\mathbb{I}(\widehat{f}_{n}^{\mathrm{PT,H}}(\bm{X})\geq0)-\mathbb{I}(\eta(\bm{X})\geq\frac{1}{2})\right)\right],\\
         \mathcal{E}_{D^{\mathrm{PT}}}^{0 / 1}\left(\widehat{f}_n^{\mathrm{PT}, \mathrm{H}}, h_{D^{\mathrm{PT}}}^o\right)&= \mathbb{E}_{\bm{X}}\left[\eta^{\mathrm{PT}}(\bm{X})\left(\mathbb{I}(\widehat{f}_{n}^{\mathrm{PT,H}}(\bm{X})<0)-\mathbb{I}(\eta^{\mathrm{PT}}(\bm{X})<\frac{1}{2})\right)\right.\\
        & \left.+(1-\eta^{\mathrm{PT}}(\bm{X}))\left(\mathbb{I}(\widehat{f}_{n}^{\mathrm{PT,H}}(\bm{X})\geq0)-\mathbb{I}(\eta^{\mathrm{PT}}(\bm{X})\geq\frac{1}{2})\right)\right].\\
    \end{aligned}
\end{equation*}
Subtracting these expressions with the convention $\operatorname{sign}(0)=1$ gives the following non-positive term. The identity also holds at posterior ties.
\begin{equation}\label{Generic PT 2}
    \begin{aligned}
        &\quad\mathcal{E}_{D^{\mathrm{PT}}}^{0 / 1}\left(\widehat{f}_n^{\mathrm{PT}, \mathrm{H}}, h_D^o\right) - \mathcal{E}_{D^{\mathrm{PT}}}^{0 / 1}\left(\widehat{f}_n^{\mathrm{PT}, \mathrm{H}}, h_{D^{\mathrm{PT}}}^o\right) \\
        & = -2\mathbb{E}_{\bm{X}}\left\{\left(\eta^{\mathrm{PT}}(\bm{X})-\frac{1}{2}\right)\left[\mathbb{I}(\eta^{\mathrm{PT}}(\bm{X})\geq\frac{1}{2})-\mathbb{I}(\eta(\bm{X})\geq\frac{1}{2})\right]\right\}\\
        & = -2\mathbb{E}_{\bm{X}}\left\{\left(\eta^{\mathrm{PT}}(\bm{X})-\frac{1}{2}\right)\left[\mathbb{I}(\eta^{\mathrm{PT}}(\bm{X})\geq\frac{1}{2})-\mathbb{I}(\eta(\bm{X})\geq\frac{1}{2})\right]\mid \bm{X}\in \mathrm{BS}\right\}P_D(\bm{X}\in \mathrm{BS})\\
        & = -2\mathbb{E}_{\bm{X}}\left\{\left|\left(\eta^{\mathrm{PT}}(\bm{X})-\frac{1}{2}\right)\left[\mathbb{I}(\eta^{\mathrm{PT}}(\bm{X})\geq\frac{1}{2})-\mathbb{I}(\eta(\bm{X})\geq\frac{1}{2})\right]\right|\mid \bm{X}\in \mathrm{BS}\right\}P_D(\bm{X}\in \mathrm{BS})\\
        & \leq 0.
    \end{aligned}
\end{equation}
Its magnitude satisfies
\begin{equation*}
    \begin{aligned}
        &\left|\mathcal{E}_{D^{\mathrm{PT}}}^{0/1}(\widehat f_n^{\mathrm{PT,H}},h_D^o)-\mathcal{E}_{D^{\mathrm{PT}}}^{0/1}(\widehat f_n^{\mathrm{PT,H}},h_{D^{\mathrm{PT}}}^o)\right|\\
        & = 2\mathbb{E}_{\bm{X}}\left\{\left|\left(\eta^{\mathrm{PT}}(\bm{X})-\frac{1}{2}\right)\left[\mathbb{I}(\eta^{\mathrm{PT}}(\bm{X})\geq\frac{1}{2})-\mathbb{I}(\eta(\bm{X})\geq\frac{1}{2})\right]\right|\mid \bm{X}\in \mathrm{BS}\right\}P_D(\bm{X}\in \mathrm{BS})\\
        & \leq 2\mathbb{E}_{\bm{X}}\left\{\left|\eta^{\mathrm{PT}}(\bm{X})-\frac{1}{2}\right|\mid \bm{X}\in \mathrm{BS}\right\}P_D(\bm{X}\in \mathrm{BS})\\
        & \leq 2P_D(\bm{X}\in \mathrm{BS})\sup_{\bm{x}\in \mathrm{BS}}\left|\eta^{\mathrm{PT}}(\bm{x})-\frac{1}{2}\right|\\
        & \leq 2P_D(\bm{X}\in \mathrm{BS})\left(\sup_{\bm{x}\in \mathrm{BS}}\left|\eta^{\mathrm{PT}}(\bm{x})-\eta(\bm{x})\right|+\sup_{\bm{x}\in \mathrm{BS}}\left|\eta(\bm{x})-\frac{1}{2}\right|\right) \\
        & \lesssim \epsilon_{n_0,n}^{q+1},
    \end{aligned}
\end{equation*}
In \eqref{Generic PT 2}, the second equality holds because $\mathbb{I}(\eta(\bm{x})\geq\frac{1}{2})\equiv\mathbb{I}(\eta^{\mathrm{PT}}(\bm{x})\geq\frac{1}{2})$ for any $\bm{x}\notin\mathrm{BS}$, the first inequality is based on $|\mathbb{I}(\cdot)-\mathbb{I}(\cdot)|\leq 1$, and the final bound uses the boundary-probability bound in \eqref{Generic PT 1} and $\eta^{\mathrm{PT}}(\bm{x})=\eta^{\mathrm{pse}}(\bm{x})$ when $\bm{x}\in\mathrm{BS}$. Based on the existing theory of various learner to control $\mathcal{E}_{D^{\mathrm{PT}}}^{0/1}\left(\widehat{f}_{n}^{\mathrm{PT}}, h_{D^{\mathrm{PT}}}^o\right)$, combining with \eqref{Generic decompose}, \eqref{Generic PT 1} and \eqref{Generic PT 2}, we have
\begin{equation*}
    \mathbb{E}_{n_0,D}\mathbb{E}_{n,D_\rho}\left\{\mathcal{E}^{0/1}_D(\widehat f_n,h_D^o)\right\}
    \lesssim \mathbb{E}_{n_0,D}\mathbb{E}_{n,D_\rho}\epsilon_{n_0,n}^{q+1}+\varepsilon_{2,n}.
\end{equation*}
\hfill Q.E.D.

\subsection{Proof of Lemma~\ref{Lem 1}}\label{sec:A.1}
Let $A=\sum_{i=1}^{n_0} I\left(\tilde{y}_{0 i}=-1, y_{0 i}=1\right)$ and $B=\sum_{i=1}^{n_0} I\left(y_{0 i}=1\right)$. Define $\widehat{\rho}_{+}=\frac{A}{B}$ and $p=P_D(Y=1)$, then we have $B \sim \operatorname{Binomial}\left(n_0, p\right)$ and $A|B \sim \operatorname{Binomial}\left(B, \rho_{+}\right)$. We analyze the mean squared error $\mathbb{E}_{n_0, D}\left|\rho_{+}-\widehat{\rho}_{+}\right|^2$ by considering two cases:

\noindent{\bf Case 1: ($B=0$).} If $B=0$, then $\widehat{\rho}_{+}=0$, and the error is $\left|\rho_{+}-\widehat{\rho}_{+}\right|=\rho_{+}$. The probability of this event is $P_D(B=0)=(1-p)^{n_0}$. Since $p$ is bounded away from zero $(p \geq C_y>0)$, so $(1-p)^{n_0}$ decays exponentially with $n_0$. Thus, the contribution to the expected error is
\begin{equation*}
    \mathbb{E}_{n_0,D}\mathbb{E}_{\bm{X}}\left[\left|\rho_{+}-\widehat{\rho}_{+}\right|^2 \mid B=0\right] P_D(B=0)=\rho_{+}^2(1-p)^{n_0} \leq \rho_{+}^2e^{-C_y n_0},
\end{equation*}
which is negligible compared to $n_0^{-1}$.

\noindent{\bf Case 2: ($B>0$).} If $B>0$, then $\widehat{\rho}_{+}=\frac{A}{B}$. Given $B, A$ follows a binomial distribution with mean $B \rho_{+}$, so:
\begin{equation*}
    \mathbb{E}_{n_0,D}\mathbb{E}_{\bm{X}}\left[\left|\rho_{+}-\widehat{\rho}_{+}\right|^2 \mid B\right] =\operatorname{Var}\left(\widehat{\rho}_{+} \mid B\right)=\frac{\rho_{+}\left(1-\rho_{+}\right)}{B}.
\end{equation*}
Then, we have
\begin{equation*}
    \mathbb{E}_{n_0,D}\mathbb{E}_{\bm{X}}\left[\left|\rho_{+}-\widehat{\rho}_{+}\right|^2 \mid B>0\right]P_D(B>0) = \rho_{+}(1-\rho_{+})\sum_{b=1}^{n_0}\frac{1}{b}P_D(B=b),
\end{equation*}
Consider the Chernoff inequality, we have $P_D(B \leq(1-\delta) n_0 p) \leq e^{-\frac{\delta^2 n_0 p}{2}}$, for $\forall\delta\in(0,1)$. By taking $\delta=\frac{1}{2}$, we have
\begin{equation*}
    \begin{aligned}
        \sum_{b=1}^{n_0}\frac{1}{b}P_D(B=b) &=\sum_{b=1}^{\lfloor n_0p/2 \rfloor}\frac{1}{b}P_D(B=b) + \sum_{b=\lfloor n_0p/2 \rfloor+1}^{n_0}\frac{1}{b}P_D(B=b)\\
        &\leq 1\cdot P_D(B<\frac{n_0p}{2}) + \frac{2}{n_0p}\cdot1 \\
        &\leq e^{-\frac{n_0p}{8}} + \frac{2}{n_0p}.
    \end{aligned}
\end{equation*}
So we have $\mathbb{E}_{n_0,D}\mathbb{E}_{\bm{X}}\left[\left|\rho_{+}-\widehat{\rho}_{+}\right|^2 \mid B>0\right]P_D(B>0) \leq \rho_{+}(1-\rho_{+})e^{-\frac{n_0p}{8}} + \frac{2\rho_{+}(1-\rho_{+})}{n_0p}$. Combining the two cases, we have
\begin{equation*}
    \begin{aligned}
        \mathbb{E}_{n_0,D}\mathbb{E}_{\bm{X}}\left|\rho_{+}-\widehat{\rho}_{+}\right|^2&=\mathbb{E}_{n_0,D}\mathbb{E}_{\bm{X}}\left[\left|\rho_{+}-\widehat{\rho}_{+}\right|^2 \mid B=0\right] P_D(B=0)\\
        &\quad +\mathbb{E}_{n_0,D}\mathbb{E}_{\bm{X}}\left[\left|\rho_{+}-\widehat{\rho}_{+}\right|^2 \mid B>0\right] P_D(B>0)\\
        &\leq \rho_{+}^2e^{-C_y n_0} + \rho_{+}(1-\rho_{+})e^{-\frac{n_0p}{8}} + \frac{2\rho_{+}(1-\rho_{+})}{n_0p}\\
        &\leq e^{-C_y n_0}+ \frac{1}{2}e^{-\frac{C_yn_0}{8}} + \frac{1}{2C_yn_0}\lesssim n_0^{-1}.
    \end{aligned}
\end{equation*}

\renewcommand{\theequation}{B.\arabic{equation}}
\renewcommand{\thetable}{B.\arabic{table}}
\renewcommand{\thefigure}{B.\arabic{figure}}
\setcounter{equation}{0} 
\setcounter{table}{0} 
\setcounter{figure}{0}

\section{Technical Details and Proofs for Neural Networks}\label{Appendix B}

Definitions and conditions for the neural-network results in Section~\ref{section 4.2} of the main text are collected in Section~\ref{sec:nn-definitions}, followed by auxiliary lemmas in Section~\ref{sec:auxiliary-lemmas} and proofs of the main-text results. The additional conditions for the sigmoid loss are used in Corollaries~\ref{Cor S2} and~\ref{Cor S3}, stated in Section~\ref{Appendix A.2}.

\subsection{Additional Definitions and Conditions}\label{sec:nn-definitions}

We first define the hierarchical composition class $\mathcal{H}_{l,\mathcal{P}}$ and the neural-network class $\mathcal{F}_{\mathrm{NN}}$. The definition of $\mathcal{H}_{l,\mathcal{P}}$ uses the following notion of $(p,C)$-smoothness.
\begin{Def}\label{Def S1}
    Let $p=r+s$ for some $r \in \mathbb{N}_0$ and $0<s \leq 1$. A function $f: \mathbb{R}^d \rightarrow \mathbb{R}$ is called $(p, C)$-smooth, if for every $\bm{\alpha}=\left(\alpha_1, \ldots, \alpha_d\right) \in \mathbb{N}_0^d$ with $\sum_{j=1}^d \alpha_j=r$ the partial derivative $\partial^r f /\left(\partial x_1^{\alpha_1} \ldots \partial x_d^{\alpha_d}\right)$ exists and satisfies
    \begin{equation*}
        \left|\frac{\partial^r f}{\partial x_1^{\alpha_1} \ldots \partial x_d^{\alpha_d}}(\bm{x})-\frac{\partial^r f}{\partial x_1^{\alpha_1} \ldots \partial x_d^{\alpha_d}}(\bm{z})\right| \leq C\|\bm{x}-\bm{z}\|_2^s.
    \end{equation*}
\end{Def} 
\begin{Def}\label{Def S2}
    Let $d \in \mathbb{N}$ and $f: \mathbb{R}^d \rightarrow \mathbb{R}$ and let $\mathcal{P}$ be a subset of $(0, \infty) \times \mathbb{N}$.

    (a) We say that $f$ satisfies a hierarchical composition model of level 0 with order and smoothness constraint $\mathcal{P}$, if there exists a $k \in\{1, \ldots, d\}$ such that
    \begin{equation*}
        f(\bm{x})=x_k \quad \text{ for all } \bm{x}=\left(x_1, \ldots, x_d\right)^{\top} \in \mathbb{R}^d.
    \end{equation*}
    
    (b) We say that $f$ satisfies a hierarchical composition model of level $l+1$ with order and smoothness constraint $\mathcal{P}$, if there exist $(p, k) \in \mathcal{P}, C>0, g: \mathbb{R}^k \rightarrow \mathbb{R}$ and $f_1, \ldots, f_k$ : $\mathbb{R}^d \rightarrow \mathbb{R}$, such that $g$ is $(p, C)$-smooth, $f_1, \ldots, f_k$ satisfy a hierarchical composition model of level $l$ with order and smoothness constraint $\mathcal{P}$ and
    \begin{equation*}
        f(\bm{x})=g\left(f_1(\bm{x}), \ldots, f_k(\bm{x})\right) \quad \text{ for all } \bm{x} \in \mathbb{R}^d,
    \end{equation*}
    For $l=1$ and some order and smoothness constraint $\mathcal{P} \subseteq(0, \infty) \times \mathbb{N}$, our space of hierarchical composition models becomes
    \begin{equation*}
        \begin{aligned}
            \mathcal{H}_{1, \mathcal{P}}= & \left\{f: \mathbb{R}^d \rightarrow \mathbb{R}: f(\bm{x})=g\left(x^{(\pi(1))}, \ldots, x^{(\pi(k))}\right),\right. \text{ where } \\
        & g: \mathbb{R}^k \rightarrow \mathbb{R} \text{ is }(p, C) \text{-smooth for some }(p, k) \in \mathcal{P}, \\
        & C>0 \text{ and } \pi:\{1, \ldots, k\} \rightarrow\{1, \ldots, d\}\} .
        \end{aligned}
    \end{equation*}
    For $l>1$, we recursively define
    \begin{equation*}
        \begin{aligned}
            \mathcal{H}_{l, \mathcal{P}}:= & \left\{f: \mathbb{R}^d \rightarrow \mathbb{R}: f(\bm{x})=g\left(f_1(\bm{x}), \ldots, f_k(\bm{x})\right),\right. \text{ where } \\
            & g: \mathbb{R}^k \rightarrow \mathbb{R} \text{ is }(p, C) \text{-smooth for some }(p, k) \in \mathcal{P}, \\
            & \left.C>0 \text{ and } f_i \in \mathcal{H}(l-1, \mathcal{P})\right\} .
        \end{aligned}
    \end{equation*}
\end{Def}
The hierarchical composition class $\mathcal{H}_{l,\mathcal{P}}$ includes the H\"older class as a special case when $l=1$ and accommodates more general composition structures \cite{Johannes2020Nonparametric}. With this notation, we restate Condition~\ref{Con 3} of the main text in full.

\noindent {\bf Condition~\ref{Con 3}. (Smooth Conditional Distribution).} 
{\it Let the corresponding conditional distribution $\eta(\bm{x}), \eta_\rho(\bm{x}), \rho_{-}(\bm{x}), \rho_{+}(\bm{x})$ be contained in the function class $\mathcal{H}_{l, \mathcal{P}}$ for some $l \in \mathbb{N}$ and $\mathcal{P} \subseteq[1, \infty) \times \mathbb{N}$. Each function $g$ in the definition of $\mathcal{H}_{l, \mathcal{P}}$ can be of different smoothness $p_g=q_g+s_g\left(q_g \in \mathbb{N}_0\right.$ and $\left.s_g \in(0,1]\right)$ and of different input dimension $k_g$, where $\left(p_g, k_g\right) \in \mathcal{P}$. Denote by $k_{\max }$ the maximal input dimension and by $p_{\max }$ the maximal smoothness of one of the functions $g$. Assume that $p_{\max }, k_{\max }<\infty$. Let each function $g$ be Lipschitz continuous with Lipschitz constant $L \geq 1$.}

We next define the neural-network class $\mathcal{F}(L_n,N_n,S_n,B_n,F_n)$.
\begin{Def}\label{Def S3}
    We consider DNNs that take $d$-dimensional inputs and produce one-dimensional outputs. A DNN with $L$ layers, and $\left\{N^{(l)}, l \in[L]\right\}$ nodes at each layer, is defined as
    \begin{equation*}
        h_j^{(l)}(\bm{x})=\sigma\left(b_j^{(l)}+\sum_{k=1}^{N^{(l-1)}} W_{j, k}^{(l)} h_k^{(l-1)}(\bm{x})\right),
    \end{equation*}
    for $l=1, \ldots, L$ and
    \begin{equation*}
        f(\bm{x})=b^{(L+1)}+\sum_{k=1}^{N^{(L)}} W_{1, k}^{(L+1)} h_k^{(L)}(\bm{x}),
    \end{equation*}
    with $N^{(0)}=d$ and $h_k^{(0)}(\bm{x})=x_k$, $\sigma(x)$ is the non-linear activation function such as $\mathrm{ReLU}(x)=\max(x,0)$ and $\mathrm{Sigmoid}(x)=(1+\exp(-x))^{-1}$. We denote $f(\bm{x})$ as $f(\bm{x} \mid \bm{\Theta})$, where $\bm{\Theta}=\left(\left(\bm{W}^{(l)}, \bm{b}^{(l)}\right)\right)_{l=1, \ldots, L+1}$ is the parameter set including all weights and biases.

    For the given $\bm{\Theta}$, let $|\bm{\Theta}|$ be the number of layers in $\bm{\Theta}$. Let $N_{\max }(\bm{\Theta})$ be the maximum number of nodes, that is, $f(\cdot \mid \bm{\Theta})$ has at most $N_{\max }(\bm{\Theta})$ nodes at each layer. We define $\|\bm{\Theta}\|_0$ as the number of nonzero parameters in $\bm{\Theta}$,
    \begin{equation*}
        \|\bm{\Theta}\|_0=\sum_{l=1}^{L+1}\left(\left\|\operatorname{vec}\left(\bm{W}^{(l)}\right)\right\|_0+\left\|\bm{b}^{(l)}\right\|_0\right),
    \end{equation*}
    where $\operatorname{vec}\left(\bm{W}^{(l)}\right)$ transforms the matrix $\bm{W}^{(l)}$ into the corresponding vector by concatenating the column vectors. Similarly, we define $\|\bm{\Theta}\|_{\infty}$ as the largest absolute value of the parameters in $\bm{\Theta}$,
    \begin{equation*}
        \|\bm{\Theta}\|_{\infty}=\max \left\{\max _{1 \leq l \leq L+1}\left\|\operatorname{vec}\left(\bm{W}^{(l)}\right)\right\|_{\infty}, \max _{1 \leq l \leq L+1}\left\|\bm{b}^{(l)}\right\|_{\infty}\right\}.
    \end{equation*}
    For a given $n$, let $\mathcal{F}_{\mathrm{NN}}$ be
    \begin{equation*}
        \begin{aligned}
            \mathcal{F}_{\mathrm{NN}}= & \mathcal{F}\left(L_n, N_n, S_n, B_n, F_n\right) \\
            = & \left\{f(\bm{x} \mid \bm{\Theta}):|\bm{\Theta}| \leq L_n, N_{\max }(\bm{\Theta}) \leq N_n,\|\bm{\Theta}\|_0 \leq S_n,\right. \\
            & \left.\|\bm{\Theta}\|_{\infty} \leq B_n,\|f(\cdot \mid \bm{\Theta})\|_{\infty} \leq F_n\right\}
        \end{aligned}
    \end{equation*}    
\end{Def}
Definition~\ref{Def S3} describes the network class through its depth, width, sparsity, parameter bounds, and output bound. Condition~\ref{Con S1} specifies the network configurations used in the analysis \cite{karimi2020deep,kim2021fast}.
\begin{Con}\label{Con S1}
    {\bf (Neural network structure)}. Let $\widehat{\eta}_\rho$ be the neural-network estimator described in Section~\ref{section 4.2} of the main text, with the network class in Definition~\ref{Def S3}, for some $L_n, N_n, F_n\in \mathbb{N}$. We choose $c_3,c_4, c_5>0$ sufficiently large and set
    \begin{equation*}
        F_n =c_3\log(n), \quad L_n=\left\lceil c_4 \cdot \log n\right\rceil \quad \text{ and } \quad N_n=\left\lceil c_5 \cdot \max _{(p, k) \in \mathcal{P}} n^{\frac{k}{2(2 p+k)}}\right\rceil.
    \end{equation*}
    or choose $c_7, c_8>0$ sufficiently large and set
    \begin{equation*}
        F_n =c_3\log(n), \quad L_n=\left\lceil c_7 \cdot \max _{(p, k) \in \mathcal{P}} n^{\frac{k}{2(2 p+k)}} \cdot \log n\right\rceil \quad \text{ and } \quad N_n=\left\lceil c_8\right\rceil.
    \end{equation*}
    The definitions of $\widehat{\rho}_{-}(\bm{x})$ and $\widehat{\rho}_{+}(\bm{x})$ can be adapted by replacing the sample size $n$ with the specific counts $n_{0-}$ and $n_{0+}$, there exists a constant $C_\rho^{\prime}>0$ s.t. $\widehat{\rho}_{-}(\bm{x})+\widehat{\rho}_{+}(\bm{x})\leq 1-C_\rho^{\prime}$. We assume that the truncated nuisance estimators satisfy a uniform Lipschitz bound on $\mathcal{X}$. All estimators are truncated to $[0,1]$. Similarly, let $\widehat{f}_n^{\mathrm{AT}}$ and $\widehat{f}_n^{\mathrm{PT}}$ be the classifiers described in Section~\ref{section 4.2} of the main text, with the network class in Definition~\ref{Def S3}, for some $L_n, N_n, F_n\in \mathbb{N}$ \cite{kim2021fast,li2024error}. 
\end{Con}
We next define the H\"older and classifier classes used in Condition~\ref{Con 5} of the main text, following Kim et al.~\cite{kim2021fast}.
\begin{Def}\label{Def 1}
     Let $d \in \mathbb{N}_{+}$ be the dimension and $p=q+s>0$, where $q \in \mathbb{N}$ and $s \in(0,1]$. Then Hölder function class $\mathcal{H}_C^{p}\left([0,1]^d\right)$ bounded by $C \in \mathbb{R}_{+}$ is defined as
    \begin{equation*}
         \resizebox{\linewidth}{!}{$\displaystyle\mathcal{H}_C^{p}\left([0,1]^d\right)=\left\{f:[0,1]^d \rightarrow \mathbb{R}, \max _{\|\bm{\alpha}\|_1 \leq q} \sup _{\bm{x} \in[0,1]^d}\left|D^{\bm{\alpha}} f(\bm{x})\right|+\max _{\|\bm{\alpha}\|_1=q} \sup _{\substack{\bm{x}, \bm{x}^{\prime} \in[0,1]^d \\ \bm{x} \neq \bm{x}^{\prime}}} \frac{\left|D^{\bm{\alpha}} f(\bm{x})-D^{\bm{\alpha}} f\left(\bm{x}^{\prime}\right)\right|}{\left\|\bm{x}-\bm{x}^{\prime}\right\|_{\infty}^s} \leq C\right\},$}
    \end{equation*}
    where $D^{\bm{\alpha}} f\left(\bm{x}\right):=\frac{\partial^{\alpha_1}}{\partial x_1^{\alpha_1}} \cdots \frac{\partial^{\alpha_d}}{\partial x_d^{\alpha_d}}f(\bm{x})$, and the multi-index $\bm{\alpha}=\left(\alpha_1, \cdots, \alpha_d\right) \in \mathbb{N}^d$.
\end{Def}
This Hölder function class $\mathcal{H}_C^{p}$ is a special case of $\mathcal{H}_{l,\mathcal{P}}$, which is common in literature. To distinguish the smoothness parameter $p$ from its usage in other definitions, we will use the notation $p_h$ in the subsequent Definition \ref{Def 2} to represent the smoothness of the boundary.
\begin{Def}\label{Def 2}
    For $g \in \mathcal{H}^{p_h}_C\left([0,1]^{d-1}\right)$ and $j \in[d]$, we define a horizon function $\Psi_{g, j}:[0,1]^d \rightarrow\{0,1\}$ as $\Psi_{g, j}(\bm{x})=\mathbb{I}\left(x_j \geq g\left(\bm{x}_{-j}\right)\right)$, where $\bm{x}_{-j}=\left(x_1, \ldots, x_{j-1}, x_{j+1}, \ldots, x_d\right)$. For each horizon function, we define the corresponding basis piece $I_{g, j}$ as $I_{g, j}=\left\{\bm{x} \in[0,1]^d: \Psi_{g, j}(\bm{x})=1\right\}$. We define a piece by the intersection of $K$ basis pieces. The set of pieces is denoted by
    \begin{equation}
        \mathcal{A}^{p_h, C, K}=\left\{A \subset[0,1]^d: A=\bigcap_{k=1}^K I_{g_k, j_k}, g_k \in \mathcal{H}^{p_h}_C\left([0,1]^{d-1}\right), j_k \in[d]\right\}
    \end{equation}
    Let $\mathcal{C}^{p_h, C, K, T}$ be the set of classifiers of the form $C(\bm{x})=2 \sum_{t=1}^T \mathbb{I}\left(\bm{x} \in A_t\right)-1$, for $T \in \mathbb{N}$, and disjoint subsets $A_1, \ldots, A_T$ of $\mathcal{X}$ in $\mathcal{A}^{p_h, C, K}$.
\end{Def} 
The hinge loss, while motivated by robustness of classification, presents a practical drawback versus the sigmoid loss. It does not yield a natural probabilistic estimate, as $\widehat{\eta}(\bm{x})$ cannot be expressed in the form $\sigma(f(\bm{x}))$, where $\sigma(z)=\left(1+e^{-z}\right)^{-1}$. Thus, we also consider sigmoid loss.\footnote{we discuss the functional equivalence between the sigmoid loss and the standard binary cross-entropy loss, $\mathcal{L}_{\text{BCE}}(y, f(\bm{x}))=-\log (\sigma(yf(\bm{x})))$. The specified sigmoid loss can be expressed as $\mathcal{L}_{\text{Sigmoid}}(y, f(\bm{x}))=2/(e^{2yf(\bm{x})}+1)=2\sigma(-2yf(\bm{x}))$. From a statistical learning perspective, the two losses are equivalent. Both are consistent surrogates for the 0-1 loss, and their shared optimization goal is to maximize the margin $yf(\bm{x})$. Their primary distinction arises during optimization, specifically from the different magnitudes of their gradients.} In order to obtain the conclusion under the sigmoid loss, we likewise do not need the continuity property of $\eta^{\mathrm{AT}}(\bm{x})$. Instead, we have requirements for the optimal function of the expected loss. We denote the optimal function under the Sigmoid loss as
\begin{equation*}
    \begin{aligned}
        f_{D^{\mathrm{AT}}}^{\star}(\bm{x})=\arg\min _{f \text{ is measurable }} R_{D^{\mathrm{AT}}}^{\mathrm{Sigmoid}}(f), \quad R_{D^{\mathrm{AT}}}^{\mathrm{Sigmoid}}(f)=\mathbb{E}_{D^{\mathrm{AT}}} \mathcal{L}_{\mathrm{Sigmoid}}(Y, f(\bm{X})), \\
        f_{D^{\mathrm{PT}}}^{\star}(\bm{x})=\arg\min _{f \text{ is measurable }} R_{D^{\mathrm{PT}}}^{\mathrm{Sigmoid}}(f), \quad R_{D^{\mathrm{PT}}}^{\mathrm{Sigmoid}}(f)=\mathbb{E}_{D^{\mathrm{PT}}} \mathcal{L}_{\mathrm{Sigmoid}}(Y, f(\bm{X})).
    \end{aligned}
\end{equation*}
The following additional conditions are used for the sigmoid-loss results in Corollaries~\ref{Cor S2} and~\ref{Cor S3}; their proofs are given in Sections~\ref{sec:proof-cor-s2} and~\ref{sec:proof-cor-s3}.
\begin{Con}\label{Con S2} 
    {\bf (Massart Margin).} There exists a constant $\epsilon \in\left(0, \frac{1}{2}\right]$ such that
    \begin{equation*}
        P_D\left(\left|\eta(\bm{X})-\frac{1}{2}\right| \leqslant \epsilon\right) =0.
    \end{equation*}
\end{Con}
\begin{Con}\label{Con S3} 
    Suppose $f_{D^{\mathrm{AT}}}^{\star}(\bm{x})$ and $ f_{D^{\mathrm{PT}}}^{\star}(\bm{x})$ belong to Hölder function class $\mathcal{H}_C^{p_f}\left([0,1]^d\right)$.
\end{Con}
This margin condition \ref{Con S2} means that $\eta(\bm{x})$ is bounded away from $\frac{1}{2}$, namely $\eta(\bm{x}) \in\left[0, \frac{1}{2}-\epsilon\right] \cup\left[\frac{1}{2}+\epsilon, 1\right]$. In many cases, it is stronger than Condition~\ref{Con 2}. The relationship between these two types of margin conditions have been extensively discussed in the literature \cite{locatelli2017adaptivity, kpotufe2022nuances}. Condition~\ref{Con S3} is a special condition that avoids the smoothness assumption on $\eta^{\mathrm{AT}}(\bm{x})$ and $\eta^{\mathrm{PT}}(\bm{x})$, which is used in the literature \cite{li2024error}. To distinguish the smoothness parameter $p$ from its usage in other definitions, we will use the notation $p_f$ in the Condition~\ref{Con S3}.

\subsection{Auxiliary Lemmas}\label{sec:auxiliary-lemmas}

\begin{Lem}\label{Lem S1} 
    \cite{Michael2021On}. Let $\left(\bm{x}_1, y_1\right), \ldots,\left(\bm{x}_n, y_n\right)$ be independent and identically distributed random values such that $\operatorname{supp}(\bm{X})$ is bounded and $\mathbb{E}_Y\left\{\exp \left(c_1 \cdot Y^2\right)\right\}<\infty$, for some constant $c_1>0$. Let the corresponding regression function $f(\bm{x})=\mathbb{E}[Y\mid \bm{X}=\bm{x}]$ satisfies Condition~\ref{Con 3}. Let $\widehat{f}$ be defined for some $L_n, N_n \in \mathbb{N}$, and define $\widehat{f}\leftarrow T_{c_3 \cdot \log (n)} \widehat{f}$ for some $c_3>0$ sufficiently large, where $\left(T_\beta f\right)(\mathbf{x})=T_\beta(f(\mathbf{x}))$ and $T_\beta u=\max \{-\beta, \min \{\beta, u\}\}$ for some $u \in \mathbb{R}$ and $\beta>0$. The outer expectation below is taken over the training sample.

    (a) Choose $c_4, c_5>0$ sufficiently large and set $L_n=\left\lceil c_4 \cdot \log n\right\rceil$ and \\ $N_n=\left\lceil c_5 \cdot \max _{(p, k) \in \mathcal{P}} n^{\frac{k}{2(2 p+k)}}\right\rceil$. Then
    \begin{equation*}
        \mathbb{E}\mathbb{E}_{\bm{X}} \left(\widehat{f}(\bm{X})-f(\bm{X})\right)^2 \leq c_6 \cdot(\log n)^6 \cdot \max _{(p, k) \in \mathcal{P}} n^{-\frac{2 p}{2 p+k}},
    \end{equation*}
    holds for sufficiently large $n$, which $p$,$k$ are defined in Condition~\ref{Con 3}.
    
    (b) Choose $c_7, c_8>0$ sufficiently large and set $L_n=\left\lceil c_7 \cdot \max _{(p, k) \in \mathcal{P}} n^{\frac{k}{2(2 p+k)}} \cdot \log n\right\rceil$ and $N_n=\left\lceil c_8\right\rceil$. Then
    \begin{equation*}
        \mathbb{E}\mathbb{E}_{\bm{X}} \left(\widehat{f}(\bm{X})-f(\bm{X})\right)^2 \leq c_9 \cdot(\log n)^6 \cdot \max _{(p, k) \in \mathcal{P}} n^{-\frac{2 p}{2 p+k}},
    \end{equation*}
    holds for sufficiently large $n$, which p,k are defined in Condition~\ref{Con 3}.
\end{Lem}
\begin{Rem}
    In the most general case, the convergence rate degrades to $n^{-\frac{2p}{2p+d}}$ for $(p,C)$-smooth function $f$, which exhibits the curse of dimensionality. In the restricted underlying function class $\mathcal{H}_{l,\mathcal{P}}$, we have $k<<d$.
\end{Rem}

\begin{Lem}\label{Lem S2}
    \cite{kim2021fast}. Let $\widehat{f}_n$ be the estimator constructed with suitably chosen parameters $L_n, N_n, S_n, B_n, F_n$ and Hinge loss. First, under Condition~\ref{Con 2} with a noise exponent $q \in[0, \infty]$, the excess 0-1 risk satisfies: 
    \begin{equation*}
        \sup _{h_D^o  \in \mathcal{C}^{p_h, C, K, T}} \mathbb{E}\left[\mathcal{E}_D^{0/1}\left(\widehat{f}_n, h_D^o\right)\right] \lesssim\left(\frac{\log ^3 n}{n}\right)^{\frac{p_h(q+1)}{p_h(q+2)+(d-1)(q+1)}}.
    \end{equation*}
    Furthermore, assume the margin condition holds. That is, for the decision boundary $DB=\{\bm{x}: \eta(\bm{x})=1 / 2\}$, there exist constants $C_{\gamma}>0$, and a margin exponent $\gamma \in[1, \infty]$ such that for any $t\geq 0$,
    \begin{equation}\label{Lem S2 condition}
        P_D(\mathrm{dist}(\bm{X}, \mathrm{DB}) \leq t) \leq C_{\gamma} t^\gamma,
    \end{equation}
    where $\operatorname{dist}\left(\bm{x}, \mathrm{DB}\right)=\inf _{\bm{x}^{\prime} \in \mathrm{DB}}\left\|\bm{x}-\bm{x}^{\prime}\right\|_2$. Under this additional assumption \eqref{Lem S2 condition}, the excess 0-1 risk is bounded by:
    \begin{equation*}
        \sup _{h_D^o \in C^{p_h, C, K, T}} \mathbb{E}\left[\mathcal{E}_D^{0/1}\left(\widehat{f}_n, h_D^o \right)\right] \lesssim\left(\frac{\log ^3 n}{n}\right)^{\frac{p_h(q+1)}{p_h(q+2)+(d-1)(q+1) / \gamma}}.
    \end{equation*}
\end{Lem}

\begin{Lem}\label{Lem S3}
    \cite{li2024error}. Let $\widehat{f}_n$ be the estimator constructed with suitably chosen parameters $L_n, N_n, S_n, B_n, F_n$ and Sigmoid loss. Under Conditions~\ref{Con S2}--\ref{Con S3}, the excess 0-1 risk satisfies:
    \begin{equation*}
        \mathbb{E}\left[\mathcal{E}_D^{0/1}\left(\widehat{f}_n, h_D^o \right)\right] \lesssim n^{\frac{-2p_f}{2p_f+d}} \log ^7 n,
    \end{equation*}
    where $p_f$ is the smooth constant in Condition~\ref{Con S3}.
\end{Lem}

\subsection{Proof of Theorem~\ref{Mth 1}}\label{sec:proof-extraction}
We now discuss $\epsilon_{n_0,n}^q$ for a given noise exponent $q$ from Condition~\ref{Con 2}, and 
\begin{equation*}
    \epsilon_{n_0,n} = \frac{1}{2}\left(\sup _{\bm{x} \in \mathcal{X}}\left|\rho_{+}(\bm{x})-\widehat{\rho}_{+}(\bm{x})\right|+\sup _{\bm{x} \in \mathcal{X}}\left|\rho_{-}(\bm{x})-\widehat{\rho}_{-}(\bm{x})\right|\right)+\sup _{\bm{x} \in \mathcal{X}}\left|\eta_\rho(\bm{x})-\widehat{\eta}_\rho(\bm{x})\right|.
\end{equation*}
Under Condition~\ref{Con 1}(a), by Lemma~\ref{Lem 1} and Lemma~\ref{Lem 2}, we have
\begin{equation*}
    \mathbb{E}_{n_0,D}\mathbb{E}_{n, D_\rho} \epsilon_{n_0,n}^2 \lesssim n_0^{-1}+ (\log n)^6 \max _{(p, k) \in \mathcal{P}} n^{-\frac{2p}{(2p+k)(d+1)}}.
\end{equation*}
Since $\epsilon_{n_0, n}$ is bounded by a constant $C$, we have the following results:

\noindent{(i)}. For $q\leq2$, Jensen's inequality for concave functions yields
\begin{equation}\label{Mth 2 1}
    \mathbb{E}_{n_0,D}\mathbb{E}_{n,D_\rho}\epsilon_{n_0, n}^q \leq (\mathbb{E}_{n_0,D}\mathbb{E}_{n,D_\rho}\epsilon_{n_0, n}^2)^{\frac{q}{2}}\lesssim n_0^{-\frac{q}{2}}+(\log n)^6\max _{(p, k) \in \mathcal{P}} n^{-\frac{pq}{(2p+k)(d+1)}}.
\end{equation}
(ii). For $q>2$, the uniform boundedness implies
\begin{equation}\label{Mth 2 2}
    \mathbb{E}_{n_0,D}\mathbb{E}_{n,D_\rho}\epsilon_{n_0, n}^q \lesssim C^{q-2}\mathbb{E}_{n_0,D}\mathbb{E}_{n,D_\rho}\epsilon_{n_0, n}^2\lesssim n_0^{-1}+(\log n)^6\max _{(p, k) \in \mathcal{P}} n^{-\frac{2p}{(2p+k)(d+1)}}.
\end{equation}
Combining the two cases, we have 
\begin{equation*}
    \mathbb{E}_{n_0,D}\mathbb{E}_{n,D_\rho}\epsilon_{n_0, n}^q\leq n_0^{-\frac{q\wedge2}{2}}+(\log n)^6\max _{(p, k) \in \mathcal{P}} n^{-\frac{p(q\wedge2)}{(2p+k)(d+1)}}.
\end{equation*} 
Furthermore, under Condition~\ref{Con 1}(b) and by Lemma~\ref{Lem 2}, we have:
\begin{equation*}
    \begin{aligned}
        \mathbb{E}_{n_0,D}\mathbb{E}_{n, D_\rho} \epsilon_{n_0,n}^2 &\lesssim (\log n_{0-})^6 \max _{(p, k) \in \mathcal{P}} n_{0-}^{-\frac{2p}{(2p+k)(d+1)}}+ (\log n_{0+})^6 \max _{(p, k) \in \mathcal{P}} n_{0+}^{-\frac{2p}{(2p+k)(d+1)}}\\
        &\quad + (\log n)^6 \max _{(p, k) \in \mathcal{P}} n^{-\frac{2p}{(2p+k)(d+1)}}\lesssim \left(\log \left(n_0\right)\right)^6 \max _{(p, k) \in \mathcal{P}}\left(n_0\right)^{-\frac{2p}{(2p+k)(d+1)}},
    \end{aligned}
\end{equation*}
Similar to \eqref{Mth 2 1} and \eqref{Mth 2 2}, we also obtain the bound
\begin{equation*}
    \mathbb{E}_{n_0,D}\mathbb{E}_{n,D_\rho}\epsilon_{n_0, n}^q\lesssim \left(\log n_0\right)^6 \max _{(p, k) \in \mathcal{P}}n_0^{-\frac{p(q\wedge 2)}{(2p+k)(d+1)}}.
\end{equation*} 
\hfill Q.E.D.

\subsection{Proof of Theorem~\ref{Mth 2}}\label{sec:proof-at}
Based on the Theorem~\ref{Mth generic}, we only need to analyze $\mathcal{E}^{0/1}_{D^{\operatorname{\mathrm{AT}}}}\left(\widehat{f}_{n}^{\operatorname{\mathrm{AT,H}}}, h_D^o\right)$. This part is challenging because $\eta^{\operatorname{\mathrm{AT}}}(\bm{x})$ is a discontinuous function, which is incompatible with many existing theoretical results. Fortunately, $D^{\mathrm{AT}}$ has the same decision boundary with $D$ based on the properties of $\tau$ implied by \eqref{drift}, that is, 
\begin{equation}\label{Mth 3 conclusion}
    \begin{aligned}
        \mathrm{DB}=\{\bm{x}:\eta(\bm{x})=\frac{1}{2}\}&=\{\bm{x}:\eta^{\mathrm{AT}}(\bm{x})=\frac{1}{2}\}, \\
        \mathrm{Sign}(\eta(\bm{x})-\frac{1}{2})&=\mathrm{Sign}(\eta^{\mathrm{AT}}(\bm{x})-\frac{1}{2}), \quad \forall \bm{x}\in\mathcal{X},
    \end{aligned}
\end{equation}
for the same reason for $h_D^o=h_{D^{\mathrm{AT}}}^o$. Thus, we can take Conditions~\ref{Con 2} and \ref{Con 5} to characterize the smooth decision boundary $\mathrm{DB}$ and $\mathcal{E}^{0/1}_{D^{\operatorname{\mathrm{AT}}}}\left(\widehat{f}_{n}^{\operatorname{\mathrm{AT,H}}}, h_D^o\right)$. 

First, we discuss a special case, which is $q>1$ in Condition~\ref{Con 2}. If $\forall C_\gamma, \gamma>0$, there exists $t^{\prime}\geq 0$, s.t.
\begin{equation}\label{Mth 3 gamma vs q 1}
    P_D(\mathrm{dist}(\bm{X}, \mathrm{DB}) \leq t^{\prime}) > C_{\gamma} t^{\prime \gamma}.
\end{equation}
Then, consider the Lipschitz constant $L$ of $\eta(\bm{x})$, we can obtain
\begin{equation*}
    P_D(|\eta(\bm{X})-\frac{1}{2}|\leq Lt^{\prime})>C_\gamma L^{\gamma}t^{\prime \gamma}.
\end{equation*}
Setting $\gamma=q, C_q= C_\gamma L^{\gamma}$ contradicts Condition~\ref{Con 2}. By contradiction, there exists constants $C_\gamma=C_qL^{\gamma}$, and a margin exponent $\gamma=q \in[1, \infty]$ such that for any $t\geq 0$,
\begin{equation}\label{Mth 3 gamma vs q 2}
    P_D(\operatorname{dist}(\bm{X}, \mathrm{DB}) \leq t) \leq C_\gamma t^\gamma.
\end{equation}
Thus, under Condition~\ref{Con 5} and $q>1$, taking Lemma~\ref{Lem S2}, we have
\begin{equation*}
    \mathbb{E}_{D^{\mathrm{AT}}_n}\left[\mathcal{E}^{0/1}_{D^{\mathrm{AT}}}\left(\widehat{f}_{n}^{\mathrm{AT,H}}, h_D^o\right)\right] \lesssim\left(\frac{\log ^3 n}{n}\right)^{\frac{p_h(q+1)}{p_h(q+2)+(d-1)(q+1) / q}}.
\end{equation*}
Conditional on the nuisance-fitting data, the sample-split construction gives i.i.d. observations from $\mathcal D^{\mathrm{AT}}$. Applying Lemma~\ref{Lem S2} conditionally, with constants uniform over the nuisance estimates, and then taking expectations gives the stated bound.
\begin{equation*}
    \mathbb{E}_{n,D_\rho}\mathbb{E}_{n_0,D}\left[\mathcal{E}^{0/1}_D\left(\widehat{f}_{n}^{\mathrm{AT,H}}, h_D^o\right)\right] \lesssim\left(\frac{\log ^3 n}{n}\right)^{\frac{p_h(q+1)}{p_h(q+2)+(d-1)(q+1) / q}}.
\end{equation*}
Similarly, under Condition~\ref{Con 5} and $q\in (0,1]$, taking Lemma~\ref{Lem S2}, we have
\begin{equation*}
    \mathbb{E}_{n,D_\rho}\mathbb{E}_{n_0,D}\left[\mathcal{E}^{0/1}_D\left(\widehat{f}_{n}^{\mathrm{AT,H}}, h_D^o\right)\right] \lesssim\left(\frac{\log ^3 n}{n}\right)^{\frac{p_h(q+1)}{p_h(q+2)+(d-1)(q+1)}}.
\end{equation*}
\hfill Q.E.D.

\subsection{Proof of Theorem~\ref{Mth 3}}\label{sec:proof-pt}

Based on Theorem~\ref{Mth generic}, we only need to analyze $\mathcal{E}^{0/1}_{D^{\operatorname{\mathrm{PT}}}}\left(\widehat{f}_{n}^{\mathrm{PT,H}}, h_{D^\mathrm{PT}}^o\right)$. 

We first condition on the nuisance-fitting data, so that the sample-split observations are i.i.d. from $\mathcal D^{\mathrm{PT}}$. Consider $q>1$, and denote the decision boundary by $\mathrm{DB}^{\mathrm{PT}}=\{\bm x:\eta^{\mathrm{PT}}(\bm x)=1/2\}$. Since $\eta^{\mathrm{PT}}$ may be discontinuous, Condition~\ref{Con 2} does not automatically imply the required mixed-distribution margin conditions. We first examine the following relation between the
boundary regions of $\eta^{\mathrm{PT}}$ and $\eta$.

For any $0<u<1/2$, by the construction of $D^{\mathrm{PT}}$, $\eta^{\mathrm{PT}}(\bm{x})$ equals either $0$ or $1$ outside the pseudo-tagged boundary set $\mathrm{BS}$, while $\eta^{\mathrm{PT}}(\bm{x})=\eta^{\mathrm{pse}}(\bm{x})$ on $\mathrm{BS}$. Therefore,
\begin{equation*}
    \left\{\left|\eta^{\mathrm{PT}}(\bm{X})-\frac{1}{2}\right|\leq u\right\}=\left\{\bm{X}\in\mathrm{BS}:\left|\eta^{\mathrm{pse}}(\bm{X})-\frac{1}{2}\right|\leq u\right\}.
\end{equation*}
Since $\sup_{\bm{x}\in\mathrm{BS}}|\eta^{\mathrm{pse}}(\bm{x})-\eta(\bm{x})|\lesssim\epsilon_{n_0,n}$, there exists constant $c$, such that
\begin{equation*}
    \left\{\left|\eta^{\mathrm{PT}}(\bm{X})-\frac{1}{2}\right|\leq u\right\}\subseteq\left\{\left|\eta(\bm{X})-\frac{1}{2}\right|\leq u+c\epsilon_{n_0,n}\right\}.
\end{equation*}
Hence, by Condition~\ref{Con 2}, $P_D\left(\left|\eta^{\mathrm{PT}}(\bm{X})-\frac{1}{2}\right|\leq u\right)\leq C_q\left(u+c\epsilon_{n_0,n}\right)^q$. In particular, for scales satisfying $u\geq c\epsilon_{n_0,n}$, this gives
\begin{equation*}
    P_D\left(\left|\eta^{\mathrm{PT}}(\bm{X})-\frac{1}{2}\right|\leq u\right)\leq C_q2^q u^q.
\end{equation*}
The preceding bound only controls scales $u\ge c\epsilon_{n_0,n}$. We therefore use the additional mixed-distribution margin assumptions stated in Section~\ref{section 4.2} of the main text, with posterior-margin exponent $q$ and geometric-margin exponent $\gamma=q$. In particular, for a constant $C_\gamma>0$ uniform over the nuisance estimates and all $t\ge0$,
\begin{equation*}
    P_D(\operatorname{dist}(\bm{X}, \mathrm{DB}^{\mathrm{PT}}) \leq t) \leq C_\gamma t^\gamma.
\end{equation*}
Under Condition~\ref{Con 5} and these margin assumptions,
applying Lemma~\ref{Lem S2} conditionally yields
\begin{equation*}
    \mathbb{E}_{D_n^{\mathrm{PT}}}\left[\mathcal{E}_{D^{\mathrm{PT}}}^{0 / 1}\left(\widehat{f}_n^{\mathrm{PT}, \mathrm{H}}, h_{D^{\mathrm{PT}}}^o\right)\right] \lesssim\left(\frac{\log ^3 n}{n}\right)^{\frac{p_h(q+1)}{p_h(q+2)+(d-1)(q+1) / q}}.
\end{equation*}
Taking expectation over the nuisance-fitting data and
applying Theorem~\ref{Mth 1}, together with the moment argument
in Section~\ref{sec:proof-extraction} for the power $q+1$, gives, under
Condition~\hyperref[Con 1]{1(a)},
\begin{equation*}
    \mathbb{E}_{n, D_\rho} \mathbb{E}_{n_0, D}\left[\mathcal{E}_D^{0 / 1}\left(\widehat{f}_n^{\mathrm{PT}, \mathrm{H}}, h_D^o\right)\right] \lesssim n_0^{-\frac{(q+1) \wedge 2}{2}}+(\log n)^6 \max _{(p, k) \in \mathcal{P}} n^{-\frac{p((q+1) \wedge 2)}{(2 p+k)(d+1)}}+\left(\frac{\log ^3 n}{n}\right)^{\frac{p_h(q+1)}{p_h(q+2)+(d-1)(q+1) / q}}.
\end{equation*}
Similarly, if $q \in(0,1]$, taking Lemma~\ref{Lem S2}, we have
\begin{equation*}
    \mathbb{E}_{n, D_\rho} \mathbb{E}_{n_0, D}\left[\mathcal{E}_D^{0 / 1}\left(\widehat{f}_n^{\mathrm{PT}, \mathrm{H}}, h_D^o\right)\right] \lesssim n_0^{-\frac{(q+1) \wedge 2}{2}}+(\log n)^6 \max _{(p, k) \in \mathcal{P}} n^{-\frac{p((q+1) \wedge 2)}{(2 p+k)(d+1)}}+\left(\frac{\log ^3 n}{n}\right)^{\frac{p_h(q+1)}{p_h(q+2)+(d-1)(q+1)}}.    
\end{equation*}
Under Condition~\ref{Con 1}(b), we denote $\gamma=q\vee 1$, taking Theorem~\ref{Mth 1}, then
\begin{equation*}
    \begin{aligned}
        \mathbb{E}_{n, D_\rho}& \mathbb{E}_{n_0, D}\left[\mathcal{E}_D^{0 / 1}\left(\widehat{f}_n^{\mathrm{PT}, \mathrm{H}}, h_D^o\right)\right] \lesssim \left(\log n_0\right)^6 \max_{(p, k) \in \mathcal{P}}n_0^{-\frac{p((q+1) \wedge 2)}{(2 p+k)(d+1)}} 
        +\left(\frac{\log ^3 n}{n}\right)^{\frac{p_h(q+1)}{p_h(q+2)+(d-1)(q+1)/\gamma}}.
    \end{aligned}
\end{equation*}
\hfill Q.E.D.

\subsection{Proof of Lemma~\ref{Lem 2}}\label{sec:proof-lemma2}
We build the error bound for $\widehat{\eta}_\rho(\bm{x})$ as an example. First, we establish the Lipschitz continuity of the squared error term. Under Condition~\ref{Con 3}, which states the Lipschitz continuity and boundedness of $\widehat{\eta}_\rho(\bm{x})$ and $\eta_\rho(\bm{x})$, we have the following result. For any $\bm{x}, \bm{z} \in \mathcal{X}$, we have
\begin{equation*}
    \begin{aligned}
    & \left|\left(\widehat{\eta}_\rho(\bm{x})-\eta_\rho(\bm{x})\right)^2 - \left(\widehat{\eta}_\rho(\bm{z})-\eta_\rho(\bm{z})\right)^2\right| \\
    = &\left| \left(\widehat{\eta}_\rho(\bm{x})-\eta_\rho(\bm{x})\right) - \left(\widehat{\eta}_\rho(\bm{z})-\eta_\rho(\bm{z})\right) \right| \cdot \left| \left(\widehat{\eta}_\rho(\bm{x})-\eta_\rho(\bm{x})\right) + \left(\widehat{\eta}_\rho(\bm{z})-\eta_\rho(\bm{z})\right) \right| \\
    \leq & \left( |\widehat{\eta}_\rho(\bm{x})-\widehat{\eta}_\rho(\bm{z})| + |\eta_\rho(\bm{x})-\eta_\rho(\bm{z})| \right) \cdot \left( |\widehat{\eta}_\rho(\bm{x})-\eta_\rho(\bm{x})| + |\widehat{\eta}_\rho(\bm{z})-\eta_\rho(\bm{z})| \right) \\
    \leq & \left( L\|\bm{x}-\bm{z}\|_2 + L\|\bm{x}-\bm{z}\|_2 \right) \cdot (1 + 1) = 4L \|\bm{x}-\bm{z}\|_2.
    \end{aligned}
\end{equation*}
Then the function $s(\bm{x}):=\left(\widehat{\eta}_\rho(\bm{x})-\eta_\rho(\bm{x})\right)^2$ is $4L$-Lipschitz continuous. Since $0\leq s\leq1$, we may enlarge the fixed Lipschitz bound to $L\geq1/(4r_{\bm{x}})$. Every nontrivial cone below then has radius $N/(4L)\leq r_{\bm{x}}$. Next, we bound the supremum of $s(\bm{x})$ by bounding its expectation. Specifically, we claim that if $\sup_{\bm{x}\in\mathcal{X}} s(\bm{x})\geq N$ for some constant $N>0$, then $\mathbb{E}_{\bm{X}} s(\bm{X})\geq \frac{C^{\prime}_{\bm{x}} \mu_{-} V_d}{(4L)^d(d+1)} N^{d+1}$, where $V_d$ is the volume of the $d$-dimensional unit ball. The proof proceeds by lower-bounding the expectation over a specific ball $B\left(\bm{x}_0, \frac{N}{4L}\right)$, where $\bm{x}_0$ is a point satisfying $s\left(\bm{x}_0\right) \geq N$. We have
\begin{equation}\label{eq:Lemma 2 1}
    \mathbb{E}_{\bm{X}}s(\bm{X})\geq\mathbb{E}_{\bm{X}}s(\bm{X})\mathbb{I}(\bm{X}\in B(\bm{x}_0, \frac{N}{4L}))\geq \mathbb{E}_{\bm{X}}R(\bm{X};\bm{x}_0)\mathbb{I}(\bm{X}\in B(\bm{x}_0, \frac{N}{4L})),
\end{equation}
where $R(\bm{x};\bm{x}_0)=N-4L\|\bm{x}-\bm{x}_0\|_2$, which satisfies $R(\bm{x})<s(\bm{x})$. Then, under Condition~\ref{Con 4}, we have
\begin{equation}\label{eq:Lemma 2 2}
    \begin{aligned}
        \mathbb{E}_{\bm{X}}&R(\bm{X};\bm{x}_0)\mathbb{I}(\bm{X}\in B(\bm{x}_0, \frac{N}{4L})) \geq\mu_{-}\int_{\mathcal{X}\cap B(\bm{x}_0, \frac{N}{4L})}R(\bm{x};\bm{x}_0)\mathrm{d}\bm{x}\\
        &\geq\mu_{-}C^{\prime}_{\bm{x}}\int_{B(\bm{x}_0, \frac{N}{4L})}R(\bm{x};\bm{x}_0)\mathrm{d}\bm{x}=\frac{C^{\prime}_{\bm{x}} \mu_{-} V_d}{(4L)^d(d+1)} N^{d+1}.
    \end{aligned}
\end{equation}
By contraposition, for any fixed realization of the training sample $D_n$,
\begin{equation}\label{Lemma 2 equ 1}
    \sup_{\bm{x}\in\mathcal{X}}s(\bm{x})\leq \left(\dfrac{(4L)^d(d+1)}{C^{\prime}_{\bm{x}}\mu_{-} V_d}\mathbb{E}_{\bm{X}}s(\bm{X})\right)^{\frac{1}{d+1}}.
\end{equation}
Recalling the definitions of $\widehat{\eta}_\rho$ and $\eta_\rho$, we have $R_{D_\rho}^{\mathrm{MSE}}\left(\widehat{\eta}_{\rho}\right)-R_{D_\rho}^{\mathrm{MSE}}\left(\eta_\rho\right)=\mathbb{E}_{\bm{X}}s(\bm{X})$. By Lemma~\ref{Lem S1}, after taking expectation over the training sample, we have
\begin{equation}\label{Lemma 2 equ 2}
    \mathbb{E}_{n,D_\rho}\mathbb{E}_{\bm{X}}s(\bm{X})\lesssim(\log n)^6\max_{(p,k)\in\mathcal{P}}n^{-\frac{2p}{2p+k}}.
\end{equation}
Taking expectation with respect to the training sample on both sides of \eqref{Lemma 2 equ 1} yields
\begin{equation*}
    \mathbb{E}_{n,D_\rho}\sup_{\bm{x}\in\mathcal{X}}s(\bm{x})\lesssim \mathbb{E}_{n,D_\rho}\left[\left\{\mathbb{E}_{\bm{X}}s(\bm{X})\right\}^{\frac{1}{d+1}}\right].
\end{equation*}
 Since $u\mapsto u^{1/(d+1)}$ is concave on $[0,\infty)$, Jensen's inequality gives
\begin{equation*}
    \mathbb{E}_{n,D_\rho}\left[\left\{\mathbb{E}_{\bm{X}}s(\bm{X})\right\}^{\frac{1}{d+1}}\right]\leq \left\{\mathbb{E}_{n,D_\rho}\mathbb{E}_{\bm{X}}s(\bm{X})\right\}^{\frac{1}{d+1}}.
\end{equation*}
Combining this inequality with \eqref{Lemma 2 equ 2}, we obtain
\begin{equation*}
    \mathbb{E}_{n,D_\rho}\sup_{\bm{x}\in\mathcal{X}}\left(\widehat{\eta}_{\rho}(\bm{x})-\eta_\rho(\bm{x})\right)^2=\widetilde{O}\left(\max_{(p,k)\in\mathcal{P}}n^{-\frac{2p}{(2p+k)(d+1)}}\right).
\end{equation*}
The analyses for $\widehat{\rho}_{+}(\bm{x})$ and $\widehat{\rho}_{-}(\bm{x})$ are analogous, with the effective sample sizes being $n_{0+}$ and $n_{0-}$, respectively. Let us focus on the error bound for $\widehat{\rho}_{+}(\bm{x})$. Conditioning on the value of $n_{0+}$, the uniform convergence result from the previous steps yields:
\begin{equation}\label{eq:conditional_rate}
    \mathbb{E}_{n_0,D}\left[\sup_{\bm{x}\in\mathcal{X}}\left(\widehat{\rho}_+(\bm{x})-\rho_+(\bm{x})\right)^2 \bigg| n_{0+}=t \right] =\widetilde{O}\left(\max_{(p,k)\in \mathcal{P}}t^{-\frac{2p}{(2p+k)(d+1)}}\right).
\end{equation}
To obtain the unconditional expectation, we use the law of total expectation. Let $p_+ = P_D(Y=1)$. Note that $n_{0+}$ is a random variable following a Binomial distribution $\operatorname{Binomial}(n_0, p_+)$. We know $p_+ \geq C_y > 0$. Let $t_0 = \lfloor n_0 p_+ / 2 \rfloor$. We split the expectation into two cases based on whether $n_{0+}$ is smaller than $t_0$:
\begin{equation*}
    \begin{aligned}
        \mathbb{E}_{n_0,D}\left[\sup_{\bm{x}\in\mathcal{X}}\left(\widehat{\rho}_+(\bm{x})-\rho_+(\bm{x})\right)^2\right] = & \sum_{t=0}^{n_0} \mathbb{E}_{n_0,D}\left[\sup_{\bm{x}\in\mathcal{X}}\left(\widehat{\rho}_+(\bm{x})-\rho_+(\bm{x})\right)^2 \bigg| n_{0+}=t \right] P_D(n_{0+}=t) \\
        = & \sum_{t=0}^{t_0-1} \mathbb{E}_{n_0,D}\left[\sup_{\bm{x}\in\mathcal{X}}\left(\widehat{\rho}_+(\bm{x})-\rho_+(\bm{x})\right)^2 \bigg| n_{0+}=t \right] P_D(n_{0+}=t)\\
        &+ \sum_{t=t_0}^{n_0} \mathbb{E}_{n_0,D}\left[\sup_{\bm{x}\in\mathcal{X}}\left(\widehat{\rho}_+(\bm{x})-\rho_+(\bm{x})\right)^2 \bigg| n_{0+}=t \right] P_D(n_{0+}=t).
    \end{aligned}
\end{equation*}
For the first term, we use the fact that the squared error is always bounded by 1. For small $t$, we can use a trivial bound:
\begin{equation*}
    \sum_{t=0}^{t_0-1} \mathbb{E}_{n_0,D}\left[\sup_{\bm{x}\in\mathcal{X}}\left(\widehat{\rho}_+(\bm{x})-\rho_+(\bm{x})\right)^2 | n_{0+}=t\right] P_D(n_{0+}=t) \leq P_D(n_{0+} < t_0).
\end{equation*}
Since $t_0 = \lfloor n_0 p_+ / 2 \rfloor$, this is a large deviation event. By Hoeffding's inequality, $P(n_{0+} < t_0) \leq \exp(-n_0 p_+^2/2)$, which decays exponentially fast with $n_0$.

For the second term, we have $t \geq t_0$. Therefore, the conditional rate from \eqref{eq:conditional_rate} can be bounded in terms of $n_0$:
\begin{equation*}
    \begin{aligned}
        &\sum_{t=t_0}^{n_0} \mathbb{E}_{n_0,D}\left[\sup_{\bm{x}\in\mathcal{X}}\left(\widehat{\rho}_+(\bm{x})-\rho_+(\bm{x})\right)^2 | n_{0+}=t\right] P(n_{0+}=t)=\sum_{t=t_0}^{n_0} \widetilde{O}\left(\max_{(p,k)\in \mathcal{P}}t^{-\frac{2p}{(2p+k)(d+1)}}\right) P(n_{0+}=t) \\
        &=\widetilde{O}\left(\max_{(p,k)\in \mathcal{P}}t_0^{-\frac{2p}{(2p+k)(d+1)}}\right) \sum_{t=t_0}^{n_0} P(n_{0+}=t) =\widetilde{O}\left(\max_{(p,k)\in \mathcal{P}}t_0^{-\frac{2p}{(2p+k)(d+1)}}\right) \cdot P(n_{0+} \geq t_0)\\ &=\widetilde{O}\left(\max_{(p,k)\in \mathcal{P}}n_0^{-\frac{2p}{(2p+k)(d+1)}}\right).
    \end{aligned}
\end{equation*}
Combining both terms, the exponentially decaying part is dominated by the polynomially decaying part for sufficiently large $n_0$. Thus, we conclude that
\begin{equation*}
    \mathbb{E}_{n_0,D}\sup_{\bm{x}\in\mathcal{X}}\left(\widehat{\rho}_+(\bm{x})-\rho_+(\bm{x})\right)^2 = \max_{(p,k)\in \mathcal{P}}\widetilde{O}\left(n_0^{-\frac{2p}{(2p+k)(d+1)}}\right).
\end{equation*}
The same argument holds for $\widehat{\rho}_{-}(\bm{x})$ by replacing $n_{0+}$ with $n_{0-}$, which completes the proof.

\hfill Q.E.D.

\renewcommand{\theequation}{C.\arabic{equation}}
\renewcommand{\thetable}{C.\arabic{table}}
\renewcommand{\thefigure}{C.\arabic{figure}}
\setcounter{equation}{0} 
\setcounter{table}{0} 
\setcounter{figure}{0}

\section{Additional Results and Methodological Details}\label{Appendix C}

Section~\ref{sec:covariate-shift} considers covariate shift, and Section~\ref{sec:pseudo-label-resampling} discusses pseudo-label resampling. Section~\ref{Appendix A.2} states additional corollaries, with proofs in Sections~\ref{sec:proof-cor-s1}--\ref{sec:proof-cor-s3}. Section~\ref{sec:bspline} gives a B-spline implementation, with proofs in Sections~\ref{sec:proof-bspline-extraction} and~\ref{sec:proof-bspline-classification}.

\subsection{Extension to Covariate Shift}\label{sec:covariate-shift}
We have assumed for notational simplicity that the small clean dataset and the large noisy dataset are drawn from the same marginal feature distribution. We now relax this assumption to address the more general and practical setting of \textbf{covariate shift}.

Let the small clean dataset, $D^{\mathrm{cl}}_{n_0}:=\{(\bm{x}_{0i}, \tilde{y}_{0i}, y_{0i})\}_{i=1}^{n_0}$, be drawn from a distribution $D^{\mathrm{cl}}$, where its marginal feature distribution $P_{D^{\mathrm{cl}}}(\bm{x})$ may differ from $P_D(\bm{x})$. We also denote
\begin{equation*}
    \begin{aligned}
        &\eta^{\mathrm{cl}}_{\rho}(\bm{x})=P_{D^{\mathrm{cl}}}(\widetilde{Y}=1|\bm{X}=\bm{x}),\\
        &\rho^{\mathrm{cl}}_+(\bm{x})=P_{D^{\mathrm{cl}}}(\widetilde{Y}=-1|Y=1,\bm{X}=\bm{x}),\\
        &\rho^{\mathrm{cl}}_-(\bm{x})=P_{D^{\mathrm{cl}}}(\widetilde{Y}=1|Y=-1,\bm{X}=\bm{x}).
    \end{aligned}
\end{equation*}
We continue to use $P_D(\bm{x})$ to denote the target distribution. Under this covariate shift setting, the primary modification to our theoretical framework is in the Condition~\ref{Con 4}. Accordingly, we replace Condition~\ref{Con 4} from the main text with the following Condition~\hyperref[Con:4_star]{4*}.

\begin{Constar}[4*]\label{Con:4_star}
    Let $\mathcal{X} \subset[0,1]^d$ be the closed support of the feature vector $\bm{X}$ and $P_D(\bm{x}),P_{D^{\mathrm{cl}}}(\bm{x})$ are absolutely continuous with respect to the Lebesgue measure $\lambda$ on $[0,1]^d$. Write $R(\bm{x};\bm{x}_0)=N-L\left\|\bm{x}-\bm{x}_0\right\|_2$. There exist constants $r_{\bm{x}}, C_{\bm{x}}, C^{\prime}_{\bm{x}}, C_y, \mu_{-}>0$ such that:
    \begin{align}
        \lambda[\mathcal{X} \cap B(\bm{x}, r)] & \geq C_{\bm{x}} \lambda[B(\bm{x}, r)], \quad \forall 0<r \leq r_{\bm{x}}, \forall \bm{x} \in \mathcal{X}, \\
        \int_{\mathcal{X}\cap B(\bm{x}_0, \frac{N}{L})}R(\bm{x};\bm{x}_0)\mathrm{d}\bm{x} &\geq C^{\prime}_{\bm{x}}\int_{B(\bm{x}_0, \frac{N}{L})}R(\bm{x};\bm{x}_0)\mathrm{d}\bm{x},\quad \forall \bm{x}_0\in\mathcal{X},\quad 0<N/L\leq r_{\bm{x}}, \\
        p_{D^{\mathrm{cl}}}(\bm{x})\geq \mu_{-},&\quad p_D(\bm{x})=p_{D_{\rho}}(\bm{x}) \geq \mu_{-}, \quad \forall \bm{x} \in \mathcal{X}, \label{eq:density_condition_cs} \\
         \operatorname{Supp}(p_{D^{\mathrm{cl}}}) = \operatorname{Supp}&(p_D) = \mathcal{X}, \quad \eta^{\mathrm{cl}}_{\rho}=\eta_{\rho},\quad \rho_{+}^{\mathrm{cl}}=\rho_{+},\quad \rho_{-}^{\mathrm{cl}}=\rho_{-}.\label{eq:support_condition_cs}
    \end{align}
    where $\operatorname{Supp}(\cdot)$ represents the support of a function.
\end{Constar}
The cone inequality on this local range follows from the ball-volume condition. Indeed, for $r=N/L\leq r_{\bm{x}}$, the layer-cake identity gives
\begin{equation*}
    \begin{aligned}
    \int_{\mathcal X\cap B(\bm x_0,r)}(N-L\|\bm x-\bm x_0\|_2)\,\mathrm d\bm x
    &=L\int_0^r\lambda[\mathcal X\cap B(\bm x_0,t)]\,\mathrm dt\\
    &\geq C_{\bm x}L\int_0^r\lambda[B(\bm x_0,t)]\,\mathrm dt
    =C_{\bm x}\int_{B(\bm x_0,r)}(N-L\|\bm x-\bm x_0\|_2)\,\mathrm d\bm x.
    \end{aligned}
\end{equation*}
Condition~\hyperref[Con:4_star]{4*} allows the audited and noisy samples to have different covariate marginals while retaining common support, invariant conditional mechanisms, and density lower bounds. The analysis does not require estimating an importance weight, but it relies on these support and design conditions. With the modified condition in place, we can now state a summary theorem confirming that all of our main theoretical results remain valid under the covariate shift setting.
\begin{Mth}[Validity of Main Results under Covariate Shift]\label{Thm:CovariateShiftSummary}
    Let Condition~\hyperref[Con:4_star]{4*} replace Condition~\ref{Con 4}. Under the same remaining conditions as stated in the main text, Lemmas~\ref{Lem S1}--\ref{Lem S2}, Theorems~\ref{Mth 1}--\ref{Mth 3}, and Corollaries~\ref{Cor S1}--\ref{Cor S3} all continue to hold.
\end{Mth}
\noindent {\bf Proof.} The core of the argument rests on the validity of Lemma~\ref{Lem 2}.

The only place where the change in the marginal distribution from $p_D(\bm{x})$ to $p_{D^{\mathrm{cl}}}(\bm{x})$ has an effect is in the estimation of $\widehat{\rho}_+(\bm{x}),\widehat{\rho}_-(\bm{x})$ from the clean dataset $D^{\mathrm{cl}}_{n_0}$. Specifically, this affects the derivation of the uniform error bounds for these estimators in Lemma~\ref{Lem 2}, particularly in steps leading from \eqref{eq:Lemma 2 1} and \eqref{eq:Lemma 2 2} to \eqref{Lemma 2 equ 1} in Section~\ref{sec:proof-lemma2}. In these steps, the expectation is taken with respect to $\bm{X} \sim p_{D^{\mathrm{cl}}}(\bm{x})$. However, \eqref{eq:Lemma 2 1} and \eqref{eq:Lemma 2 2} depend on the boundedness of density and not on the specific marginal distribution over which the expectation is taken, as long as the support is fixed.

Crucially, the final conclusion of Lemma~\ref{Lem 2} provides a uniform convergence bound, i.e., a supremum over all $\bm{x} \in \mathcal{X}$. Our fundamental assumption for transfer learning is that the conditional distributions are invariant, meaning $\eta^{\mathrm{cl}}_{\rho}(\bm{x}) = \eta_{\rho}(\bm{x})$, $\rho_{+}^{\mathrm{cl}}(\bm{x}) = \rho_{+}(\bm{x})$, and $\rho_{-}^{\mathrm{cl}}(\bm{x}) = \rho_{-}(\bm{x})$ for all $\bm{x} \in \mathcal{X}$. Since Condition~\hyperref[Con:4_star]{4*} ensures that the support $\mathcal{X}$ is identical for both $D$ and $D^{\mathrm{cl}}$, the uniform convergence bound remains sufficient for all subsequent theoretical derivations.

Given that Lemma~\ref{Lem 2} is the foundational result for all subsequent analyses, its validity under covariate shift naturally implies the other results also hold without modification.

\hfill Q.E.D.

\subsection{Pseudo-label Resampling in Pro-PT}\label{sec:pseudo-label-resampling}
As described in Section~\ref{section 3.2} of the main text, the random resampling of pseudo-labels at every training epoch is a crucial design choice that provides a regularization effect against overfitting. To fully appreciate this, we can analyze a natural but suboptimal variant of our method, which we refer to as ``Fixed-Sampling Pro-PT". In this alternative approach, a pseudo-label is sampled only once for each boundary sample and then remains fixed throughout the entire training process. This subsection provides an intuitive explanation for why such a fixed scheme is prone to overfitting by cementing initial sampling noise, and how our dynamic resampling approach mitigates this pitfall. 

\begin{figure}[ht]
    \centering
    \includegraphics[width=0.7\linewidth]{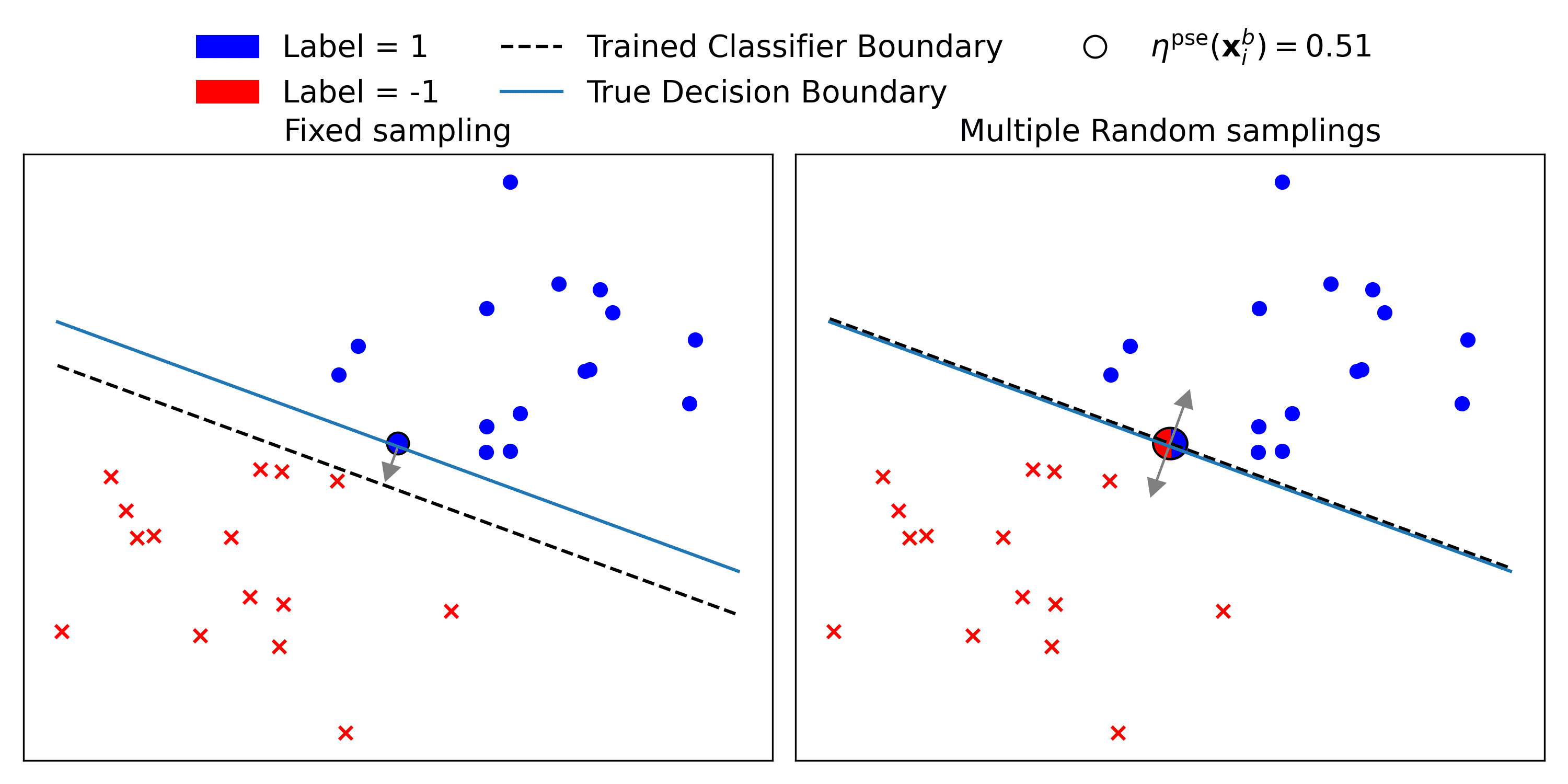} 
    \caption{An intuitive illustration of the benefits of multiple random sampling.}
    \label{fig:tug_of_war}
\end{figure}

As illustrated in Figure~\ref{fig:tug_of_war}, consider a boundary sample with an estimated posterior of $\eta^{\text{pse}}(\bm{x}_i^b) = 0.51$. A fixed pseudo-labeling scheme might, by chance, assign it a permanent label of $+1$. The model would then be forced to treat this highly uncertain point as a confident positive example, potentially distorting the decision boundary far into the negative region to accommodate this data point (left). Our method avoids this, where the process resembles a ``tug-of-war" at the decision boundary. In one epoch, the sample might be labeled as $+1$, pulling the boundary slightly towards it. In the next epoch, it might be labeled $-1$, pulling the boundary back (right). Over many epochs, the influence of the random sampling averages out, and the decision boundary remains correctly positioned near this point of high uncertainty, leading to a more robust and better-generalized model.

\subsection{Additional Corollaries}\label{Appendix A.2}
Corollary~\ref{Cor S1} considers a classifier trained only on the extracted BO data. Corollaries~\ref{Cor S2} and~\ref{Cor S3} give the sigmoid-loss results for Pro-AT and Pro-PT, respectively. We begin with the classifier $\widehat{f}_{n}^{\mathrm{BO,H}}$ trained directly on the extracted BO data.
\begin{Cor}\label{Cor S1}
    Suppose that Conditions~\ref{Con 1}--\ref{Con 5} and \ref{Con S1} hold. Use the independent nuisance-fitting and training samples specified in Section~\ref{section 4} of the main text. Let $\mathcal F_{\mathrm{fit}}$ contain the nuisance-fitting data and the fitted threshold $\tau$, and define
    \begin{equation*}
        \begin{aligned}
        A_\tau&=\left\{\bm x:\left|\widehat\eta_\rho(\bm x)-\frac12+
        \frac12\{\widehat\rho_+(\bm x)-\widehat\rho_-(\bm x)\}\right|>\tau\right\},\\
        \pi_\tau&=P_D(\bm X\in A_\tau\mid\mathcal F_{\mathrm{fit}}),\qquad
        G_\tau=\{\epsilon_{n_0,n}\leq\tau\}.
        \end{aligned}
    \end{equation*}
    Put $\gamma=q\vee1$ and $\alpha=p_h(q+1)/\{p_h(q+2)+(d-1)(q+1)/\gamma\}$. For an integer $j\geq1$, write $r(j)=1\wedge\{\log^3(ej)/j\}^{\alpha}$ and set $r(0)=1$. Conditional on the extracted sample size $m\geq1$, use the network sizes in Lemma~\ref{Lem S2} with sample size $m$; if $m=0$, return the constant classifier $+1$. Then, with expectations over both independent stages,
    \begin{equation}\label{eq:bo-conditional-risk}
        \mathbb E\!\left[\mathcal E_D^{0/1}(\widehat f_n^{\mathrm{BO,H}},h_D^o)\right]
        \leq C\mathbb E[r(m)]+\varepsilon^{\mathrm{BO}}+P(G_\tau^c),
        \qquad \varepsilon^{\mathrm{BO}}=2\mathbb E(1-\pi_\tau).
    \end{equation}
\end{Cor}
The learning term averages over the random number of extracted observations. The term $\varepsilon^{\mathrm{BO}}$ measures the covariate mass discarded by extraction. The event $G_\tau$ is the extraction guarantee in \eqref{drift}; its complement contributes at most one unit of excess risk. In particular, this last term is zero for the theoretical choice $\tau=C\epsilon_{n_0,n}$ with $C\geq1$.
\begin{Rem}\label{rem:bo-selection}
    Conditional on $\mathcal F_{\mathrm{fit}}$, the extraction set is fixed and the $n$ candidate covariates are independent. Consequently,
    \begin{equation}\label{eq:bo-selected-law}
        m\mid\mathcal F_{\mathrm{fit}}\sim\operatorname{Bin}(n,\pi_\tau),
        \qquad
        p_{D^{\mathrm{BO}}}(\bm x\mid\mathcal F_{\mathrm{fit}})
        =\frac{p_D(\bm x)\mathbb I(\bm x\in A_\tau)}{\pi_\tau}
        \quad(\pi_\tau>0).
    \end{equation}
    The selected observations are i.i.d.\ from this conditional law given $\mathcal F_{\mathrm{fit}}$ and $m$. On $G_\tau$, their labels are $h_D^o(\bm X)$. The total variation distance has the explicit form
    \begin{equation*}
        \begin{aligned}
        d_{\mathrm{TV}}(P_D,P_{D^{\mathrm{BO}}}\mid\mathcal F_{\mathrm{fit}})
        &=\frac12\int_{A_\tau}p_D(\bm x)(\pi_\tau^{-1}-1)\,\mathrm d\bm x
          +\frac12\int_{A_\tau^c}p_D(\bm x)\,\mathrm d\bm x\\
        &=1-\pi_\tau.
        \end{aligned}
    \end{equation*}
    Thus $\varepsilon^{\mathrm{BO}}$ is twice the expected boundary-sample probability. Its definition also covers $\pi_\tau=0$, when $m=0$ almost surely and a selected conditional distribution is unnecessary. For fixed nuisance estimates, this discarded probability is nondecreasing in $\tau$.
\end{Rem}

We next state the sigmoid-loss results under the additional conditions in Section~\ref{sec:nn-definitions}.
\begin{Cor}\label{Cor S2}
    Suppose that Conditions~\ref{Con 1}, \ref{Con 3}, \ref{Con 4} and \ref{Con S1}--\ref{Con S3} hold. Taking the same $\tau$ of Theorem~\ref{Mth 2}, we have
    \begin{equation*}
        \mathbb{E}_{n_0, D}\mathbb{E}_{n, D_{\rho}}\left[\mathcal{E}^{0/1}_D\left(\widehat{f}_{n}^{~\mathrm{AT,S}}, h_D^o\right)\right] = \widetilde{O}(n^{-\frac{2 p_f}{2 p_f+d}}).
    \end{equation*}
\end{Cor}
\begin{Cor}\label{Cor S3}
    Suppose that Conditions~\ref{Con 1}(a), \ref{Con 3}, \ref{Con 4} and \ref{Con S1}--\ref{Con S3} hold. Taking the same $\tau$ of Theorem~\ref{Mth 3}, we have
    \begin{equation*}
        \mathbb{E}_{n_0, D}\mathbb{E}_{n, D_{\rho}}\left[\mathcal{E}^{0/1}_D\left(\widehat{f}_{n}^{~\mathrm{PT,S}}, h_D^o\right)\right] = \widetilde{O}(n_0^{-\frac{(q+1) \wedge 2}{2}}+\max _{(p, k) \in \mathcal{P}} n^{-\frac{p((q+1) \wedge 2)}{(2 p+k)(d+1)}}+n^{-\frac{2 p_f}{2 p_f+d}}).
    \end{equation*}
    Moreover, if Condition~\ref{Con 1}(b) holds instead of \ref{Con 1}(a), we have
    \begin{equation*}
        \mathbb{E}_{n_0, D}\mathbb{E}_{n, D_{\rho}}\left[\mathcal{E}^{0/1}_D\left(\widehat{f}_{n}^{~\mathrm{PT,S}}, h_D^o\right)\right] = \widetilde{O}(\max _{(p, k) \in \mathcal{P}}\left(n_0\right)^{-\frac{p((q+1) \wedge 2)}{(2 p+k)(d+1)}}+n^{-\frac{2 p_f}{2 p_f+d}}).
    \end{equation*}
\end{Cor}
Corollaries~\ref{Cor S2} and \ref{Cor S3} establish convergence rates for our methods when employing the Sigmoid loss, complementing the results for the Hinge loss. We highlight that our framework is not restricted to these specific loss functions. The core methodology is agnostic to the final classification loss. Leveraging recent theoretical advancements in deep learning, similar guarantees can be derived for a broader class of loss functions known to be consistent for classification, which demonstrates the versatility of our approach.

\subsection{Proof of Corollary~\ref{Cor S1}}\label{sec:proof-cor-s1}
Fix $\mathcal F_{\mathrm{fit}}$. Independence of the candidate observations gives \eqref{eq:bo-selected-law}. More explicitly, conditional on any fixed selection pattern, the selected covariates have product law $P_D(\cdot\mid A_\tau)^{\otimes m}$. Averaging over patterns with the same count preserves this product law. On $G_\tau$, \eqref{drift} also identifies their labels with the BO labels.

First suppose $G_\tau$ holds, $\pi_\tau\geq1/2$, and $m\geq1$. We use the error decomposition
\begin{equation}\label{Cor 1 decompose}
    \mathcal{E}^{0/1}_D\left(\widehat{f}_{n}^{\mathrm{BO,H}}, h_D^o\right)=\mathcal{E}^{0/1}_D\left(\widehat{f}_{n}^{\mathrm{BO,H}}, h_D^o\right) - \mathcal{E}^{0/1}_{D^{\mathrm{BO}}}\left(\widehat{f}_{n}^{\mathrm{BO,H}}, h_D^o\right) + \mathcal{E}^{0/1}_{D^{\mathrm{BO}}}\left(\widehat{f}_{n}^{\mathrm{BO,H}}, h_D^o\right).
\end{equation}
The BO labels have the same CAD decision set as $h_D^o$ and satisfy every finite posterior-margin exponent. We may take the BO posterior to equal $1/2$ on the clean boundary $\mathrm{DB}=\{\bm x:\eta(\bm x)=1/2\}$ and $\eta^\star$ elsewhere, since $\mathrm{DB}$ has zero selected probability. When $q>1$, the smoothness of $\eta$ implies $|\eta(\bm x)-1/2|\leq L\,\mathrm{dist}(\bm x,\mathrm{DB})$. Condition~\ref{Con 2} therefore gives
\begin{equation*}
    P_D(\mathrm{dist}(\bm X,\mathrm{DB})\leq t\mid A_\tau,\mathcal F_{\mathrm{fit}})
    \leq\frac{C_q(Lt)^q}{\pi_\tau}\leq2C_q(Lt)^q.
\end{equation*}
For $q\leq1$, use the first bound in Lemma~\ref{Lem S2}. For $q>1$, use its geometric-margin bound with $\gamma=q$. In both cases the constants are uniform over $\pi_\tau\geq1/2$, and conditioning also on $m$ gives
\begin{equation*}
    \mathbb E\!\left[
    \mathcal E_{D^{\mathrm{BO}}}^{0/1}(\widehat f_n^{\mathrm{BO,H}},h_D^o)
    \mid\mathcal F_{\mathrm{fit}},m\right]\leq C r(m).
\end{equation*}

For the first two terms, expand conditional on $\mathcal F_{\mathrm{fit}}$:
\begin{equation*}
    \begin{aligned}
        &\mathcal{E}^{0/1}_D\left(\widehat{f}_{n}^{\mathrm{BO,H}}, h_D^o\right) - \mathcal{E}^{0/1}_{D^{\mathrm{BO}}}\left(\widehat{f}_{n}^{\mathrm{BO,H}}, h_D^o\right)\\
        = & \left(\mathbb{E}_D[\mathbb{I}(Y\operatorname{sign}(\widehat{f}_{n}^{\mathrm{BO,H}}(\bm{X}))<0)]- \mathbb{E}_D\left[\mathbb{I}\left(Y h_D^o(\bm{X})<0\right)\right]\right)\\
        & -\left(\mathbb{E}_{D^{\mathrm{BO}}}[\mathbb{I}(Y\operatorname{sign}(\widehat{f}_{n}^{\mathrm{BO,H}}(\bm{X}))<0)] - \mathbb{E}_{D^{\mathrm{BO}}}\left[\mathbb{I}\left(Y h_D^o(\bm{X})<0\right)\right]\right) \\
        = & \int_{\mathcal{X}}\left(\mathbb{I}(\widehat{f}_{n}^{\mathrm{BO,H}}(\bm{x})<0)-\mathbb{I}(\eta(\bm{x})<\frac{1}{2})\right)\left\{2\left[p_D(\bm{x})\eta(\bm{x})-p_{D^{\mathrm{BO}}}(\bm{x})\eta^{\star}(\bm{x})\right]-(p_D(\bm{x})-p_{D^{\mathrm{BO}}}(\bm{x}))\right\}\mathrm{d}\bm{x}\\
        = & 2\int_{\mathcal{X}}\left(\mathbb{I}(\widehat{f}_{n}^{\mathrm{BO,H}}(\bm{x})<0)-\mathbb{I}(\eta(\bm{x})<\frac{1}{2})\right)(\eta(\bm{x})-\eta^{\star}(\bm{x}))p_D(\bm{x})\mathrm{d}\bm{x}\\
        & + 2\int_{\mathcal{X}}\left(\mathbb{I}(\widehat{f}_{n}^{\mathrm{BO,H}}(\bm{x})<0)-\mathbb{I}(\eta(\bm{x})<\frac{1}{2})\right)\left(\eta^{\star}(\bm{x})-\frac{1}{2}\right)(p_D(\bm{x})-p_{D^{\text{BO}}}(\bm{x}))\mathrm{d}\bm{x} \\
        \leq & 2\int_{\mathcal{X}}\left(\mathbb{I}(\widehat{f}_{n}^{\mathrm{BO,H}}(\bm{x})<0)-\mathbb{I}(\eta(\bm{x})<\frac{1}{2})\right)\left(\eta^{\star}(\bm{x})-\frac{1}{2}\right)(p_D(\bm{x})-p_{D^{\text{BO}}}(\bm{x}))\mathrm{d}\bm{x} \\
        \leq & \int_{\mathcal{X}}|p_D(\bm{x})-p_{D^{\text{BO}}}(\bm{x})|\mathrm{d}\bm{x} = 2d_{\mathrm{TV}}\left(P_D,P_{D^{\mathrm{BO}}}\mid\mathcal F_{\mathrm{fit}}\right).
    \end{aligned}
\end{equation*}
where the first inequality holds because the product of  $\mathbb{I}\left(\hat{f}_n^{\mathrm{BO}, \mathrm{H}}(\bm{x})<0\right)-\mathbb{I}\left(\eta(\bm{x})<\frac{1}{2}\right)$ and $\eta(\bm{x})-\eta^{\star}(\bm{x})$ is always non-positive. The second inequality follows from a straightforward bound, since $|\mathbb{I}(\widehat{f}_{n}^{\mathrm{BO,H}}(\bm{x})<0)-\mathbb{I}(\eta(\bm{x})<\frac{1}{2})| \leq 1$ and $\left|\eta^{\star}(\bm{x})-\frac{1}{2}\right| \leq \frac{1}{2}$. 
By Remark~\ref{rem:bo-selection}, the last term equals $2(1-\pi_\tau)$. If $m=0$, the excess risk is at most $1=r(0)$. If $\pi_\tau<1/2$, it is at most $1\leq2(1-\pi_\tau)$, so no lower bound on the selection probability is needed. These cases together yield, on $G_\tau$,
\begin{equation*}
    \mathbb E\!\left[\mathcal E_D^{0/1}(\widehat f_n^{\mathrm{BO,H}},h_D^o)
        \mid\mathcal F_{\mathrm{fit}}\right]
    \leq C\sum_{j=0}^n r(j){n\choose j}\pi_\tau^j(1-\pi_\tau)^{n-j}
        +2(1-\pi_\tau).
\end{equation*}
The binomial sum is $\mathbb E[r(m)\mid\mathcal F_{\mathrm{fit}}]$. On $G_\tau^c$, use the bound $\mathcal E_D^{0/1}\leq1$. Taking the outer expectation proves \eqref{eq:bo-conditional-risk}.

\hfill Q.E.D.

\subsection{Proof of Corollary~\ref{Cor S2}}\label{sec:proof-cor-s2}
Applying the same decomposition strategy yields
\begin{equation*}
    \begin{aligned}
        \mathcal{E}^{0/1}_D&\left(\widehat{f}_{n}^{\mathrm{AT,S}}, h_D^o\right)=\mathcal{E}^{0/1}_D\left(\widehat{f}_{n}^{\mathrm{AT,S}}, h_D^o\right) - \mathcal{E}^{0/1}_{D^{\mathrm{AT}}}\left(\widehat{f}_{n}^{\mathrm{AT,S}}, h_D^o\right) + \mathcal{E}^{0/1}_{D^{\mathrm{AT}}}\left(\widehat{f}_{n}^{\mathrm{AT,S}}, h_D^o\right) \\
        &=2\mathbb{E}_{\bm{X}}\left\{(\eta(\bm{X})-\eta^{\mathrm{AT}}(\bm{X}))\left[\mathbb{I}(\eta(\bm{X})\geq\frac{1}{2})-\mathbb{I}(\widehat{f}_{n}^{\mathrm{AT,S}}(\bm{x})\geq0)\right]\right\} + \mathcal{E}_{D^{\mathrm{AT}}}^{0/1}\left(\widehat{f}_{n}^{\mathrm{AT,S}}, h_D^o\right) \\
        &\leq \mathcal{E}_{D^{\mathrm{AT}}}^{0/1}\left(\widehat{f}_{n}^{\mathrm{AT,S}}, h_D^o\right).
    \end{aligned}
\end{equation*}
Based on the conclusion \eqref{Mth 3 conclusion}, similarly, we can take Conditions~\ref{Con S2}--\ref{Con S3} and Lemma~\ref{Lem S3} to characterize $\mathcal{E}_{D^{\mathrm{AT}}}^{0/1}\left(\widehat{f}_{n}^{\mathrm{AT,S}}, h_D^o\right)$.
\begin{equation*}
    \mathbb{E}_{n,D_\rho}\mathbb{E}_{n_0,D}\left[\mathcal{E}_{D^{\mathrm{AT}}}^{0/1}\left(\widehat{f}_{n}^{\mathrm{AT,S}}, h_D^o\right)\right]\lesssim n^{-\frac{2p_f}{2p_f+d}} \log ^7 n,
\end{equation*}
or
\begin{equation*}
    \mathbb{E}_{n,D_\rho}\mathbb{E}_{n_0,D}\left[\mathcal{E}_D^{0/1}\left(\widehat{f}_{n}^{\mathrm{AT,S}}, h_D^o\right)\right]\lesssim n^{-\frac{2p_f}{2p_f+d}} \log ^7 n.
\end{equation*}
\hfill Q.E.D.

\subsection{Proof of Corollary~\ref{Cor S3}}\label{sec:proof-cor-s3}
Applying the same decomposition strategy yields
\begin{equation*}
    \begin{aligned}
        \mathcal{E}_D^{0 / 1}\left(\widehat{f}_n^{\mathrm{PT}, \mathrm{S}}, h_D^o\right)=&\mathcal{E}_D^{0 / 1}\left(\widehat{f}_n^{\mathrm{PT,S}}, h_D^o\right)-\mathcal{E}_{D^{\mathrm{PT}}}^{0 / 1}\left(\widehat{f}_n^{\mathrm{PT,S}}, h_D^o\right) \\
        &+\mathcal{E}_{D^{\mathrm{PT}}}^{0 / 1}\left(\widehat{f}_n^{\mathrm{PT,S}}, h_D^o\right) - \mathcal{E}_{D^{\mathrm{PT}}}^{0 / 1}\left(\widehat{f}_n^{\mathrm{PT,S}}, h_{D^{\mathrm{PT}}}^o\right) + \mathcal{E}_{D^{\mathrm{PT}}}^{0 / 1}\left(\widehat{f}_n^{\mathrm{PT,S}}, h_{D^{\mathrm{PT}}}^o\right)\\
        \lesssim & \epsilon^{q+1}_{n_0,n}+\mathcal{E}_{D^{\mathrm{PT}}}^{0 / 1}\left(\widehat{f}_n^{\mathrm{PT,S}}, h_{D^{\mathrm{PT}}}^o\right).
    \end{aligned}
\end{equation*}
We can take Conditions~\ref{Con S2}--\ref{Con S3} and Lemma~\ref{Lem S3} to characterize $\mathcal{E}_{D^{\mathrm{PT}}}^{0 / 1}\left(\widehat{f}_n^{\mathrm{PT}, \mathrm{S}}, h_{D^{\mathrm{PT}}}^o\right)$.
\begin{equation*}
    \mathbb{E}_{n, D_\rho} \mathbb{E}_{n_0, D}\left[\mathcal{E}_{D^{\mathrm{PT}}}^{0 / 1}\left(\widehat{f}_n^{\mathrm{PT}, \mathrm{S}}, h_{D^{\mathrm{PT}}}^o\right)\right] \lesssim n^{-\frac{2 p_f}{2 p_f+d}} \log ^7 n,
\end{equation*}
or
\begin{equation}\label{Cor 2 final}
    \mathbb{E}_{n, D_\rho} \mathbb{E}_{n_0, D}\left[\mathcal{E}_D^{0 / 1}\left(\widehat{f}_n^{\mathrm{PT}, \mathrm{S}}, h_D^o\right)\right] \lesssim \mathbb{E}_{n, D_\rho} \mathbb{E}_{n_0, D}\epsilon^{q+1}_{n_0,n}+n^{-\frac{2 p_f}{2 p_f+d}} \log ^7 n.
\end{equation}
Combining \eqref{Cor 2 final} with Theorem~\ref{Mth 1} yields the result.

\hfill Q.E.D.

\subsection{A B-spline Implementation}\label{sec:bspline}
In this subsection, we provide a fully nonparametric B-spline instantiation of the generic framework in Section~\ref{section 3} of the main text. The purpose is to show that the proposed procedure is not tied to neural networks. The first stage uses B-spline least-squares estimators for the nuisance functions $\eta_{\rho}(\bm{x}),\rho_{+}(\bm{x})$, and $\rho_{-}(\bm{x})$. The final stage directly estimates the decision boundary by a B-spline sieve classifier, instead of estimating the discontinuous posterior functions $\eta^{\mathrm{AT}}(\bm{x})$ and $\eta^{\mathrm{PT}}(\bm{x})$.
\begin{Con}\label{Con S4}
    {\bf (B-spline nuisance specification).} Suppose that Condition~\ref{Con 4} holds. As a concrete B-spline case study of Condition~\ref{Con 3}, assume each nuisance function to be estimated is a $(p,C)$-smooth function of $k$ effective variables for some $(p,k)\in\mathcal{P}$. For a sample of size $N$, let $\mathcal{S}_{N}$ be the corresponding tensor-product B-spline sieve with order larger than $p$, $J_N$ quasi-uniform intervals per coordinate, and $K_N\asymp J_N^k$ normalized nonnegative basis functions that sum to one. Assume:
    (i). The B-spline estimators are truncated to $[0,1]$. (ii). There exists a constant $C_\rho^{\prime}>0$ such that $\widehat{\rho}_{-}(\bm{x})+\widehat{\rho}_{+}(\bm{x})\leq 1-C_\rho^{\prime}$.
\end{Con}
Recall that $D_n=\{(\bm{x}_i,\tilde{y}_i)\}_{i=1}^n$ is drawn i.i.d. from $D_\rho$, and $D_{n_0}=\{(\bm{x}_{0 i}, y_{0 i}, \tilde{y}_{0 i})\}_{i=1}^{n_0}$ is the audited clean dataset. Let $n_{0,+}=\sum_{i=1}^{n_0}\mathbb{I}(y_{0i}=1)$ and $n_{0,-}=\sum_{i=1}^{n_0}\mathbb{I}(y_{0i}=-1)$. For $\eta_{\rho}(\bm{x})$, define $\widehat{\eta}_{\rho}=\argmin_{g\in\mathcal{S}_{n}}\frac{1}{n}\sum_{i=1}^n \mathcal{L}_{\mathrm{MSE}}(\tilde y_i,g(\bm{x}_i))$, followed by truncation to $[0,1]$.
Under Condition~\ref{Con 1}(a), define $\widehat{\rho}_{+}=\frac{\sum_{i=1}^{n_0}\mathbb{I}(y_{0i}=1,\tilde y_{0i}=-1)}{\sum_{i=1}^{n_0}\mathbb{I}(y_{0i}=1)}, \widehat{\rho}_{-}=\frac{\sum_{i=1}^{n_0}\mathbb{I}(y_{0i}=-1,\tilde y_{0i}=1)}{\sum_{i=1}^{n_0}\mathbb{I}(y_{0i}=-1)}$, with the convention that the ratio is set to $0$ if the denominator is $0$. In this case, $\widehat{\rho}_{+}(\bm{x})\equiv\widehat{\rho}_{+}$ and $\widehat{\rho}_{-}(\bm{x})\equiv\widehat{\rho}_{-}$. Besides, under Condition~\ref{Con 1}(b), define
\begin{equation*}
\widehat{\rho}_{+}=\argmin_{g\in\mathcal{S}_{n_{0,+}}}\frac{\sum_{i=1}^{n_0}\mathbb{I}(y_{0i}=1)\mathcal{L}_{\mathrm{MSE}}(-\tilde y_{0i},g(\bm{x}_{0i}))}{\sum_{i=1}^{n_0}\mathbb{I}(y_{0i}=1)},\quad \\
\widehat{\rho}_{-}=\argmin_{g\in\mathcal{S}_{n_{0,-}}}\frac{\sum_{i=1}^{n_0}\mathbb{I}(y_{0i}=-1)\mathcal{L}_{\mathrm{MSE}}(\tilde y_{0i},g(\bm{x}_{0i}))}{\sum_{i=1}^{n_0}\mathbb{I}(y_{0i}=-1)}.
\end{equation*}
Both estimators are truncated to $[0,1]$. For an empty class sample, set its estimate to zero. For a singular least-squares design, use the minimum-norm solution before truncation. As in Section~\ref{section 4}, nuisance fitting and mixed-data training use independent samples; their sizes have the stated orders $n_0$ and $n$.

We record explicitly the design quantities that enter the uniform estimation bound. For a target $g$ with effective covariate $\bm U$, let $Q_g$ be its fitting distribution, let $\bm b(\bm u)$ be the $K_N$-dimensional spline basis, and define
\begin{equation*}
    \widehat G_N=\frac1N\sum_{i=1}^N\bm b(\bm U_i)\bm b(\bm U_i)^\top,\qquad
    \Lambda_N=\sup_{\bm u}\frac1N\sum_{i=1}^N
       \left|\bm b(\bm u)^\top\widehat G_N^{-1}\bm b(\bm U_i)\right|.
\end{equation*}
The supremum is over the projection of $\mathcal X$ onto the effective variables. For deterministic $\lambda_N\geq1$ and $v_N>0$, put
\begin{equation*}
    H_N=\{\widehat G_N\text{ is invertible},\ \Lambda_N\leq\lambda_N,\
                  \max_j(\widehat G_N^{-1})_{jj}\leq v_N\}.
\end{equation*}
With a fixed approximation constant $C_s$ determined by the smoothness norm and spline order, define
\begin{equation}\label{eq:spline-design-error}
    r_g(N)=1\wedge\left\{C_s(1+\lambda_N)^2J_N^{-2p}
                 +\frac{v_N\{1+\log(2K_N)\}}{N}
                 +P_{Q_g^{\otimes N}}(H_N^c)\right\},\qquad r_g(0)=1.
\end{equation}
This bound retains the probability of an unstable design. For $\eta_\rho$, $Q_g$ is the marginal law of the effective covariates. For $\rho_+$ and $\rho_-$, it is their law conditional on $Y=1$ and $Y=-1$, respectively. Write
\begin{equation*}
    \bar r_\pm(n_0)=\sum_{j=0}^{n_0}P(n_{0,\pm}=j)\,r_{\rho_\pm}(j).
\end{equation*}

\begin{Mth}\label{Mth S2}
    Suppose that Conditions~\ref{Con 1}, \ref{Con 2}, and \ref{Con S4} hold. For every fixed $s>0$, under Condition~\ref{Con 1}(a),
    \begin{equation*}
        \mathbb E\epsilon_{n_0,n}^{s}
        \lesssim\{n_0^{-1}+r_{\eta_\rho}(n)\}^{(s\wedge2)/2}.
    \end{equation*}
    Under Condition~\ref{Con 1}(b),
    \begin{equation*}
        \mathbb E\epsilon_{n_0,n}^{s}
        \lesssim\{r_{\eta_\rho}(n)+\bar r_+(n_0)+\bar r_-(n_0)\}^{(s\wedge2)/2}.
    \end{equation*}
    The choices $s=q$ and $s=q+1$ give the extraction and pseudo-labeling error bounds, respectively.
\end{Mth}
For example, if the actual fitting designs satisfy $\lambda_N=O(1)$, $v_N=O(K_N)$, and $P(H_N^c)=\widetilde O(N^{-2p/(2p+k)})$, then choosing $J_N\asymp(N/\log N)^{1/(2p+k)}$ gives $r_g(N)=\widetilde O(N^{-2p/(2p+k)})$. These bounds are assessed under each target's fitting distribution. Substitution into Theorem~\ref{Mth S2} gives the explicit rates derived at the end of its proof.

We next consider a B-spline sieve classifier trained on the mixed data $D_n^{\mathrm{AT}}$ and $D_n^{\mathrm{PT}}$ constructed in Section~\ref{section 3.2} of the main text. Let $\mathcal C_J^{p_h,C,K,T}$ be the sieve version of the CAD class in Definition~\ref{Def 2}. Replace each horizon function by a tensor-product spline with order larger than $p_h$ and $J$ quasi-uniform intervals per coordinate. Each horizon has $M_J\asymp J^{d-1}$ coefficients. The positive region is the union of its $T$ pieces, each defined by $K$ horizon inequalities; taking the union also handles overlaps introduced by approximation.
\begin{Con}\label{Con S5}
    {\bf (B-spline boundary sieve learner).} Conditional on the independent nuisance-fitting data, the final classifiers are empirical risk minimizers over $\mathcal C_J^{p_h,C,K,T}$. Specifically, $\widehat{h}^{\mathrm{AT}}\in\argmin_{h\in\mathcal C_J^{p_h,C,K,T}}\frac{1}{n}\sum_{(\bm{x},y)\in D_n^{\mathrm{AT}}}\mathbb{I}(yh(\bm{x})<0)$ and $\widehat{h}^{\mathrm{PT}}$ is defined similarly on $D_n^{\mathrm{PT}}$.
\end{Con}
For $M\in\{\mathrm{AT},\mathrm{PT}\}$, define the conditional approximation error and the complexity term
\begin{equation}\label{eq:spline-classification-terms}
    \begin{aligned}
    A_{M,J}&=\inf_{h\in\mathcal C_J^{p_h,C,K,T}}
       \int_{\mathcal X}|2\eta^M(\bm x)-1|\,\mathbb I\{h(\bm x)\ne h_{D^M}^o(\bm x)\}\,\mathrm dP_D(\bm x),\\
    t_{n,J}&=1\wedge\left\{\frac{V_J\log(en)}{n}\right\}^{(q+1)/(q+2)},
       \qquad V_J=C_{K,T,d}(M_J+1).
    \end{aligned}
\end{equation}
Here $C_{K,T,d}$ is a fixed constant large enough to bound the VC dimension, as shown below.
\begin{Mth}\label{Mth S3}
    Suppose that Conditions~\ref{Con 1}--\ref{Con 5}, \ref{Con S4}, and \ref{Con S5} hold. Use the sample split above and let $G_\tau=\{\epsilon_{n_0,n}\leq\tau\}$. For Pro-PT, use the conditional posterior-margin assumption with exponent $q$ and uniform constants stated in Section~\ref{section 4.2}. Then
    \begin{itemize}
        \item[(a)] For Pro-AT,
        \begin{equation*}
            \mathbb E\!\left[\mathcal E_D^{0/1}(\widehat h^{\mathrm{AT}},h_D^o)\right]
            \lesssim\mathbb E A_{\mathrm{AT},J}+t_{n,J}+P(G_\tau^c).
        \end{equation*}
        \item[(b)] Under Condition~\ref{Con 1}(a), for Pro-PT,
        \begin{equation*}
            \mathbb E\!\left[\mathcal E_D^{0/1}(\widehat h^{\mathrm{PT}},h_D^o)\right]
            \lesssim\{n_0^{-1}+r_{\eta_\rho}(n)\}^{((q+1)\wedge2)/2}
              +\mathbb E A_{\mathrm{PT},J}+t_{n,J}+P(G_\tau^c).
        \end{equation*}
        \item[(c)] Under Condition~\ref{Con 1}(b), for Pro-PT,
        \begin{equation*}
            \mathbb E\!\left[\mathcal E_D^{0/1}(\widehat h^{\mathrm{PT}},h_D^o)\right]
            \lesssim\{r_{\eta_\rho}(n)+\bar r_+(n_0)+\bar r_-(n_0)\}^{((q+1)\wedge2)/2}
              +\mathbb E A_{\mathrm{PT},J}+t_{n,J}+P(G_\tau^c).
        \end{equation*}
    \end{itemize}
\end{Mth}
For a deterministic $\tau>0$, Markov's inequality also gives $P(G_\tau^c)\leq1\wedge\{\tau^{-s}\mathbb E\epsilon_{n_0,n}^s\}$ for any fixed $s>0$, with the moment bounded by Theorem~\ref{Mth S2}.
The approximation terms in \eqref{eq:spline-classification-terms} keep the covariate distribution explicit. Whenever $\mathbb E A_{M,J}\lesssim J^{-p_h\gamma}$, with $\gamma=q\vee1$, balancing the two classification terms gives
\begin{equation}\label{eq:spline-rate-specialization}
    J\asymp(n/\log n)^{\frac{q+1}{p_h\gamma(q+2)+(d-1)(q+1)}},
    \qquad
    \mathbb E A_{M,J}+t_{n,J}
       =\widetilde O\!\left(n^{-\frac{p_h(q+1)}{p_h(q+2)+(d-1)(q+1)/\gamma}}\right).
\end{equation}
Thus the displayed power rate is recovered when the weighted boundary-approximation error has this order. Theorem~\ref{Mth S3} applies with the actual approximation error otherwise.

\subsection{Proof of Theorem~\ref{Mth S2}}\label{sec:proof-bspline-extraction}
Under Condition~\ref{Con S4}, each conditional mean function estimated in the first stage is a smooth low-dimensional regression function. We first establish its uniform least-squares error bound under its own fitting distribution. This makes explicit the empirical projection and design-matrix terms in spline series estimation \cite{chen2015optimaluniform}.

Let $Z_i\in\{0,1\}$ be the response in the probability regression, so $\mathbb E[Z_i\mid\bm U_i]=g(\bm U_i)$, and put $\xi_i=Z_i-g(\bm U_i)$. Spline approximation gives a deterministic $g_J\in\mathcal S_N$ with $\|g-g_J\|_\infty\leq C_aJ_N^{-p}$. On an invertible design, define the empirical projection
\begin{equation*}
    P_Na(\bm u)=\bm b(\bm u)^\top\widehat G_N^{-1}
       \frac1N\sum_{i=1}^N\bm b(\bm U_i)a(\bm U_i).
\end{equation*}
Since $P_Ng_J=g_J$, the untruncated fit $\widetilde g$ satisfies
\begin{equation}\label{eq:spline-ls-decomposition}
    \begin{aligned}
    \widetilde g-g
       &=P_N(g-g_J)+(g_J-g)
          +\bm b^\top\widehat G_N^{-1}\frac1N\sum_{i=1}^N\bm b(\bm U_i)\xi_i,\\
    \|P_Ng-g\|_\infty
       &\leq(1+\Lambda_N)C_aJ_N^{-p}.
    \end{aligned}
\end{equation}
Write $\bm a=\widehat G_N^{-1}N^{-1}\sum_i\bm b(\bm U_i)\xi_i$. For its $j$th coordinate, the squared weights of the independent centered responses sum to
\begin{equation*}
    \sum_{i=1}^N\left\{\frac1N\bm e_j^\top\widehat G_N^{-1}\bm b(\bm U_i)\right\}^2
       =\frac1N(\widehat G_N^{-1})_{jj}.
\end{equation*}
Each centered Bernoulli response has range length one. Conditional Hoeffding's inequality and a union bound therefore give, on $H_N$,
\begin{equation*}
    P\!\left(\max_{1\leq j\leq K_N}|a_j|>t\mid\bm U_1,\ldots,\bm U_N\right)
       \leq 2K_N\exp(-2Nt^2/v_N).
\end{equation*}
Integrating this bound at $t^2=v_N\log(2K_N)/(2N)$ yields
\begin{equation*}
    \mathbb E\!\left[\max_j|a_j|^2\mid\bm U_1,\ldots,\bm U_N\right]
       \leq\frac{v_N\{\log(2K_N)+1\}}{2N}.
\end{equation*}
Nonnegativity and partition of unity imply $\|\bm b^\top\bm a\|_\infty\leq\max_j|a_j|$. Squaring \eqref{eq:spline-ls-decomposition}, using $(a+b)^2\leq2a^2+2b^2$, and then truncating to $[0,1]$ gives
\begin{equation*}
    \mathbb E\!\left[\|\widehat g-g\|_\infty^2\mathbb I(H_N)\right]
       \leq 2C_a^2(1+\lambda_N)^2J_N^{-2p}
          +\frac{v_N\{\log(2K_N)+1\}}{N}.
\end{equation*}
The truncation cannot enlarge the error because $g$ takes values in $[0,1]$. On $H_N^c$, including singular designs, the squared error is at most one. This proves $\mathbb E\|\widehat g-g\|_\infty^2\leq r_g(N)$ with $C_s=2C_a^2$ in \eqref{eq:spline-design-error}.

Under Condition~\ref{Con 1}(a), $\rho_{+}$ and $\rho_{-}$ are constants. Lemma~\ref{Lem 1} gives $\mathbb{E}_{n_0,D}|\widehat{\rho}_{+}-\rho_{+}|^2\lesssim n_0^{-1}$ and $\mathbb{E}_{n_0,D}|\widehat{\rho}_{-}-\rho_{-}|^2\lesssim n_0^{-1}$. The B-spline bound applied to $\widehat\eta_\rho$ gives $\mathbb E\|\widehat\eta_\rho-\eta_\rho\|_\infty^2\leq r_{\eta_\rho}(n)$. By the definition \eqref{epsilon n_0 n},
\begin{equation*}
    \mathbb E\epsilon_{n_0,n}^2\lesssim n_0^{-1}+r_{\eta_\rho}(n).
\end{equation*}
Since $\epsilon_{n_0,n}\leq2$, Jensen's inequality for $0<s\leq2$ and boundedness for $s>2$ imply
\begin{equation*}
    \mathbb E\epsilon_{n_0,n}^{s}
    \leq
    \begin{cases}
        (\mathbb E\epsilon_{n_0,n}^2)^{s/2},&0<s\leq2,\\
        2^{s-2}\mathbb E\epsilon_{n_0,n}^2,&s>2.
    \end{cases}
\end{equation*}
This proves the class-dependent assertion.

Under Condition~\ref{Con 1}(b), condition also on $n_{0,+}$ and $n_{0,-}$. Given $n_{0,+}=j$, the positive-class covariates are i.i.d.\ from their class-conditional law, and the preceding bound gives
\begin{equation*}
    \mathbb E\|\widehat\rho_+-\rho_+\|_\infty^2
       \leq\sum_{j=0}^{n_0}P(n_{0,+}=j)r_{\rho_+}(j)=\bar r_+(n_0).
\end{equation*}
The same calculation applies to $\rho_-$. Hence
\begin{equation*}
    \mathbb E\epsilon_{n_0,n}^2
       \lesssim r_{\eta_\rho}(n)+\bar r_+(n_0)+\bar r_-(n_0).
\end{equation*}
The preceding moment argument proves the instance-dependent assertion.

The count calculation yields explicit powers when the design bounds following Theorem~\ref{Mth S2} hold. Put $p_\pm=P_D(Y=\pm1)>0$. For any $0\leq r_{\rho_\pm}\leq1$, the binomial lower-tail bound gives
\begin{equation*}
    \bar r_\pm(n_0)
    \leq P(n_{0,\pm}<n_0p_\pm/2)
          +\sup_{j\geq n_0p_\pm/2}r_{\rho_\pm}(j)
    \leq e^{-n_0p_\pm/8}
          +\sup_{j\geq n_0p_\pm/2}r_{\rho_\pm}(j).
\end{equation*}
Thus $r_g(N)=\widetilde O(\max_{(p,k)\in\mathcal P}N^{-2p/(2p+k)})$ under the relevant fitting laws yields $\bar r_\pm(n_0)=\widetilde O(\max_{(p,k)\in\mathcal P}n_0^{-2p/(2p+k)})$. With $n_0\ll n$, substitution and the same moment argument give the following power expressions for $s=q$:
\begin{equation*}
    \begin{aligned}
    \mathbb E_{n_0,D}\mathbb E_{n,D_\rho}\epsilon_{n_0,n}^q
       &=\widetilde O\!\left(n_0^{-(q\wedge2)/2}
          +\max_{(p,k)\in\mathcal P}n^{-p(q\wedge2)/(2p+k)}\right)
          &&\text{under Condition~\ref{Con 1}(a)},\\
    \mathbb E_{n_0,D}\mathbb E_{n,D_\rho}\epsilon_{n_0,n}^q
       &=\widetilde O\!\left(\max_{(p,k)\in\mathcal P}
          n_0^{-p(q\wedge2)/(2p+k)}\right)
          &&\text{under Condition~\ref{Con 1}(b)}.
    \end{aligned}
\end{equation*}
The same substitution with $s=q+1$ gives the pseudo-labeling terms. The general bounds in Theorem~\ref{Mth S2} retain $\bar r_\pm$ when the class-conditional designs do not give these powers.

\hfill Q.E.D.

\subsection{Proof of Theorem~\ref{Mth S3}}\label{sec:proof-bspline-classification}
Fix the nuisance-fitting data and write $M\in\{\mathrm{AT},\mathrm{PT}\}$. We first prove the conditional learning bound for the spline boundary sieve, and then use the transfer decompositions in Section~\ref{Appendix A}.

\noindent\textbf{Boundary approximation.}
For each of the $KT$ horizon functions in the CAD representation of $h_{D^M}^o$, spline approximation gives $\|g_{tk}-g_{tk,J}\|_\infty\leq C_aJ^{-p_h}$. Form the approximate classifier by the same intersections and union. If its label differs from $h_{D^M}^o$, at least one horizon inequality has changed sign. Therefore the disagreement set is contained in
\begin{equation*}
    B_J=\bigcup_{t=1}^T\bigcup_{k=1}^K
      \{\bm x:|x_{j_{tk}}-g_{tk}(\bm x_{-j_{tk}})|\leq C_aJ^{-p_h}\},
\end{equation*}
and
\begin{equation}\label{eq:spline-boundary-strip}
    A_{M,J}\leq\int_{B_J}|2\eta^M(\bm x)-1|\,\mathrm dP_D(\bm x)
       \leq P_D(B_J),\qquad \lambda(B_J)\leq2KTC_aJ^{-p_h}.
\end{equation}
Absolute continuity in Condition~\ref{Con 4} implies that the last probability tends to zero uniformly over sets with this vanishing volume. More explicitly, if $\omega_D(u)=\sup_{\lambda(B)\leq u}P_D(B)$, then $A_{M,J}\leq\omega_D(2KTC_aJ^{-p_h})$ and $\omega_D(u)\to0$ as $u\to0$. Equation~\eqref{eq:spline-boundary-strip} retains the weighted probability needed for a quantitative rate.

\noindent\textbf{Sieve complexity and estimation error.}
On any $r$ covariates, one spline horizon is a linear threshold in at most $M_J+1$ features. Its number of label patterns is at most $(r+1)^{M_J+1}$ for a fixed coordinate. Allowing the $d$ coordinate choices, and then $KT$ horizons, bounds the number of patterns by
\begin{equation*}
    \Pi_J(r)\leq\{d(r+1)^{M_J+1}\}^{KT}.
\end{equation*}
The VC dimension of a fixed union of intersections of these threshold classes is at most $C_{K,T,d}(M_J+1)=V_J$; equivalently, apply the threshold growth bound $\sum_{\ell=0}^{M_J+1}{r\choose\ell}$ and solve $2^r\leq\{d(er/(M_J+1))^{M_J+1}\}^{KT}$ for $r\geq M_J+1$.

Write $e_M(h)=\mathcal E_{D^M}^{0/1}(h,h_{D^M}^o)$ and $\Delta_h=\mathbb I(Yh(\bm X)<0)-\mathbb I(Yh_{D^M}^o(\bm X)<0)$. For any $t>0$, the posterior-margin bound yields
\begin{equation*}
    \begin{aligned}
    \operatorname{Var}_{D^M}(\Delta_h)
       &\leq P_D(h(\bm X)\ne h_{D^M}^o(\bm X))\\
       &\leq P_D(|\eta^M(\bm X)-1/2|\leq t)
          +\frac{e_M(h)}{2t}
        \leq C_qt^q+\frac{e_M(h)}{2t}
        \lesssim e_M(h)^{q/(q+1)}.
    \end{aligned}
\end{equation*}
The last inequality follows by taking $t=e_M(h)^{1/(q+1)}$; if $e_M(h)=0$, take a limit. For Pro-AT, the posterior-margin bound follows from Condition~\ref{Con 2}, since its posterior equals $\eta$ on boundary samples and belongs to $\{0,1\}$ elsewhere. For Pro-PT, it is the mixed-distribution margin assumption stated in the theorem.

Let $\kappa=q/(q+1)$ and $u_{n,J}=V_J\log(en)/n$. The bounded-loss VC empirical-process bound applied to the variance inequality gives the localized fluctuation order $\sqrt{r^\kappa u_{n,J}}+u_{n,J}$ at excess-risk scale $r$. The ERM oracle inequality of \cite[Theorem~2 and Section~2.4]{massart2006risk} applies with this variance modulus and with approximation error $A_{M,J}$. Its fixed point solves
\begin{equation*}
    r\asymp\sqrt{r^\kappa u_{n,J}}+u_{n,J},
    \qquad r\asymp u_{n,J}^{1/(2-\kappa)}
           =u_{n,J}^{(q+1)/(q+2)}\quad(u_{n,J}\leq1).
\end{equation*}
Consequently,
\begin{equation}\label{eq:spline-erm-oracle}
    \mathbb E[e_M(\widehat h^M)\mid\mathcal F_{\mathrm{fit}}]
        \lesssim A_{M,J}+t_{n,J}.
\end{equation}
Measurable versions can be obtained from a countable dense coefficient set. A small downward shift of each horizon before rational approximation preserves its labels on any fixed finite sample, including positive ties. Each spline graph has Lebesgue measure zero, so absolute continuity also gives convergence of the risks. Finally, balancing $J^{-p_h\gamma}$ with $\{J^{d-1}\log(en)/n\}^{(q+1)/(q+2)}$ gives \eqref{eq:spline-rate-specialization}.

\noindent\textbf{Transfer to the clean distribution.}
On $G_\tau$, the extracted labels are BO labels. By the same decomposition as in the proof of Theorem~\ref{Mth 2},
\begin{equation*}
    \mathcal{E}^{0/1}_D\bigl(\widehat{h}^{\mathrm{AT}},h_D^o\bigr)=\left[\mathcal{E}^{0/1}_D\bigl(\widehat{h}^{\mathrm{AT}},h_D^o\bigr)-\mathcal{E}^{0/1}_{D^{\mathrm{AT}}}\bigl(\widehat{h}^{\mathrm{AT}},h_D^o\bigr)\right]+\mathcal{E}^{0/1}_{D^{\mathrm{AT}}}\bigl(\widehat{h}^{\mathrm{AT}},h_D^o\bigr).
\end{equation*}
The first bracket is non-positive, as shown in Section~\ref{Appendix A}, and $h_{D^{\mathrm{AT}}}^o=h_D^o$. Hence $\mathcal E_D^{0/1}(\widehat h^{\mathrm{AT}},h_D^o)\leq\mathcal E_{D^{\mathrm{AT}}}^{0/1}(\widehat h^{\mathrm{AT}},h_D^o)$. Combining this inequality with \eqref{eq:spline-erm-oracle}, and using $\mathcal E_D^{0/1}\leq1$ on $G_\tau^c$, gives
\begin{equation*}
    \mathbb E\!\left[\mathcal E_D^{0/1}(\widehat h^{\mathrm{AT}},h_D^o)\right]
       \lesssim \mathbb E A_{\mathrm{AT},J}+t_{n,J}+P(G_\tau^c).
\end{equation*}

For Pro-PT, use the same posterior-drift and boundary-shift decomposition as \eqref{Generic PT 1} and \eqref{Generic PT 2}:
\begin{equation*}
    \begin{aligned}
    \mathcal E_D^{0/1}(\widehat h^{\mathrm{PT}},h_D^o)
    ={}&\left[\mathcal E_D^{0/1}(\widehat h^{\mathrm{PT}},h_D^o)
         -\mathcal E_{D^{\mathrm{PT}}}^{0/1}(\widehat h^{\mathrm{PT}},h_D^o)\right]\\
       &+\left[\mathcal E_{D^{\mathrm{PT}}}^{0/1}(\widehat h^{\mathrm{PT}},h_D^o)
         -\mathcal E_{D^{\mathrm{PT}}}^{0/1}(\widehat h^{\mathrm{PT}},h_{D^{\mathrm{PT}}}^o)\right]
         +e_{\mathrm{PT}}(\widehat h^{\mathrm{PT}}).
    \end{aligned}
\end{equation*}
Condition~\ref{Con S4} supplies the fitted denominator bound used in the posterior calculation in Section~\ref{Appendix A}. Thus, with $\delta=\sup_{\bm x\in\mathrm{BS}}|\eta^{\mathrm{pse}}(\bm x)-\eta(\bm x)|$ and $\delta=0$ if $\mathrm{BS}$ is empty,
\begin{equation*}
    \delta\leq\frac{\|\widehat\eta_\rho-\eta_\rho\|_\infty+
                       \|\widehat\rho_+-\rho_+\|_\infty+
                       \|\widehat\rho_--\rho_-\|_\infty}{C_\rho'}
           \leq \frac{2\epsilon_{n_0,n}}{C_\rho'}.
\end{equation*}
To bound the two brackets together, set $w_D=|2\eta-1|$ and $w_{\mathrm{PT}}=|2\eta^{\mathrm{PT}}-1|$. On a boundary sample with $|\eta-1/2|>2\delta$, the two Bayes signs agree and $w_D\leq2w_{\mathrm{PT}}$. When $|\eta-1/2|\leq2\delta$, $w_D\leq4\delta$. Outside $\mathrm{BS}$, on $G_\tau$, the BO posterior has the clean Bayes sign and $w_D\leq w_{\mathrm{PT}}=1$. Therefore, for every classifier $h$,
\begin{equation*}
    \begin{aligned}
    w_D(\bm x)\mathbb I\{h(\bm x)\ne h_D^o(\bm x)\}
       &\leq2w_{\mathrm{PT}}(\bm x)\mathbb I\{h(\bm x)\ne h_{D^{\mathrm{PT}}}^o(\bm x)\}
          +4\delta\,\mathbb I\{|\eta(\bm x)-1/2|\leq2\delta\},\\
    \mathcal E_D^{0/1}(h,h_D^o)
       &\leq2e_{\mathrm{PT}}(h)+4\delta C_q(2\delta)^q
        \lesssim e_{\mathrm{PT}}(h)+\epsilon_{n_0,n}^{q+1}.
    \end{aligned}
\end{equation*}
Taking expectations and applying \eqref{eq:spline-erm-oracle} yields
\begin{equation*}
    \mathbb E\!\left[\mathcal E_D^{0/1}(\widehat h^{\mathrm{PT}},h_D^o)\right]
       \lesssim \mathbb E\epsilon_{n_0,n}^{q+1}
          +\mathbb E A_{\mathrm{PT},J}+t_{n,J}+P(G_\tau^c).
\end{equation*}
Repeating the proof of Theorem~\ref{Mth S2} with the power $q+1$ completes the proof. Under the design specialization stated after that theorem, the same substitution gives
\begin{equation*}
    \begin{aligned}
    \mathbb E\epsilon_{n_0,n}^{q+1}
       &=\widetilde O\!\left(n_0^{-((q+1)\wedge2)/2}
          +\max_{(p,k)\in\mathcal P}n^{-p((q+1)\wedge2)/(2p+k)}\right)
          &&\text{under Condition~\ref{Con 1}(a)},\\
    \mathbb E\epsilon_{n_0,n}^{q+1}
       &=\widetilde O\!\left(\max_{(p,k)\in\mathcal P}
          n_0^{-p((q+1)\wedge2)/(2p+k)}\right)
          &&\text{under Condition~\ref{Con 1}(b)}.
    \end{aligned}
\end{equation*}
Combining these expressions with \eqref{eq:spline-rate-specialization} recovers the stated power orders whenever the corresponding design and weighted approximation bounds hold, with $P(G_\tau^c)$ retained for the chosen threshold.

\hfill Q.E.D.

\renewcommand{\theequation}{D.\arabic{equation}}
\renewcommand{\thetable}{D.\arabic{table}}
\renewcommand{\thefigure}{D.\arabic{figure}}
\setcounter{equation}{0} 
\setcounter{table}{0} 
\setcounter{figure}{0}

\section{Additional Experimental Details}\label{Appendix D}

\subsection{Simulation Details}\label{Appendix D.1}

\noindent\textbf{Data generation.} We give the full covariate-generation mechanisms for the three examples in Section~\ref{section 5} of the main text.

\begin{figure}[ht]
    \centering
    \subfloat[Distribution of $\eta(\bm{X})$]{
    \includegraphics[width=0.3\linewidth]{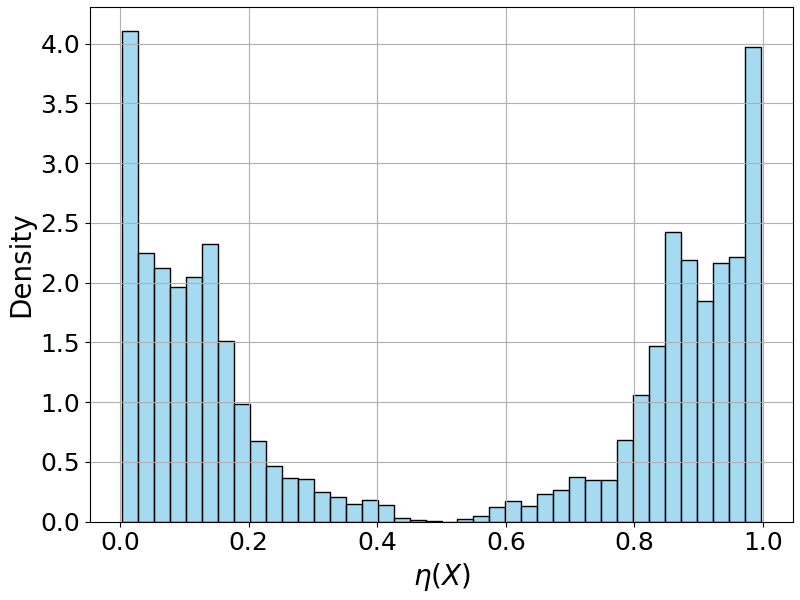}
    \label{Fig: eta}
    }
    \hfill
    \subfloat[Dataset 1 (Class-dependent High noise)]{
    \includegraphics[width=0.65\linewidth]{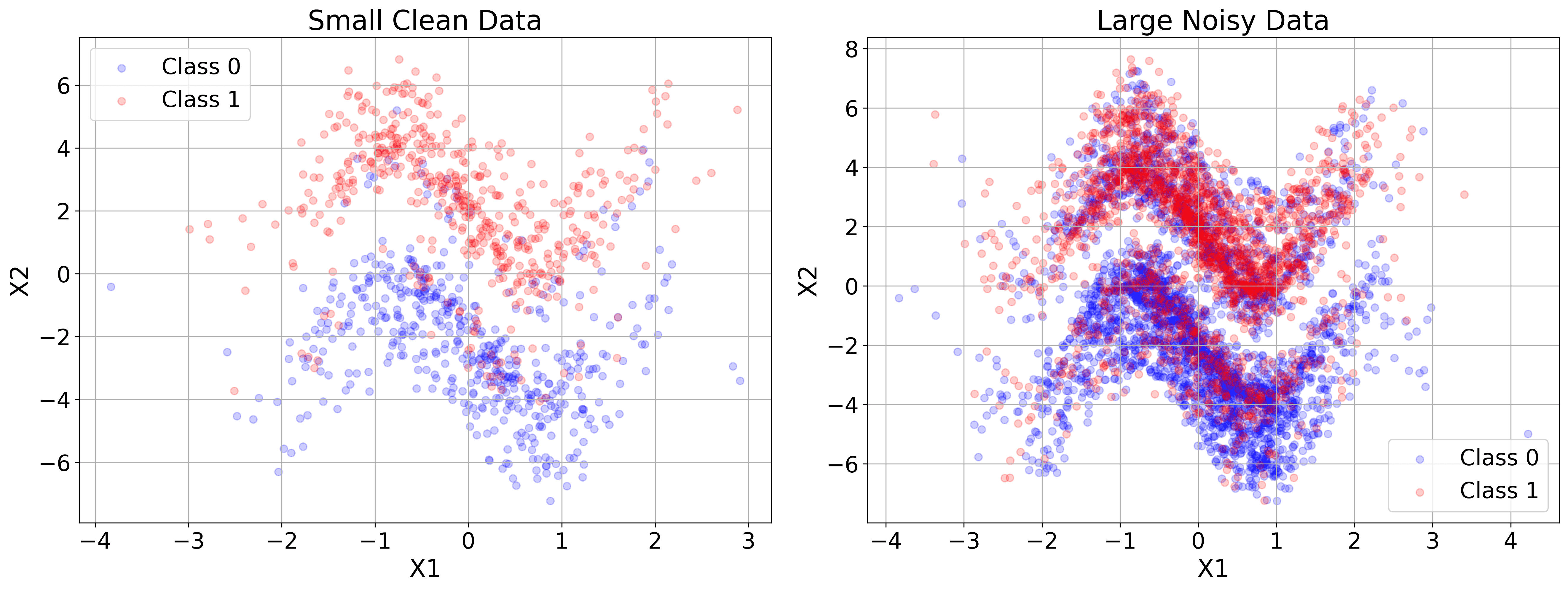}
    \label{Fig: dataset1}
    }
    \caption{A simple visualization of data generation.}
\end{figure}

\noindent\textbf{Example 1.} (Low-dimensional model) Let $\bm{X} = (X_1, X_2,\ldots,X_{10}) \in \mathbb{R}^{10}$. We generate samples from the following distribution:
\begin{equation*}
    \begin{aligned}
        & \eta(\bm{X})=\sigma\left(2 \sin \left(2 X_1\right)+X_2\right),\quad X_1 \sim \mathcal{N}(0,1), \quad X_2|X_1 \sim -2\sin(2X_1) + U + V + W, \\
        & \text{where}\quad U\sim F(u)=(8+u^3)\mathbb{I}(-2<u<2)/16,\quad Z\sim \operatorname{Bernoulli(1/2)},\\
        & W|U,V\sim\mathcal{U}(0,1)\mathbb{I}(|U|+V>1,Z=1),\quad V|U,Z \sim \mathcal{U}(0,3)\mathbb{I}(|U|>1/2, Z=1),
    \end{aligned}
\end{equation*}
where $\eta(\bm{X})$ satisfies Condition~\ref{Con 2}, yielding the distribution shown in Figure~\ref{Fig: eta}. The components $X_3,\ldots,X_{10}$ are sampled from $\mathcal{N}(0,1)$. 

\noindent{\bf Example 2.}  (High-dimensional sparse model) The generation process of Example 1 remains fundamentally consistent, with the modification that the dimensionality of $\bm{X}$ is extended to 100 , i.e., $\bm{X}=\left(X_1, X_2, \ldots, X_{100}\right) \in \mathbb{R}^{100}$. The components $X_3,\ldots,X_{100}$ are independently sampled from $\mathcal{N}(0,1)$.

\noindent{\bf Example 3.} (High-dimensional model) Let $\bm{X} = (X_1, X_2, \ldots, X_{100}) \in \mathbb{R}^{100}$. We generate samples from the following distribution:
\begin{equation*}
    \begin{aligned}
        & \eta(\bm{X})=\sigma\left(6(\|\bm{X}\|_2-1)\right),\quad \bm{X}=(\bm{\Theta}R)/\|\bm{\Theta}\|_2,\text{where} \quad\bm{\Theta}\sim\mathcal{N}(\bm{0},\bm{I}_{100}),\quad \\
        &  R = (6 + U + V + W)/6, \text{ where } U,V,W \text{ as defined in Example 1}.
    \end{aligned}
\end{equation*}

Figures~\ref{Fig: eta} and~\ref{Fig: dataset1} illustrate the simulated data, and Table~\ref{tab:result_Simulation_class_all} reports the complete results under class-dependent noise.
\noindent\textbf{Neural-network implementation.} The neural networks used in our simulation studies were fully-connected feed-forward networks with ReLU activation functions in the hidden layers. The output layer for probability estimation tasks used a Sigmoid activation to ensure the output is in $[0,1]$, while the final classifier's output layer had no activation. The specific architectures were tailored to the complexity of each example. (1) For \textbf{Examples 1 and 2}, which are based on low-dimensional manifolds, we employed a network with $L_n=2$ hidden layers, each containing $N_n=64$ neurons. (2) For \textbf{Example 3}, which involves a higher-dimensional feature space, we utilized a more complex architecture with $L_n=3$ hidden layers, each containing $N_n=256$ neurons.

The optimization process for all methods was conducted using the Adam optimizer with an initial learning rate of $10^{-4}$. Each trial was trained for a total of 100 epochs. An exception was made for the NCR baseline, which, following its original implementation, was trained for 200 epochs, including a 100-epoch warm-up phase. The final classification models for all methods were trained using the sigmoid loss function.
\begin{table}[ht]
    \centering
    \caption{\textbf{Performance comparison (Class-dependent Noise)}: Reported values represent the Mean(Standard deviation) across 500 independent experimental trials. All best results {\it (excluding the oracle)} are highlighted in {\bf bold}.}
    \renewcommand{\arraystretch}{1.4} 
    \begin{adjustbox}{width=1.0\linewidth}{
    \begin{tabular}{l|lcccccc}
        \toprule  
        \multirow{2}{*}{Example} & Noise type & \multicolumn{3}{c}{Low} & \multicolumn{3}{c}{High} \\ \cline{3-8} 
        & Metric & $\widehat{\mathcal{E}}^{0/1}_D(f, h_D^o)$ (↓) & AUC (↑) & $F_1$ (↑) & $\widehat{\mathcal{E}}^{0/1}_D(f, h_D^o)$  (↓) & AUC (↑) & $F_1$ (↑)\\
        \hline
        \multirow{9}{*}{Example 1} 
        & Vanilla-$D$  & 0.1383(0.0713) & 0.8481(0.0505) & 0.7403(0.1116) & 0.1383(0.0713) & 0.8481(0.0505) & 0.7403(0.1116)\\ 
        & Vanilla-$D_\rho$  & 0.0749(0.0068) & 0.9047(0.0023) & 0.7905(0.0104)  & 0.0920(0.0099) & 0.9033(0.0024) & 0.7641(0.0157)\\ 
        & Co-teaching & 0.0598(0.0068) & 0.9042(0.0024) & 0.8142(0.0104) & 0.0674(0.0091) & 0.9023(0.0027) & 0.8034(0.0141)\\
        & NCR & 0.0714(0.0063) & 0.9061(0.0022) & 0.7955(0.0095)  & 0.0881(0.0093) & 0.9055(0.0023) & 0.7696(0.0145)\\
        & ALC(10\%) & 0.0713(0.0063) & 0.9054(0.0022) & 0.7959(0.0097)  & 0.0849(0.0088) & 0.9047(0.0023) & 0.7750(0.0138)\\
        \cline{2-8}
        & \cellcolor{LightPink}Pro-PT & \cellcolor{LightPink}0.0425(0.0040) & \cellcolor{LightPink}0.9061(0.0022) & \cellcolor{LightPink}0.8465(0.0047)  & \cellcolor{LightPink}0.0456(0.0063) & \cellcolor{LightPink}0.9056(0.0023)  & \cellcolor{LightPink}0.8442(0.0066)\\
        & \cellcolor{LightPink}Pro-AT(10\%)  & \cellcolor{LightPink}0.0398(0.0030)  & \cellcolor{LightPink}{\bf 0.9066(0.0021)} & \cellcolor{LightPink}0.8487(0.0038)  & \cellcolor{LightPink}0.0416(0.0044) & \cellcolor{LightPink}0.9063(0.0022)  & \cellcolor{LightPink}0.8487(0.0038)\\
        & \cellcolor{LightPink}Pro-AT  & \cellcolor{LightPink}{\bf 0.0391(0.0023)}  & \cellcolor{LightPink}{\bf 0.9066(0.0021)} & \cellcolor{LightPink}{\bf 0.8488(0.0032)}  & \cellcolor{LightPink}{\bf 0.0392(0.0024)} & \cellcolor{LightPink}{\bf 0.9066(0.0021)}  & \cellcolor{LightPink}{\bf 0.8487(0.0029)}\\
        \cline{2-8}
        & Oracle & 0.0385(0.0021)  & 0.9067(0.0021) & 0.8492(0.0029) & 0.0385(0.0021) & 0.9067(0.0021) & 0.8492(0.0029)\\
        \hline
        \multirow{9}{*}{Example 2} 
        & Vanilla-$D$  & 0.2505(0.0506) & 0.6997(0.0619) & 0.6350(0.0794)  & 0.2505(0.0506) & 0.6997(0.0619) & 0.6350(0.0794)\\ 
        & Vanilla-$D_\rho$  & 0.1115(0.0086) & 0.8842(0.0039) & 0.7393(0.0141) & 0.1349(0.0108) & 0.8754(0.0053) & 0.7012(0.0188)\\ 
        & Co-teaching & 0.0997(0.0075) & 0.8854(0.0038) & 0.7599(0.0121) & 0.1164(0.0090) & 0.8777(0.0048) & 0.7350(0.0151)\\
        & NCR & 0.1090(0.0083) & 0.8841(0.0040) & 0.7441(0.0133) & 0.1333(0.0105) & 0.8741(0.0055) & 0.7051(0.0179)\\
        & ALC(10\%) & 0.0938(0.0067)  & 0.8908(0.0030) & 0.7665(0.0106) & 0.1131(0.0088) & 0.8872(0.0034) & 0.7354(0.0147)\\
        \cline{2-8}
        & \cellcolor{LightPink}Pro-PT  & \cellcolor{LightPink}0.0604(0.0041)  & \cellcolor{LightPink}0.8959(0.0027) & \cellcolor{LightPink}0.8247(0.0077) & \cellcolor{LightPink}0.0667(0.0052) & \cellcolor{LightPink}0.8923(0.0031)  & \cellcolor{LightPink}0.8193(0.0098)\\
        & \cellcolor{LightPink}Pro-AT(10\%)  & \cellcolor{LightPink}0.0563(0.0035) & \cellcolor{LightPink}0.8981(0.0026) & \cellcolor{LightPink}0.8294(0.0061)  & \cellcolor{LightPink}0.0609(0.0045) & \cellcolor{LightPink}0.8955(0.0028)  & \cellcolor{LightPink}0.8253(0.0078)\\
        & \cellcolor{LightPink}Pro-AT  & \cellcolor{LightPink}{\bf 0.0533(0.0033)}  & \cellcolor{LightPink}{\bf 0.8992(0.0025)} & \cellcolor{LightPink}{\bf 0.8333(0.0045)}  & \cellcolor{LightPink}{\bf 0.0542(0.0035)} & \cellcolor{LightPink}{\bf 0.8987(0.0026)}  & \cellcolor{LightPink}{\bf 0.8332(0.0047)}\\       
        \cline{2-8}
        & Oracle & 0.0541(0.0027)  & 0.8985(0.0025) & 0.8332(0.0037) & 0.0541(0.0027) & 0.8985(0.0025) & 0.8332(0.0037)\\
        \hline
        \multirow{9}{*}{Example 3} 
        & Vanilla-$D$  &  0.2109(0.1084) & 0.8995(0.0190)  & 0.7475(0.0652) & 0.2109(0.1084) & 0.8995(0.0190) & 0.7475(0.0652)\\ 
        & Vanilla-$D_\rho$  & 0.1538(0.0235) & 0.9161(0.0040) & 0.6551(0.0441)  & 0.2611(0.0402) & 0.8980(0.0096) & 0.4133(0.1088)\\ 
        & Co-teaching & 0.1249(0.0234) & 0.9179(0.0044) & 0.7068(0.0402)  & 0.1913(0.0389) & 0.9030(0.0128) & 0.5794(0.0815)\\
        & NCR & 0.1022(0.0116) & 0.9194(0.0029) & 0.7454(0.0186)  & 0.1664(0.0452) & 0.9014(0.0444) & 0.6267(0.1153)\\
        & ALC(10\%) & 0.1273(0.0161)  & 0.9194(0.0029) & 0.7031(0.0279) & 0.1932(0.0215) & 0.9170(0.0041) & 0.5764(0.0458)\\
        \cline{2-8}
        & \cellcolor{LightPink}Pro-PT  & \cellcolor{LightPink}0.0458(0.0228)  & \cellcolor{LightPink}0.9170(0.0060) & \cellcolor{LightPink}0.8297(0.0321)  & \cellcolor{LightPink}0.1014(0.0505)  & \cellcolor{LightPink}0.8867(0.0223)  & \cellcolor{LightPink}0.7612(0.0745)\\
        & \cellcolor{LightPink}Pro-AT(10\%)  & \cellcolor{LightPink}0.0248(0.0144)  & \cellcolor{LightPink}0.9234(0.0035) & \cellcolor{LightPink}0.8567(0.0193)  & \cellcolor{LightPink}0.0646(0.0418) & \cellcolor{LightPink}0.9071(0.0150)  & \cellcolor{LightPink}0.8079(0.0582)\\
        & \cellcolor{LightPink}Pro-AT & \cellcolor{LightPink}{\bf 0.0080(0.0048)} & \cellcolor{LightPink}{\bf 0.9273(0.0024)} & \cellcolor{LightPink}{\bf 0.8783(0.0066)}  & \cellcolor{LightPink}{\bf 0.0062(0.0028)} & \cellcolor{LightPink}{\bf 0.9278(0.0023)}  & \cellcolor{LightPink}{\bf 0.8807(0.0039)}\\
        \cline{2-8}
        & Oracle & 0.0054(0.0019)  & 0.9282(0.0021) & 0.8815(0.0030)  & 0.0054(0.0019) & 0.9282(0.0021) & 0.8815(0.0030)\\
        \bottomrule
    \end{tabular}
    }\end{adjustbox} 
    \label{tab:result_Simulation_class_all}
\end{table}

\noindent\textbf{Alternative learners.} To further examine the model-agnostic nature of the proposed framework, we conduct an additional backend experiment on Example 2 under class-dependent label noise. In this experiment, both the estimation of $\eta_\rho(\bm{x})$ and the final classifier are implemented using the same non-neural backend: B-spline, random forest, or gradient boosting. The complete results are summarized in Table~\ref{tab:result_Simulation_backend_dataset2}.
\begin{table}[ht]
    \centering
    \caption{\textbf{Performance comparison with different classifier backends (Example 2, class-dependent noise)}: Reported values represent the Mean(Standard deviation) across 500 independent experimental trials. All best results {\it (excluding the oracle)} are highlighted in {\bf bold}.}
    \renewcommand{\arraystretch}{1.25}
    \begin{adjustbox}{width=1.0\linewidth}{
    \begin{tabular}{l|lcccccc}
        \toprule
        \multirow{2}{*}{Backend} & \multirow{2}{*}{Method} & \multicolumn{3}{c}{Low} & \multicolumn{3}{c}{High} \\ \cline{3-8}
        & & $\widehat{\mathcal{E}}^{0/1}_D(f, h_D^o)$ ($\downarrow$) & AUC ($\uparrow$) & $F_1$ ($\uparrow$) & $\widehat{\mathcal{E}}^{0/1}_D(f, h_D^o)$ ($\downarrow$) & AUC ($\uparrow$) & $F_1$ ($\uparrow$)\\
        \hline
        \multirow{9}{*}{B-Spline} & Vanilla-$D$ & 0.0151(0.0062) & 0.9156(0.0056) & 0.8724(0.0068) & 0.0151(0.0062) & 0.9156(0.0056) & 0.8724(0.0068)\\
        & Vanilla-$D_\rho$ & 0.0139(0.0050) & 0.9220(0.0033) & 0.8698(0.0067) & 0.0269(0.0092) & 0.9175(0.0038) & 0.8532(0.0122)\\
        & Co-teaching & 0.0218(0.0050) & 0.9217(0.0040) & 0.8598(0.0067) & 0.0425(0.0083) & 0.9189(0.0042) & 0.8330(0.0113)\\
        & NCR & 0.0142(0.0050) & 0.9246(0.0029) & 0.8692(0.0067) & 0.0258(0.0086) & 0.9211(0.0034) & 0.8544(0.0115)\\
        & ALC(10\%) & 0.0232(0.0072) & 0.9236(0.0030) & 0.8582(0.0093) & 0.0373(0.0112) & 0.9203(0.0035) & 0.8403(0.0149)\\
        \cline{2-8}
        & \cellcolor{LightPink}Pro-PT & \cellcolor{LightPink}0.0036(0.0017) & \cellcolor{LightPink}0.9210(0.0046) & \cellcolor{LightPink}0.8837(0.0031) & \cellcolor{LightPink}0.0060(0.0026) & \cellcolor{LightPink}0.9194(0.0048) & \cellcolor{LightPink}0.8813(0.0037)\\
        & \cellcolor{LightPink}Pro-AT(10\%) & \cellcolor{LightPink}0.0030(0.0012) & \cellcolor{LightPink}{\bf 0.9288(0.0026)} & \cellcolor{LightPink}0.8846(0.0026) & \cellcolor{LightPink}0.0036(0.0015) & \cellcolor{LightPink}{\bf 0.9269(0.0029)} & \cellcolor{LightPink}0.8839(0.0028)\\
        & \cellcolor{LightPink}Pro-AT & \cellcolor{LightPink}{\bf 0.0028(0.0012)} & \cellcolor{LightPink}0.9248(0.0055) & \cellcolor{LightPink}{\bf 0.8847(0.0027)} & \cellcolor{LightPink}{\bf 0.0035(0.0015)} & \cellcolor{LightPink}0.9253(0.0049) & \cellcolor{LightPink}{\bf 0.8839(0.0028)}\\
        \cline{2-8}
        & Oracle & 0.0011(0.0007) & 0.9325(0.0020) & 0.8864(0.0025) & 0.0011(0.0007) & 0.9325(0.0020) & 0.8864(0.0025)\\
        \hline
        \multirow{9}{*}{Random Forest} & Vanilla-$D$ & 0.0511(0.0033) & 0.8800(0.0047) & {\bf 0.8365(0.0059)} & {\bf 0.0511(0.0033)} & 0.8800(0.0047) & {\bf 0.8365(0.0059)}\\
        & Vanilla-$D_\rho$ & 0.0836(0.0127) & 0.8868(0.0034) & 0.7804(0.0194) & 0.1379(0.0235) & 0.8812(0.0039) & 0.6931(0.0414)\\
        & Co-teaching & 0.0609(0.0082) & 0.8690(0.0071) & 0.8122(0.0125) & 0.1167(0.0411) & 0.8674(0.0074) & 0.7228(0.0728)\\
        & NCR & 0.1113(0.0130) & 0.8751(0.0032) & 0.7403(0.0207) & 0.1584(0.0191) & 0.8670(0.0036) & 0.6606(0.0350)\\
        & ALC(10\%) & 0.0882(0.0146) & 0.8887(0.0032) & 0.7734(0.0225) & 0.1519(0.0261) & 0.8840(0.0036) & 0.6680(0.0480)\\
        \cline{2-8}
        & \cellcolor{LightPink}Pro-PT & \cellcolor{LightPink}0.0808(0.0135) & \cellcolor{LightPink}0.8869(0.0034) & \cellcolor{LightPink}0.7846(0.0203) & \cellcolor{LightPink}0.1335(0.0254) & \cellcolor{LightPink}0.8814(0.0040) & \cellcolor{LightPink}0.7006(0.0439)\\
        & \cellcolor{LightPink}Pro-AT(10\%) & \cellcolor{LightPink}0.0514(0.0046) & \cellcolor{LightPink}0.8900(0.0030) & \cellcolor{LightPink}0.8278(0.0074) & \cellcolor{LightPink}0.0531(0.0050) & \cellcolor{LightPink}0.8851(0.0038) & \cellcolor{LightPink}0.8256(0.0083)\\
        & \cellcolor{LightPink}Pro-AT & \cellcolor{LightPink}{\bf 0.0503(0.0055)} & \cellcolor{LightPink}{\bf 0.8904(0.0031)} & \cellcolor{LightPink}0.8302(0.0089) & \cellcolor{LightPink}0.0534(0.0103) & \cellcolor{LightPink}{\bf 0.8860(0.0039)} & \cellcolor{LightPink}0.8259(0.0161)\\
        \cline{2-8}
        & Oracle & 0.0399(0.0035) & 0.8985(0.0026) & 0.8476(0.0044) & 0.0399(0.0035) & 0.8985(0.0026) & 0.8476(0.0044)\\
        \hline
        \multirow{9}{*}{\makecell{Gradient\\Boosting}} & Vanilla-$D$ & 0.0334(0.0085) & 0.9001(0.0061) & 0.8540(0.0090) & 0.0334(0.0085) & 0.9001(0.0061) & 0.8540(0.0090)\\
        & Vanilla-$D_\rho$ & 0.0382(0.0067) & 0.9105(0.0032) & 0.8401(0.0090) & 0.0678(0.0112) & 0.9028(0.0039) & 0.8006(0.0160)\\
        & Co-teaching & 0.0270(0.0062) & 0.9079(0.0039) & 0.8544(0.0083) & 0.0457(0.0098) & 0.9025(0.0049) & 0.8302(0.0137)\\
        & NCR & 0.0354(0.0067) & 0.9115(0.0030) & 0.8435(0.0089) & 0.0611(0.0107) & 0.9054(0.0035) & 0.8093(0.0152)\\
        & ALC(10\%) & 0.0323(0.0058) & 0.9161(0.0030) & 0.8473(0.0078) & 0.0584(0.0098) & 0.9109(0.0034) & 0.8127(0.0140)\\
        \cline{2-8}
        & \cellcolor{LightPink}Pro-PT & \cellcolor{LightPink}0.0155(0.0040) & \cellcolor{LightPink}0.9122(0.0036) & \cellcolor{LightPink}0.8715(0.0047) & \cellcolor{LightPink}0.0207(0.0048) & \cellcolor{LightPink}0.9080(0.0042) & \cellcolor{LightPink}0.8662(0.0058)\\
        & \cellcolor{LightPink}Pro-AT(10\%) & \cellcolor{LightPink}0.0097(0.0028) & \cellcolor{LightPink}0.9179(0.0032) & \cellcolor{LightPink}0.8775(0.0037) & \cellcolor{LightPink}0.0143(0.0040) & \cellcolor{LightPink}0.9145(0.0036) & \cellcolor{LightPink}0.8728(0.0048)\\
        & \cellcolor{LightPink}Pro-AT & \cellcolor{LightPink}{\bf 0.0077(0.0028)} & \cellcolor{LightPink}{\bf 0.9208(0.0042)} & \cellcolor{LightPink}{\bf 0.8797(0.0037)} & \cellcolor{LightPink}{\bf 0.0076(0.0032)} & \cellcolor{LightPink}{\bf 0.9225(0.0041)} & \cellcolor{LightPink}{\bf 0.8798(0.0041)}\\
        \cline{2-8}
        & Oracle & 0.0046(0.0014) & 0.9269(0.0022) & 0.8830(0.0027) & 0.0046(0.0014) & 0.9269(0.0022) & 0.8830(0.0027)\\
        \bottomrule
    \end{tabular}
    }\end{adjustbox}
    \label{tab:result_Simulation_backend_dataset2}
\end{table}

\noindent\textbf{Instance-dependent noise.} The instance-dependent noise mechanism used in Section~\ref{section 5} of the main text is specified by
\begin{equation*}
    \begin{aligned}
        \rho_{+}(\bm{x})= \begin{cases}\frac{1}{2} & \text { if } \eta(\bm{x}) \leq \frac{1}{2}, \\ \frac{1}{2}-\frac{1}{4}s\left(4\eta(\bm{x})-2\right) & \text { if } \frac{1}{2}<\eta(\bm{x}) \leq \frac{3}{4}, \\ \frac{1}{4} & \text { if } \eta(\bm{x})>\frac{3}{4},\end{cases} \\
        \rho_{-}(\bm{x})= \begin{cases}\frac{1}{8} & \text { if } \eta(\bm{x}) \leq \frac{1}{4}, \\ \frac{1}{8}+\frac{1}{8}s\left(4\eta(\bm{x})-1\right) & \text { if } \frac{1}{4}<\eta(\bm{x}) \leq \frac{1}{2}, \\ \frac{1}{4}& \text { if } \eta(\bm{x})>\frac{1}{2},\end{cases}
    \end{aligned}
\end{equation*}
where $s(x)=-20x^7 + 70x^6 - 84x^5 + 35x^4$ is a smooth, sigmoid-shaped polynomial function that maps the unit interval $[0,1]$ onto itself. This setup generates a challenging noise environment where the corruption is intrinsically linked to the underlying data distribution. For this setting, recognizing its complexity, we consider larger clean sample sizes of $n_0=1000$ and $n_0=2000$ to better estimate the underlying noise functions. The complete results are reported in Table~\ref{tab:result_Simulation_instance_all}.
\begin{table}[ht]
    \centering
    \caption{\textbf{Performance comparison (Instance-dependent Noise)}: Reported values represent the Mean(Standard deviation) across 500 independent experimental trials. All best results {\it (excluding the oracle)} are highlighted in {\bf bold}.}
    \renewcommand{\arraystretch}{1.4} 
    \begin{adjustbox}{width=1.0\linewidth}{
    \begin{tabular}{l|lcccccc}
        \toprule  
        \multirow{2}{*}{Example} & $n_0$ & \multicolumn{3}{c}{1000} & \multicolumn{3}{c}{2000} \\ \cline{3-8} 
        &Metric & $\widehat{\mathcal{E}}^{0/1}_D(f, h_D^o)$ (↓) & AUC (↑) & $F_1$ (↑) & $\widehat{\mathcal{E}}^{0/1}_D(f, h_D^o)$ (↓) & AUC (↑) & $F_1$ (↑)\\
        \hline
        \multirow{9}{*}{Example 1} 
        & Vanilla-$D$ & 0.0669(0.0171) & 0.8921(0.0081) & 0.8200(0.0209) & 0.0473(0.0038) & 0.9028(0.0027) & 0.8401(0.0047)\\ 
        & Vanilla-$D_\rho$ & 0.0568(0.0049) & 0.9053(0.0022) & 0.8182(0.0073) & 0.0568(0.0049) & 0.9053(0.0022) & 0.8182(0.0073)\\ 
        & Co-teaching & 0.0475(0.0045) & 0.9050(0.0022) & 0.8332(0.0072) & 0.0475(0.0045) & 0.9050(0.0022) & 0.8332(0.0072)\\
        & NCR & 0.0542(0.0046) & {\bf 0.9065(0.0021)} & 0.8217(0.0069) & 0.0542(0.0046) & {\bf 0.9065(0.0021)} & 0.8217(0.0069)\\
        & ALC(10\%) & 0.0546(0.0046) & 0.9059(0.0021) & 0.8215(0.0070) & 0.0546(0.0046) & 0.9059(0.0021) & 0.8215(0.0070)\\
        \cline{2-8}
        & \cellcolor{LightPink}Pro-PT& \cellcolor{LightPink}0.0602(0.0229) & \cellcolor{LightPink}0.9031(0.0035) & \cellcolor{LightPink}0.8155(0.0383) & \cellcolor{LightPink}0.0497(0.0105) & \cellcolor{LightPink}0.9041(0.0028) & \cellcolor{LightPink}0.8324(0.0169)\\
        & \cellcolor{LightPink}Pro-AT(10\%) & \cellcolor{LightPink}0.0507(0.0144) & \cellcolor{LightPink}0.9048(0.0028) & \cellcolor{LightPink}0.8292(0.0231) & \cellcolor{LightPink}0.0440(0.0074) & \cellcolor{LightPink}0.9055(0.0023) & \cellcolor{LightPink}0.8400(0.0121)\\
        & \cellcolor{LightPink}Pro-AT& \cellcolor{LightPink}{\bf 0.0460(0.0092)} & \cellcolor{LightPink}0.9056(0.0024) & \cellcolor{LightPink}{\bf 0.8361(0.0148)} & \cellcolor{LightPink}{\bf 0.0418(0.0050)} & \cellcolor{LightPink}0.9060(0.0022) & \cellcolor{LightPink}{\bf 0.8432(0.0082)}\\
        \cline{2-8}
        & Oracle & 0.0385(0.0021) & 0.9067(0.0021) & 0.8492(0.0029) & 0.0385(0.0021) & 0.9067(0.0021) & 0.8492(0.0029)\\
        \hline
        \multirow{9}{*}{Example 2} 
        & Vanilla-$D$ & 0.1563(0.0313) & 0.8059(0.0327) & 0.7294(0.0398) & 0.0859(0.0083) & 0.8744(0.0073) & 0.8016(0.0089)\\ 
        & Vanilla-$D_\rho$ & 0.0873(0.0061) & 0.8871(0.0035) & 0.7799(0.0094) & 0.0873(0.0061) & 0.8871(0.0035) & 0.7799(0.0094)\\ 
        & Co-teaching & 0.0813(0.0057) & 0.8880(0.0034) & 0.7903(0.0087) & 0.0813(0.0057) & 0.8880(0.0034) & 0.7903(0.0087)\\
        & NCR & 0.0847(0.0060) & 0.8876(0.0036) & 0.7838(0.0089) & 0.0847(0.0060) & 0.8876(0.0036) & 0.7838(0.0089)\\
        & ALC(10\%) & 0.0750(0.0047) & 0.8935(0.0028) & 0.7966(0.0072) & 0.0750(0.0047) & 0.8935(0.0028) & 0.7966(0.0072)\\
        \cline{2-8}
        & \cellcolor{LightPink}Pro-PT& \cellcolor{LightPink}0.0826(0.0226) & \cellcolor{LightPink}0.8873(0.0058) & \cellcolor{LightPink}0.7925(0.0398) & \cellcolor{LightPink}0.0720(0.0114) & \cellcolor{LightPink}0.8893(0.0045) & \cellcolor{LightPink}0.8097(0.0199)\\
        & \cellcolor{LightPink}Pro-AT(10\%) & \cellcolor{LightPink}0.0742(0.0193) & \cellcolor{LightPink}0.8918(0.0044) & \cellcolor{LightPink}0.8035(0.0341) & \cellcolor{LightPink}0.0644(0.0077) & \cellcolor{LightPink}0.8939(0.0037) & \cellcolor{LightPink}0.8189(0.0129)\\
        & \cellcolor{LightPink}Pro-AT& \cellcolor{LightPink}{\bf 0.0665(0.0146)} & \cellcolor{LightPink}{\bf 0.8947(0.0042)} & \cellcolor{LightPink}{\bf 0.8138(0.0247)} & \cellcolor{LightPink}{\bf 0.0599(0.0064)} & \cellcolor{LightPink}{\bf 0.8984(0.0029)} & \cellcolor{LightPink}{\bf 0.8244(0.0104)}\\ 
        \cline{2-8}
        & Oracle & 0.0541(0.0027) & 0.8985(0.0025) & 0.8332(0.0037) & 0.0541(0.0027) & 0.8985(0.0025) & 0.8332(0.0037)\\
        \hline
        \multirow{9}{*}{Example 3} 
        & Vanilla-$D$ & 0.1280(0.0840) & 0.9226(0.0027) & 0.8032(0.0495) & 0.0299(0.0336) & 0.9258(0.0022) & 0.8658(0.0224)\\ 
        & Vanilla-$D_\rho$ & 0.0791(0.0134) & 0.9206(0.0028) & 0.7808(0.0200) & 0.0791(0.0134) & 0.9206(0.0028) & 0.7808(0.0200)\\ 
        & Co-teaching & 0.0666(0.0115) & 0.9209(0.0042) & 0.7995(0.0163) & 0.0666(0.0115) & 0.9209(0.0042) & 0.7995(0.0163)\\
        & NCR & 0.0592(0.0079) & 0.9213(0.0023) & 0.8101(0.0113) & 0.0592(0.0079) & 0.9213(0.0023) & 0.8101(0.0113)\\
        & ALC(10\%) & 0.0762(0.0104) & 0.9216(0.0025) & 0.7851(0.0154) & 0.0762(0.0104) & 0.9216(0.0025) & 0.7851(0.0154)\\
        \cline{2-8}
        & \cellcolor{LightPink}Pro-PT& \cellcolor{LightPink}0.0747(0.0520) & \cellcolor{LightPink}0.9156(0.0094) & \cellcolor{LightPink}0.7835(0.0799) & \cellcolor{LightPink}0.0281(0.0157) & \cellcolor{LightPink}0.9230(0.0037) & \cellcolor{LightPink}0.8527(0.0210)\\
        & \cellcolor{LightPink}Pro-AT(10\%) & \cellcolor{LightPink}0.0463(0.0387) & \cellcolor{LightPink}0.9215(0.0071) & \cellcolor{LightPink}0.8258(0.0553) & \cellcolor{LightPink}0.0156(0.0096) & \cellcolor{LightPink}0.9264(0.0026) & \cellcolor{LightPink}0.8684(0.0127)\\
        & \cellcolor{LightPink}Pro-AT& \cellcolor{LightPink}{\bf 0.0232(0.0155)} & \cellcolor{LightPink}{\bf 0.9261(0.0025)} & \cellcolor{LightPink}{\bf 0.8583(0.0204)} & \cellcolor{LightPink}{\bf 0.0114(0.0050)} & \cellcolor{LightPink}{\bf 0.9272(0.0021)} & \cellcolor{LightPink}{\bf 0.8739(0.0068)}\\
        \cline{2-8}
        & Oracle & 0.0054(0.0019) & 0.9282(0.0021) & 0.8815(0.0030) & 0.0054(0.0019) & 0.9282(0.0021) & 0.8815(0.0030)\\
        \bottomrule
    \end{tabular}
    }\end{adjustbox}
    \label{tab:result_Simulation_instance_all}
\end{table}

\newpage
\subsection{NoisyCXR Details}\label{Appendix D.2}
\noindent\textbf{Dataset and labels.} Table~\ref{tab:dataset_NoisyCXR} summarizes the label counts for the NoisyCXR dataset used in Section~\ref{section 6} of the main text. Table~\ref{tab:NoisyCXRFig} provides example images for the four combinations of NoisyCXR and RSNA labels. The counts in Table~\ref{tab:dataset_NoisyCXR} indicate substantial class imbalance. More importantly, the construction of this dataset is designed to simulate the label noise found in real-world scenarios. The label conflicts arise from disagreements between different annotation sources (NIH and RSNA), creating a complex noise structure.
\begin{table}[ht]
    \centering
    \caption{The disagreements between two label sets are highlighted in bold font. Fields in italic indicate an unclear agreement between both sets of labels. NoisyCXR labels are collected from the original NIH labels where possible, and the remainder are uniformly sampled (10\%) from the ``Consolidation/infiltration" category to increase the noise rate further in the experiments as the work of Bernhardt et al.~\cite{bernhardt2022active}.}
    \begin{tabular}{lrr}
        \toprule & \multicolumn{2}{c}{ RSNA labels} \\
        \cline { 2 - 3 } & \begin{tabular}{r} 
        Pneumonia-like \\
        opacity
        \end{tabular} & \begin{tabular}{r} 
        No pneumonia-like \\
        opacity
        \end{tabular} \\
        \midrule NIH labels : & & \\
        \quad Pneumonia & 367 & {\bf 441} \\
        \begin{tabular}{l} 
        \quad Consolidation/infiltration \\
        \quad (not pneumonia)
        \end{tabular} & 3988 & 8551 \\
        \quad Other diseases only & {\bf 1101} & 5567 \\
        \quad No finding & {\bf 556} & 6,113 \\
        \midrule NoisyCXR labels: & & \\
        \quad Pneumonia-like opacity & 3,956 & {\bf 1296} \\
        \quad No pneumonia-like opacity &  {\bf 2056} & 19376 \\
        \midrule Total & 6012 & 20672 \\
        \bottomrule
    \end{tabular}
    \label{tab:dataset_NoisyCXR}
\end{table}

\noindent\textbf{Implementation.} Motivated by Li et al.~\cite{li2024GMM} and Yu et al.~\cite{yu2025spatialGMM}, we start with a well-pretrained deep learning model, which is capable of leveraging a sufficient amount of information from images without noisy labels or even true labels. In this regard, various self-supervised deep learning models can be considered. Among those choices, the EVA-X model of Yao et al.~\cite{yao2024eva} demonstrates the state-of-the-art (SOTA) performance in terms of predictive accuracy. Therefore, we are motivated to apply this pretrained SOTA model to the Chest X-ray images, so that each image can be converted into a feature vector. Subsequently, we employ a neural network architecture denoted as $\mathcal{F}(L_n=2, N_n=512)$ as the fine-tuning classifier head. The model is optimized using Adam with an initial learning rate of 0.0001 in 100 epochs.

\begin{table}[ht]
    \centering
    \caption{An Overview of NoisyCXR and RSNA Annotation.}
    \resizebox{0.8\textwidth}{!}{  
        \begin{tabular}{ccc}
            \toprule
             & 
            \multicolumn{2}{c}{RSNA Labels} \\
            \cmidrule(lr){2-3}
            & Pneumonia-like & No pneumonia-like \\
            \midrule
            NoisyCXR Labels & & \\
            \cmidrule(lr){1-1}
            \centering{Pneumonia-like}
            & \includegraphics[width=0.3\linewidth]{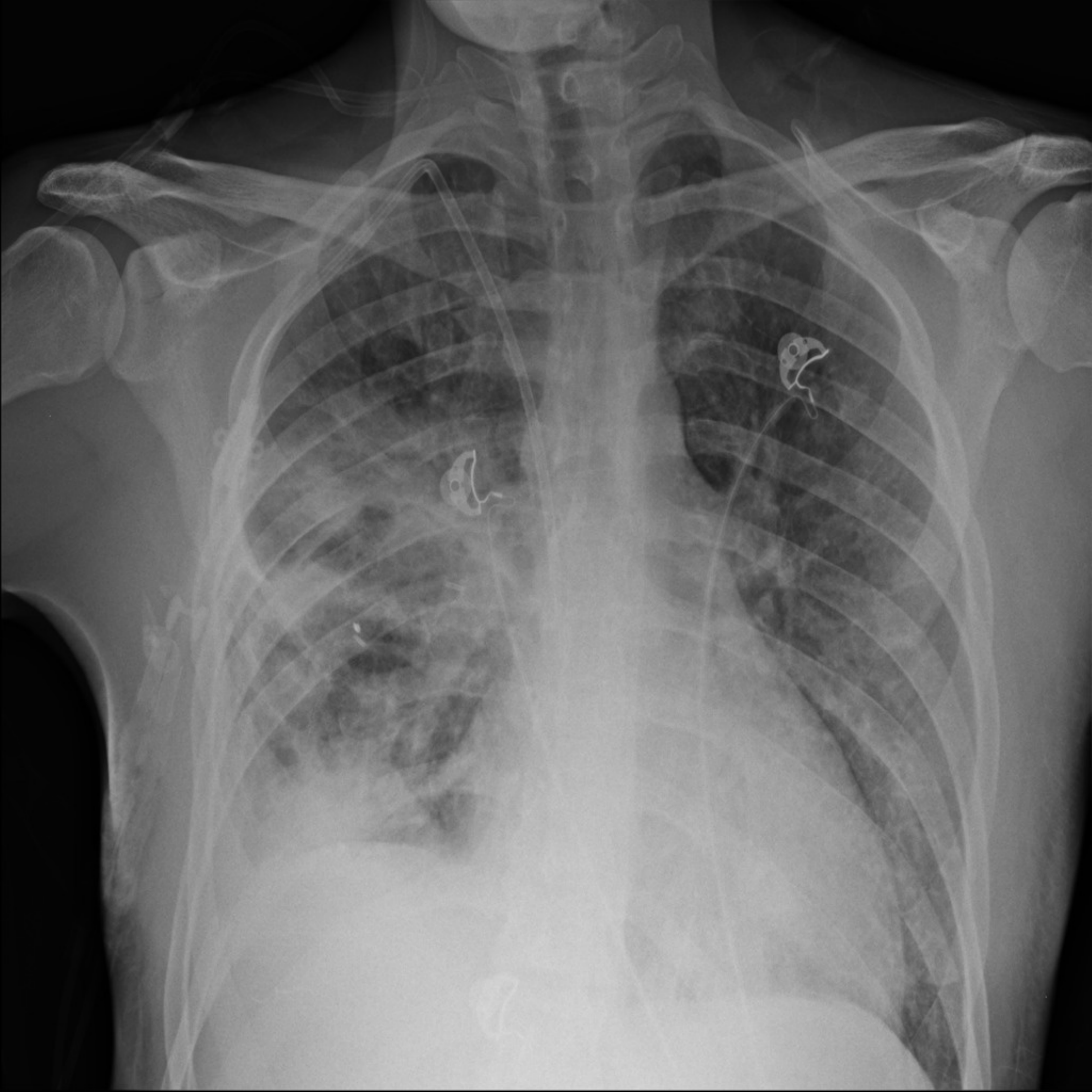} 
            & \includegraphics[width=0.3\linewidth]{pics/kaggle_0_nih_1.pdf} \\
            \midrule
            \centering{No pneumonia-like}
            & \includegraphics[width=0.3\linewidth]{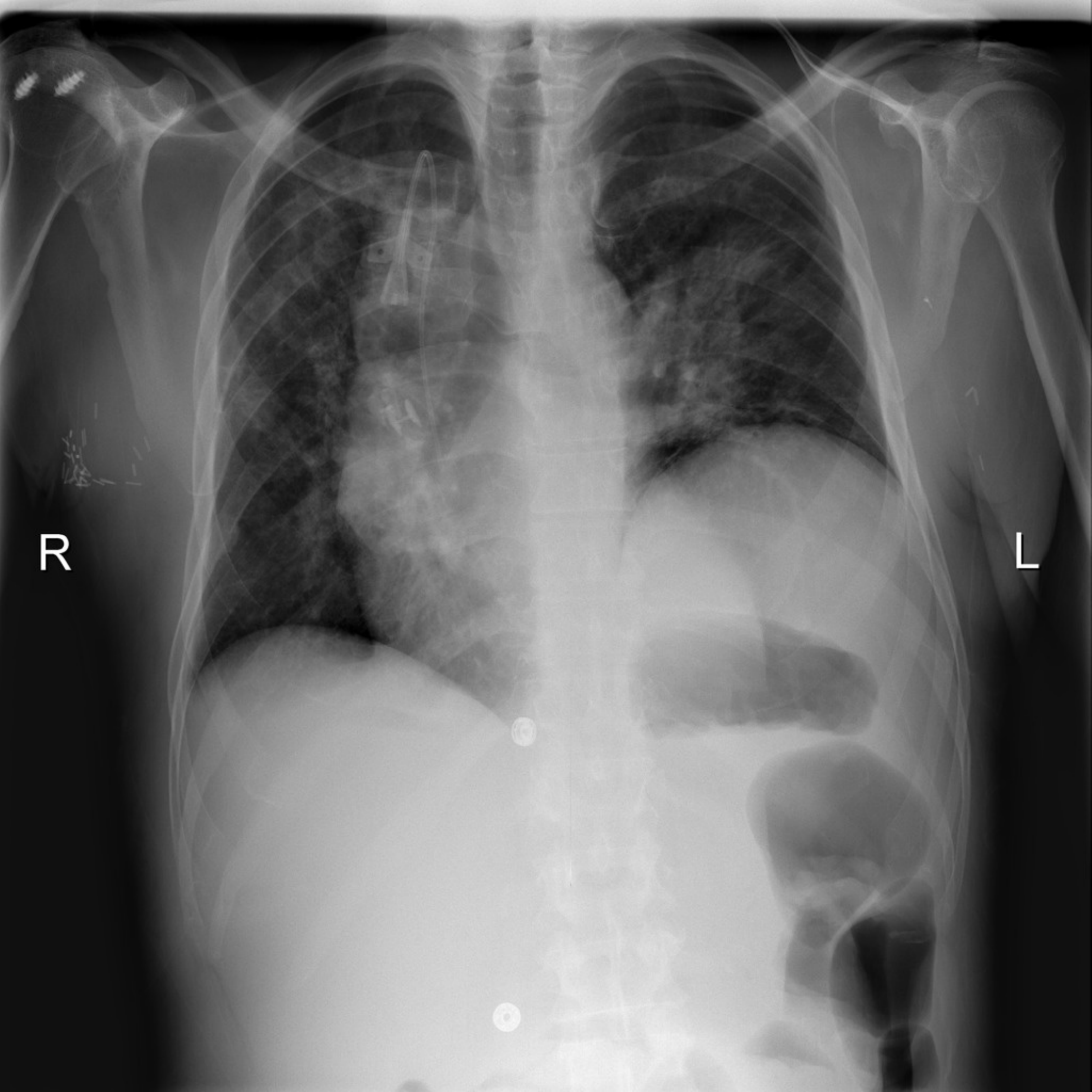} 
            & \includegraphics[width=0.3\linewidth]{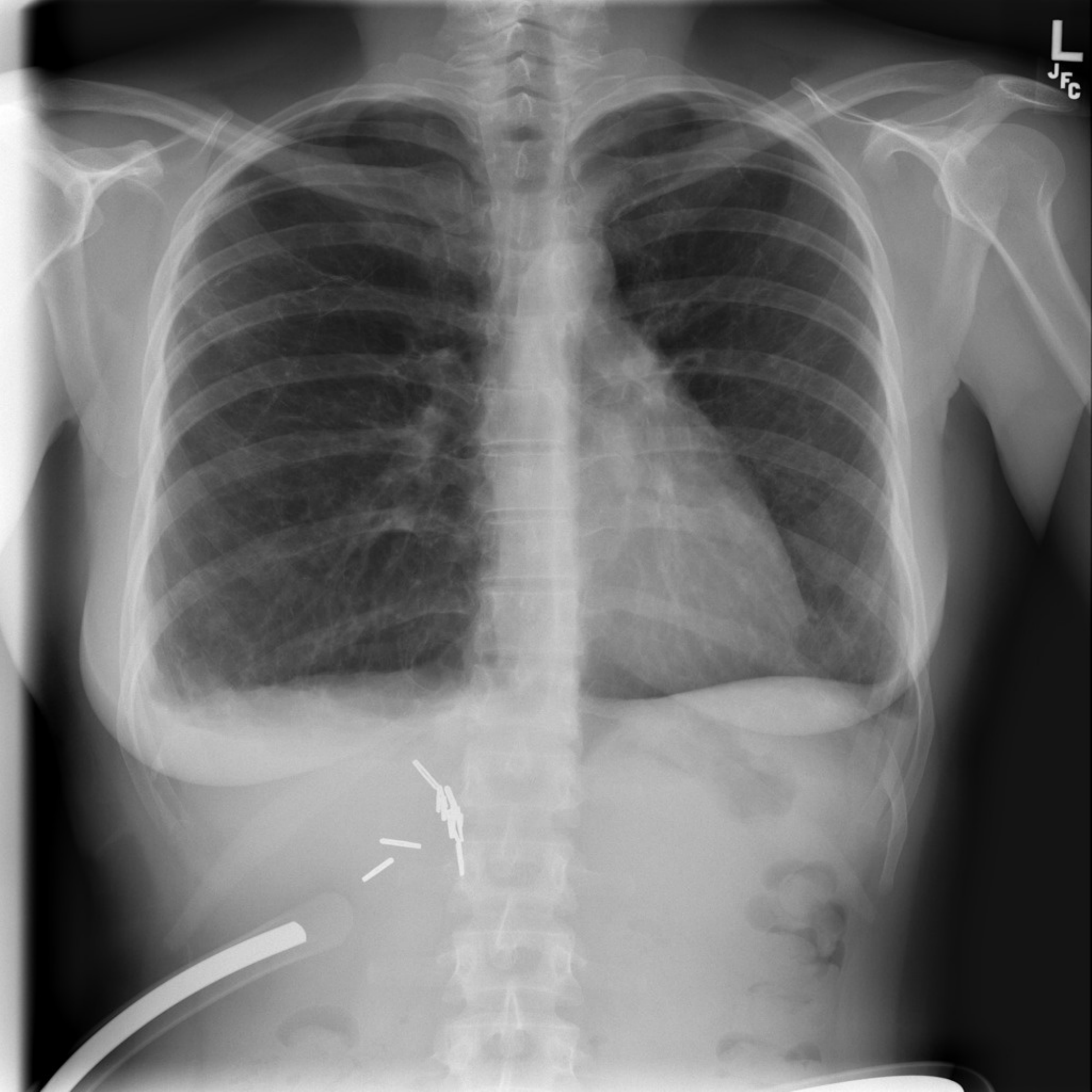} \\
            \bottomrule
        \end{tabular}
    }
    \label{tab:NoisyCXRFig}
\end{table}

\noindent\textbf{Additional results.} Table~\ref{tab:result_NoisyCXR_ind} reports the NoisyCXR results obtained with instance-dependent noise estimation under Condition~\ref{Con 1}(b), complementing the class-dependent specification in Table~\ref{tab:result_NoisyCXR} of the main text.
\begin{table}[ht] 
     \centering 
     \caption{\textbf{Performance Comparison in NoisyCXR.} The best results, excluding the Oracle baseline, are highlighted in \textbf{bold}. Models are evaluated with 500 trials.} 
     
     \begin{tabular}{l|cccc} 
         \toprule  Method & ACC (↑) & AUC (↑) & $F_1$ (↑) & PR-AUC (↑)\\ 
         \hline 
         Vanilla-$D$ & 0.8248(0.0057) & 0.8151(0.0070) & 0.5043(0.0306) & 0.6275(0.0159) \\ 
         vanilla-$D_\rho$ & 0.8183(0.0090) & 0.8654(0.0049) & 0.3976(0.0680) & 0.6582(0.0137) \\ 
         Co-teaching  & 0.8177(0.0087) & 0.8646(0.0049) & 0.3936(0.0643) & 0.6564(0.0138) \\ 
         NCR  & 0.8160(0.0082) & 0.8642(0.0049) & 0.3781(0.0636) & 0.6557(0.0138) \\ 
         ALC(10\%)  & 0.8219(0.0088) & 0.8665(0.0048) & 0.4259(0.0639) & 0.6609(0.0136) \\ 
         \cline{2-5}  
         Pro-PT& \cellcolor{LightPink}0.8263(0.0079) & \cellcolor{LightPink}0.8586(0.0053) & \cellcolor{LightPink}0.5482(0.0602) & \cellcolor{LightPink}0.6430(0.0142)\\ 
         Pro-AT(10\%)  & \cellcolor{LightPink}0.8298(0.0061) & \cellcolor{LightPink}0.8611(0.0052) & \cellcolor{LightPink}0.5585(0.0444) & \cellcolor{LightPink}0.6502(0.0140)\\ 
         Pro-AT& \cellcolor{LightPink}{\bf 0.8340(0.0050)}  &\cellcolor{LightPink}{\bf 0.8675(0.0057)} & \cellcolor{LightPink}{\bf 0.5691(0.0348)} & \cellcolor{LightPink}{\bf 0.6628(0.0147)}\\ 
         \cline{2-5} 
         Oracle & 0.8350(0.0050) & 0.8703(0.0047) & 0.5739(0.0318) & 0.6682(0.0134)\\ 
         \bottomrule 
     \end{tabular} 
     \label{tab:result_NoisyCXR_ind} 
 \end{table}

\end{document}